\documentclass[12pt]{article}
\usepackage{latexsym,graphicx,a4,epsfig,psfrag,here}

\newcommand{\bda}{\begin{\displaymath}\begin{array}{rl}}
\newcommand{\eda}{\end{array}\end{displaymath}}
\newcommand{\be}{\begin{equation}}
\newcommand{\ee}{\end{equation}}
\newcommand{\bdm}{\begin{displaymath}}
\newcommand{\edm}{\end{displaymath}}
\newcommand{\bea}{\begin{eqnarray}}
\newcommand{\eea}{\end{eqnarray}}
\newcommand{\no}{\nonumber \\}
\newcommand{\fs}{\,.}
\newcommand{\co}{\,,}
\newcommand{\al}{&\!\!\!}
\newcommand{\eff}{{e\hspace{-0.1em}f\hspace{-0.18em}f}}
\newcommand{\ind}{\scriptscriptstyle}

\newcommand{\ubar}{\overline{\rule[0.42em]{0.4em}{0em}}\hspace{-0.5em}u}
\newcommand{\dbar}{\,\overline{\rule[0.65em]{0.4em}{0em}}\hspace{-0.6em}d}
\newcommand{\lbar}{\,\overline{\rule[0.75em]{0.4em}{0em}}\hspace{-0.45em}\ell}
\newcommand{\bbar}{\,\overline{\rule[0.72em]{0.4em}{0em}}\hspace{-0.4em}b}
\newcommand{\sbar}{\hspace{0.2em}\overline{\rule[0.42em]{0.4em}{0em}}
\hspace{-0.5em}s\hspace{0.1em}} 
\newcommand{\Wbar}{\,\overline{\rule[0.75em]{0.9em}{0em}}\hspace{-1em}W}
\newcommand{\Pbar}{\,\overline{\rule[0.75em]{0.5em}{0em}}\hspace{-0.7em}P}

\newcommand{\pbar}{\overline{\rule[0.5em]{0.4em}{0em}}\hspace{-0.5em}p}
\newcommand{\Ibar}{\,\overline{\rule[0.75em]{0.3em}{0em}}\hspace{-0.5em}
I\hspace{0.05em}}

\newcommand{\R}{{\mbox{\tiny R}}}
\renewcommand{\L}{{\mbox{\tiny L}}}
\newcommand{\SP}{\hspace{-0.03em}\rule[-0.2em]{0em}{0em}_{\ind SP}}
\newcommand{\lsim}{\,\raisebox{-0.3em}{$\stackrel{\raisebox{-0.1em}{$<$}}{\sim}$

}\,}

\newcommand{\lvac}{\langle 0|\,}
\newcommand{\rvac}{\,|0\rangle}

\newcommand{\Etwo}{\sqrt{\rule[0.1em]{0em}{0.5em}s_2}}
\newcommand{\rs}{\langle r^2\rangle\rule[-0.2em]{0em}{0em}_s}
\newcommand{\ltilde}{\tilde{\ell}}
\newcommand{\rtilde}{\tilde{r}}
\newcommand{\lhat}{\hat{\ell}}
\newcommand{\Ltilde}{\tilde{L}}
\newcommand{\dao}{\Delta a_0^0}
\newcommand{\dat}{\Delta a_0^2}
\newcommand{\rS}{r_{S_2}}
\newcommand{\rStilde}{\rtilde_{S_2}}
\newcommand{\g}{\rule{0.5em}{0em}}
\begin{document}

\thispagestyle{empty}
\begin{flushright}ZU-TH 3/01\\
BUTP--01/1\end{flushright}

\vspace{3cm}
\begin{center}{\LARGE {\Huge{\boldmath $\pi\pi$\unboldmath}} \bf
scattering}

\vspace{0.5cm}
March 8, 2001

\vspace{0.5cm}
G.~Colangelo$^a$,
J.~Gasser$^b$ and  
H.~Leutwyler$^b$
\vspace{2em}
\footnotesize{\begin{tabular}{c}
$^a\,$Institute for Theoretical Physics, University of 
Z\"urich\\
Winterthurerstr. 190, CH-8057 Z\"urich, Switzerland\\
$^b\,$Institute for Theoretical Physics, University of 
Bern\\   
Sidlerstr. 5, CH-3012 Bern, Switzerland
\end{tabular}  }

\vspace{1cm}

\begin{abstract}
We demonstrate that, together with the available experimental information, 
chiral symmetry determines the low energy behaviour of the $\pi\pi$
scattering
amplitude to within very small uncertainties. 
In particular, the threshold
parameters of the $S$--,  $P$--,  $D$-- and  $F$--waves are predicted, as
well
as the mass and width of the $\rho$ and of the broad bump in the $S$--wave. 
The implications for the coupling constants that occur in the effective
Lagrangian beyond leading order and also show up in other
processes, are discussed. Also, we analyze the dependence of various 
observables on the mass of the two lightest
quarks in some detail, in view of the extrapolations required to
reach the small physical masses on the lattice. The analysis  
relies on the standard hypothesis, according to which the quark condensate is
the leading order parameter of the spontaneously broken symmetry. Our results
provide the basis for an experimental test of this hypothesis, in particular
in the framework of the ongoing DIRAC experiment: The   
prediction for the lifetime of the ground state of a $\pi^+\pi^-$ atom 
reads $\tau=(2.90\pm 0.10)\,10^{-15}\,\mbox{sec}$.
\end{abstract}

\footnotesize{\begin{tabular}{ll}
{\bf{Pacs:}}$\!\!\!\!$& 11.30.Rd, 12.38.Aw, 12.39.Fe, 13.75.Lb\\
{\bf{Keywords:}}$\!\!\!\!$& Roy equations, 
Meson-meson interactions, Pion-pion scattering,\\& Chiral symmetries
\end{tabular}}

\end{center}
\newpage
\renewcommand{\baselinestretch}{0.6}

\tableofcontents

\renewcommand{\baselinestretch}{1}

\setcounter{equation}{0}
\renewcommand{\theequation}{\thesection.\arabic{equation}}
\section{Introduction}
\label{sec:intro}
The study of $\pi \pi$ scattering is a classical subject in the field of
strong interactions. The properties of the pions are intimately related to
an approximate symmetry of QCD. In the chiral limit, where $m_u$ and $m_d$
vanish, this symmetry becomes exact, the Lagrangian being invariant under
the group SU(2)$_\R\times$SU(2)$_\L$ of chiral rotations. The symmetry is
spontaneously broken to the isospin subgroup SU(2)$_{\mbox{\tiny V}}$. 
The pions represent the corresponding Goldstone bosons.

In reality, the quarks are not massless. The theory only possesses an
approximate chiral symmetry, because $m_u$ and $m_d$ happen to be very small.
The consequences of the fact that the symmetry breaking is small may
be worked out by means of an effective field theory \cite{Weinberg
Physica}. The various quantities of interest are expanded in powers 
of the momenta and quark masses. In the case of the pion mass, for instance,
the expansion starts with \cite{GMOR}  
\bea\label{eq:Mpi} M_{\pi^+}^2=(m_u+m_d)\,B+O(m^2)\co
\hspace{2em}B=\frac{1}{F^2}\,|\lvac 
\ubar u \rvac|\co\eea
where $F$ is the value of the pion decay constant in the chiral limit,
$m_u,m_d\rightarrow 0$.
The formula shows that the square of the pion mass is proportional to the
product of $m_u+m_d$ with the order parameter $\lvac 
\ubar u \rvac$. The two factors represent quantitative measures for explicit 
and spontaneous symmetry breaking, respectively. If the explicit symmetry
breaking is turned off, the pions do become massless, as they
should: The symmetry is then exact, so that 
the spectrum contains three massless Goldstone bosons, while all 
other levels form massive, degenerate isospin multiplets. 

The properties of the Goldstone bosons are
strongly constrained by chiral symmetry: In the chiral limit, 
the scattering amplitude
vanishes when the momenta of the pions tend to zero.  
To first order in the symmetry breaking,
the $S$--wave scattering lengths are proportional to the square of the pion
mass \cite{Weinberg 1966}: \be
\label{eq:weinberg}
a_0^0=\frac{7 M_\pi^2}{32\, \pi F_\pi^2}\co\hspace{2em} a_0^2=-
\frac{M_\pi^2}{16\,  \pi F_\pi^2} \co \ee 
where $a_\ell^I$ stands for the scattering length in
the isospin $I$ channel with angular momentum $\ell$.
The two low energy theorems (\ref{eq:weinberg}) are valid only at leading
order in a series expansion in
powers of the quark masses: The next--to--leading order corrections 
were calculated in \cite{GL 1983}, and even the
next--to--next--to--leading order corrections are now known \cite{BCEGS}. 

In the following, we exploit the fact that analyticity, unitarity and
crossing
symmetry impose further constraints on the scattering amplitude. 
These were analyzed in detail in \cite{ACGL}, on the basis of the Roy
equations \cite{Roy} and of the experimental data available at intermediate
energies. The upshot of that analysis is that 
$a_0^0$ and $a_0^2$ are the essential low energy parameters: Once these are
known, the available experimental data determine the low energy
behaviour of the $\pi\pi$ scattering amplitude to within
remarkably small uncertainties. As discussed above, chiral symmetry predicts 
exactly these two parameters. Hence 
the low energy behaviour of the scattering amplitude 
is fully determined by the experimental data in the intermediate
energy region and the theoretical properties just mentioned: 
analyticity, unitarity,
crossing symmetry and chiral symmetry. 

The resulting predictions for the $S$--wave scattering lengths were
presented already \cite{CGL}. The purpose of the present paper is to (i) 
discuss the analysis that underlies these predictions in more detail, (ii)
present the results for the 
threshold parameters of the $P$--, $D$--, and $F$--waves, (iii) give an
explicit representation for the $S$-- and $P$--wave phase shifts and
(iv) extract the information about the coupling
constants of the effective Lagrangian. 
 
Several authors have
performed a comparison of the chiral perturbation theory predictions
with the data, in particular also in view of a determination of the
effective 
coupling constants $\ell_1$ and $\ell_2$ \cite{GL 1984}--\cite{ABT_Ke4}. 
Stern and collaborators
\cite{Knecht Moussallam Stern Fuchs,Girlanda:1997ed} investigate the problem 
from a
different point of view, referred to as ``Generalized Chiral Perturbation
Theory''.  These authors treat the $S$--wave scattering lengths as
free parameters and investigate the possibility that their values
strongly deviate from those predicted by Weinberg.
In the language of the effective chiral Lagrangian, this scenario
would arise if the standard estimates for
the effective coupling constant $\ell_3$ were entirely wrong:
The quark condensate would then fail to represent the leading order 
parameter of the spontaneously broken chiral symmetry.
Indeed, these estimates rely on a theoretical picture that has not been 
tested experimentally. 

On the experimental side, the situation is the following.  
As shown in early numerical analyses of the Roy equations
\cite{roy-num}, only data
sufficiently close to threshold can provide significant bounds on the
scattering lengths. The often quoted values
$a_0^0=0.26\pm0.05$, $a_0^2=-0.028\pm 0.012$ \cite{Froggatt:1977hu,Nagels}
mainly rely on the $3\cdot 10^4$ $K\rightarrow\pi\pi e\nu$ decays
collected by the Geneva-Saclay collaboration, which 
provided its final
results in 1977 \cite{rosselet}. There are new data from
Brookhaven \cite{e865,Truol}, where more than $4\cdot 10^5$ $K_{e_4}$ decays
are being analyzed, and the low energy behaviour of the relevant form
factors is now also known much better \cite{BCG,ABT_Ke4}.
As will be discussed in section \ref{sec:GCHPT},
the preliminary results 
of this experiment indeed reduce the
uncertainties significantly. A similar experiment is proposed by the NA48
collaboration at CERN \cite{Batley}.
Unfortunately, the data taking at the
DA$\Phi$NE facility is delayed, due to technical problems with the 
accelerator. 
A beautiful experiment is under way at CERN \cite{Nemenov}, which is
based on the fact that $\pi^+\pi^-$ atoms decay into a pair of neutral pions,
through the strong transition $\pi^+\pi^-\!\rightarrow\!\pi^0\pi^0$. Since
the
momentum transfer nearly vanishes, only the scattering lengths are relevant:
At leading order in isospin breaking, the transition amplitude is 
proportional to
$a_0^0\!-\!a_0^2$. The corrections at
next--to--leading order are now also known~\cite{GLR}, as a result of
which a measurement of the
lifetime of a $\pi^+\pi^-$ atom amounts to a measurement of
this combination of scattering lengths.  Finally,
we mention the new data on pion production off nucleons, obtained by the
CHAOS collaboration at Triumf \cite{Triumf}. The scattering lengths may be
extracted from these data by means of a Chew-Low extrapolation
procedure.
Chiral symmetry, however, suppresses the one-pion exchange contribution with
a 
factor of $t$, so that a careful data selection is required to arrive at a
coherent Chew-Low fit.
It yet remains to be seen whether these data permit a significant
reduction of the uncertainties in the experimental determination of 
$a_0^0$ and $a_0^2$.

The experiments mentioned above are of particular interest,
because they offer a test of the hypothesis that the quark condensate
represents the leading order parameter of the spontaneously broken symmetry:
If the predictions obtained in the present paper should turn
out to be in contradiction with the outcome of these experiments,
the commonly accepted theoretical picture would require
thorough revision.

\setcounter{equation}{0}
\renewcommand{\theequation}{\thesection.\arabic{equation}}
\section{Chiral representation}
\label{sec:chiral representation}
Throughout the present paper
we work in the isospin limit: We disregard the e.m.~interaction and set
$m_u=m_d=m$. The various elastic reactions
among two pions may then be represented by a single scattering amplitude
$A(s,t,u)$. Only two of the Mandelstam variables are independent,
$s+t+u=4M_\pi^2$ and, as a consequence of Bose statistics, the amplitude is
invariant under an interchange of $t$ and $u$.

As discussed in detail in ref.~\cite{GL 1984}, chiral perturbation theory 
allows one to study the properties of the $\pi\pi$ scattering amplitude 
that follow from the occurrence of a spontaneously broken approximate 
symmetry.
The method is based on a systematic expansion 
in powers of the momenta and of the light quark masses.  
We refer to this as the chiral expansion 
and use the standard bookkeeping, which counts the quark masses like two 
powers of momentum, $m=O(p^2)$. 

The two loop representation of the scattering amplitude given in 
\cite{BCEGS}
yields the first three terms in the chiral expansion of the partial waves:
\be\label{eq:seriespw} t^I_\ell(s)=t^I_\ell(s)_2+ t^I_\ell(s)_4+
t^I_\ell(s)_6+O(p^8) \ee
At leading order, only the $S$-- and $P$--waves are different from zero:
\bea\label{eq:CA}  t^0_0(s)_2=\frac{2s - M_\pi^2}{32 \pi
  F_\pi^2}\,,\hspace{2em} 
     t^1_1(s)_2=\frac{s - 4M_\pi^2}{96 \pi F_\pi^2}\,,\hspace{2em}
     t^2_0(s)_2=-\frac{s -2 M_\pi^2}{32 \pi F_\pi^2}\,.
\eea
In the low energy expansion, inelastic reactions start showing up only at
$O(p^8)$. The unitarity condition therefore reads:
\bea \mbox{Im}\,t^I_\ell(s)=\sigma(s)\,|\hspace{0.04em}t^{I}_\ell(s)
\hspace{0.03em}|^2+O(p^8)\co\hspace{2em}
\sigma(s)= \sqrt{1-\frac{4M_\pi^2}{s}}
\fs\eea
The condition immediately implies that the imaginary parts of the two loop
amplitude may be worked out from the one-loop representation:
\be\label{eq:imchiral} \mbox{Im}\,t_\ell^I(s)=\sigma(s)\,t_\ell^I(s)_2\,
\left\{t_\ell^I(s)_2+ 2\,
\mbox{Re} \,t_\ell^I(s)_4\right\}+O(p^8)\ee
The formula shows that, at low energies,
the imaginary parts of
the partial waves with $\ell\geq 2$ are of order $p^8$
and hence beyond the accuracy of the two loop calculation.

Stated differently, the imaginary
part of the two loop representation is due exclusively to the $S$-- and 
$P$--waves.
This implies that, up to and including $O(p^6)$, the chiral representation 
of the scattering amplitude
only involves three functions of a single variable: 
\bea\label{eq:chiral decomposition} A(s,t,u)
\al=\al C(s,t,u)+32\pi\left\{\mbox{$\frac{1}{3}$}\,U^0(s)+
\mbox{$\frac{3}{2}$}\,(s-u)\,U^1(t)
+\mbox{$\frac{3}{2}$}\,(s-t)\,U^1(u)\right.\no
\al\al \left.+\mbox{$\frac{1}{2}$}\,U^2(t)+\mbox{$\frac{1}{2}$}\,U^2(u)
-\mbox{$\frac{1}{3}$}\, U^2(s) \right\}+O(p^8)\fs\eea
The first term is a crossing symmetric polynomial in $s$, $t$, $u$,
\be\label{eq:poly}
C(s,t,u)=c_1+s\,c_2+s^2\,c_3
+(t-u)^2\,c_4+s^3\,c_5
+s\,(t-u)^2\,c_6\fs \ee
The functions $U^0(s)$, $U^1(s)$ and $U^2(s)$ describe the 
``unitarity corrections'' associated with $s$--channel isospin $I=0,1,2$,
respectively. 
In view of the fact that the chiral perturbation theory representation
for the imaginary parts of the partial waves grows with the power
$\mbox{Im}\,t^I_\ell(s)_6\propto s^3$, we need to apply several subtractions
for the dispersive representation of these functions to converge. 
It is convenient to subtract at $s=0$ and to write the dispersion integrals 
in the form
\bea\label{eq:disp U}U^0(s)\al=\al \frac{s^4}{\pi}\int_{4M_\pi^2}^\infty
ds'\;\frac{\sigma(s')\,t_0^0(s')_2\,\{t_0^0(s')_2+ 2\,
\mbox{Re} \,t_0^0(s')_4\} }
{s^{\prime\,4}(s'-s )}
\co \no  U^1(s)\al=\al \frac{s^3}{\pi}\int_{4M_\pi^2}^\infty
ds'\;\frac{\sigma(s')\,t_1^1(s')_2\,
\{t_1^1(s')_2+ 2\,\mbox{Re} \,t_1^1(s')_4\}}
{s^{\prime\,3}(s'-4M_\pi^2)\,(s' - s ) }\co\\
U^2(s)\al=\al \frac{s^4}{\pi}\int_{4M_\pi^2}^\infty
ds'\;\frac{\sigma(s')\,t_0^2(s')_2\,\{t_0^2(s')_2+ 2\,
\mbox{Re} \,t_0^2(s')_4\}  }
{s^{\prime\,4}(s'-s)}\fs
\nonumber\eea
The subtraction constants are collected in the polynomial $C(s,t,u)$.
Alternatively, we could set $C(s,t,u)=0$ and book the subtraction
terms as polynomial contributions to $U^0(s)$, $U^1(s)$, $U^2(s)$.
The decomposition of $C(s,t,u)$ into a set of three
polynomials of a single variable is not unique, however, so that
we would have to adopt a convention for this splitting -- we find it more
convenient to work with the above representation of the amplitude.

The specific structure of the unitarity correction given above was noted
already in \cite{Knecht Moussallam Stern Fuchs}. It is straightforward 
to check that
the explicit result of the full two loop calculation described in 
\cite{BCEGS} is indeed of this structure. The essential result of that
calculation is the expression for the polynomial part of the amplitude,
in terms of the effective coupling constants. The corresponding formulae,
which specify how the coefficients $c_1,\ldots,\,c_6$ depend on the quark 
masses, are given in appendix \ref{sec:bcegs}. These, in particular
contain Weinberg's low energy theorem, which in this
language states that the expansion of the coefficients $c_1$ and $c_2$
starts with
\bea\label{eq:cW} c_1=-\frac{M_\pi^2}{F_\pi^2}\left\{1+O(M_\pi^2)\right\}\co
\hspace{2em}
c_2=\frac{1}{F_\pi^2}\left\{1+O(M_\pi^2)\right\}\fs\eea
The two loop calculation specifies the expansion of these two coefficients
up to and including next--to--next--to--leading order.

\setcounter{equation}{0}
\renewcommand{\theequation}{\thesection.\arabic{equation}}
\section{Phenomenological representation}
\label{sec:phenomenological representation}
As shown by Roy \cite{Roy}, the fixed-$t$ dispersion relations for the
isospin
amplitudes can
be written in such a form that they express the $\pi\pi$ scattering 
amplitude
in terms of the imaginary parts in the physical region of the $s$--channel.
The resulting representation for $A(s,t,u)$ contains 
two subtraction constants, which may be identified with the scattering
lengths $a^0_0$ and $a^2_0$. Unitarity converts this representation
into a set of coupled integral equations, which we recently examined
in great detail \cite{ACGL}. In the present context, the main
result of interest is that the representation allows us to determine the 
imaginary parts of the scattering amplitude in terms of $a_0^0$ and
$a_0^2$. Since the resulting representation is based on the available 
experimental information, we refer to it as the phenomenological
representation. 

In the following, we treat the imaginary parts of the
partial wave amplitudes as if they were completely known from phenomenology
-- we will discuss the uncertainties in these quantities as well as their
dependence on $a_0^0$ and $a_0^2$ in detail, once we have identified
the manner in which they enter our predictions for the scattering lengths.

The chiral representation shows that the singularities 
generated by the imaginary parts of the partial waves with $\ell\geq2$ 
start 
manifesting themselves only at $O(p^8)$. Accordingly, we may
expand the corresponding contributions to the dispersion integrals
into a Taylor series of the momenta. The singularities due to the imaginary
parts of the $S$-- and $P$--waves, on the other hand, start manifesting
themselves already at $O(p^4)$ -- these cannot be replaced
by a polynomial. The corresponding contributions to the amplitude
are of the same structure as the unitarity corrections and also involve
three functions of a single variable. It is convenient to
subtract the relevant dispersion integrals in the same manner as for the
chiral representation:
\bea\label{eq:dispWbar}\Wbar^0(s)\al=\al 
\frac{s^4}{\pi}\int_{4M_\pi^2}^{\infty}
ds'\;\frac{\mbox{Im}\,t_0^0(s')}
{s^{\prime\,4}(s'-s )}
\co \no  \Wbar^1(s)\al=\al \frac{s^3}{\pi}\int_{4M_\pi^2}^{\infty}
ds'\;\frac{\mbox{Im}\,t_1^1(s')}
{s^{\prime\,3}(s'-4M_\pi^2)\,(s' - s ) }\co\\
\Wbar^2(s)\al=\al \frac{s^4}{\pi}\int_{4M_\pi^2}^{\infty}
ds'\;\frac{\mbox{Im}\,t_0^2(s')}
{s^{\prime\,4}(s'-s)}\fs
\nonumber\eea
Since all other contributions can be replaced by a polynomial, the
phenomenological amplitude takes the form
\bea\label{eq:phenrep} A(s,t,u)\al=\al 16\pi a_0^2+
\frac{4\pi }{3M_\pi^2}\,(2a_0^0-5a_0^2)\,s+
\Pbar(s,t,u) \\\al\al +32\pi\left\{\mbox{$\frac{1}{3}$}
\Wbar^0(s)+\mbox{$\frac{3}{2}$}(s-u)\Wbar^1(t)
+\mbox{$\frac{3}{2}$}(s-t)\Wbar^1(u)\right.\no
\al\al \left.\hspace{2.3em}+\mbox{$\frac{1}{2}$}\Wbar^2(t)+
\mbox{$\frac{1}{2}$}\Wbar^2(u)
-\mbox{$\frac{1}{3}$} \Wbar^2(s) \right\}+O(p^8)\,.\nonumber\eea
We have explicitly displayed the contributions from the subtraction 
constants
$a_0^0$ and $a_0^2$.
The term $\Pbar(s,t,u)$ is a crossing symmetry polynomial
\bea \Pbar(s,t,u)\al=\al \pbar_1+\pbar_2\,s+\pbar_3\,s^2+\pbar_4\,(t-u)^2+
\pbar_5\,s^3+\pbar_6\,s(t-u)^2\fs\eea
As demonstrated in the appendix, its coefficients can be
expressed in terms of the following integrals over the imaginary parts of 
the partial waves:
\bea\label{eq:Inbar} \Ibar_n^I\al=\al
\sum_{\ell=0}^\infty \,\frac{(2l+1)}{\pi}\int_{4M_\pi^2}^{\infty}\!
ds\;\frac{\mbox{Im}\,t^I_\ell(s)}{s^{n+2}(s-4M_\pi^2)}\co\\
H\al=\al
\sum_{\ell=2}^\infty\,
(2l+1)\,\ell(\ell+1)\,\frac{1}{\pi}\int_{4M_\pi^2}^\infty\!
\!ds\;\frac{2\,\mbox{Im}\,t^0_\ell(s)+4\,\mbox{Im}\,t^2_\ell(s)}{9\,s^{3}
(s-4M_\pi^2)}\fs\nonumber\eea
The explicit expressions read
\bea\label{eq:pbar}\al\al\hspace{-0.4em}\pbar_1=-128\pi M_\pi^4\left(\Ibar^1_0+
\Ibar^2_0 +2 M_\pi^2\Ibar^1_1+
2 M_\pi^2\Ibar^2_1+
8 M_\pi^4\Ibar^2_2\right),\no
\al\al\hspace{-0.4em}\pbar_2= -\frac{64\pi M_\pi^2}{3}
\left(2 \Ibar^0_0- 6 \Ibar^1_0- 2 \Ibar^2_0-15 M_\pi^2 \Ibar^1_1-
3 M_\pi^2 \Ibar^2_1 - 36 M_\pi^4 \Ibar^2_2+6 M_\pi^2 H\right),\no
\al\al\hspace{-0.4em}\pbar_3=
\frac{8\pi}{3}\!\left(4\Ibar^0_0-9\Ibar^1_0-\Ibar^2_0
      -16 M_\pi^2\Ibar^0_1 - 42 M_\pi^2 \Ibar^1_1 + 
22 M_\pi^2\Ibar^2_1     -72 M_\pi^4\Ibar^2_2   +
24 M_\pi^2 H\right),\no
\al\al\hspace{-0.4em}\pbar_4= 8 \pi \left(\Ibar^1_0+\Ibar^2_0+
   2 M_\pi^2 \Ibar^1_1 +2 M_\pi^2 \Ibar^2_1-
24 M_\pi^4 \Ibar^2_2\right),\\
\al\al\hspace{-0.4em}\pbar_5=
\frac{4 \pi}{3}\left(8 \Ibar^0_1+9 \Ibar^1_1-11 \Ibar^2_1-
32 M_\pi^2 \Ibar^0_2+44 M_\pi^2 \Ibar^2_2-6 H\right),\no
\al\al\hspace{-0.4em}\pbar_6= 4 \pi \left(\Ibar^1_1-3 \Ibar^2_1+
12 M_\pi^2 \Ibar^2_2+2 H\right) .
\nonumber\eea
The fact that, at low energies, the scattering amplitude may be represented
in terms of integrals over the imaginary parts that can be evaluated
phenomenologically, was noted earlier, by Stern and collaborators 
\cite{Knecht Moussallam Stern Fuchs}. These authors also worked out the 
implications for the threshold parameters and the effective
coupling constants of the chiral Lagrangian and we will compare their results
with ours, but we first need to specify the framework we are using.

\setcounter{equation}{0}
\renewcommand{\theequation}{\thesection.\arabic{equation}}
\section{Matching conditions}
\label{sec:matching}
In the preceding sections, we have set up two different 
representations of 
the scattering amplitude: One based on chiral perturbation theory
and one relying on the Roy equations.
The purpose of the present section is to show
that, in their common domain of validity, the two representations agree,
provided the parameters occurring therein
are properly matched.

The chiral and phenomenological representations are of the same 
structure.
The coefficients of the polynomials $C(s,t,u)$ and $\Pbar(s,t,u)$ are 
defined differently and, instead of the functions $U^I(s)$ occurring 
in the chiral
representation, the phenomenological one involves the 
functions $\Wbar^I(s)$. The latter are defined in eq.~(\ref{eq:dispWbar}), as
integrals over the imaginary parts of the physical $S$-- and $P$--waves. 

The key observation is that, in the integrals (\ref{eq:dispWbar}), 
only the region where $s'$ is of order $p^2$ matters for
the comparison of the two representations. The remainder generates 
contributions to the amplitude that are most of order $p^8$. 
Moreover, for small values of $s'$, the quantities $\mbox{Im}\,t^I_\ell(s')$
are given by the chiral representation in eq.~(\ref{eq:imchiral})
except for contributions 
that again only manifest themselves at $O(p^8)$. This implies that
the differences between the functions $\Wbar^I(s)$ and $U^I(s)$
are beyond the accuracy of the chiral representation:
\bea \Wbar^0(s)\al=\al U^0(s)+O(p^8)\co\no 
\Wbar^1(s)\al=\al U^1(s)+O(p^6)\co\\
\Wbar^2(s)\al=\al U^2(s)+O(p^8)\fs\nonumber\eea
Hence the two representations agree if and only if the polynomial parts
do,
\bea C(s,t,u)=16\pi a_0^2+
\frac{4\pi }{3M_\pi^2}\,(2a_0^0-5a_0^2)\,s+\Pbar(s,t,u)+O(p^8)\fs
\nonumber\eea 
This implies that the coefficients of $C(s,t,u)$ and $\Pbar(s,t,u)$ are
related by 
\bea\label{eq:mc}\begin{array}{lcllcl} 
c_1\al=\al  16\pi a_0^2+\pbar_1+O(p^8)\co\hspace{2em}\al
c_2\al=\al {\displaystyle \frac{4\pi }{3M_\pi^2}}\,(2a_0^0-5a_0^2)+
\pbar_2+O(p^6)\co\\
c_3\al=\al \rule{0em}{1.4em}\,\pbar_3+O(p^4)\co\hspace{2em}\al c_4\al=\al
\,\pbar_4+O(p^4)\co\hspace{2em}\\
c_5\al=\al \rule{0em}{1.8em}\,\pbar_5+O(p^2)\co\hspace{2em}\al 
c_6\al=\al\,\pbar_6+O(p^2)\fs
\end{array}
 \eea

The chiral representation specifies the coefficients 
$c_1,\ldots,\,c_6$ in terms of the effective coupling constants,
while the quantities $\pbar_1,\ldots,\,\pbar_6$ are experimentally
accessible.
Since the main uncertainties in the latter arise from the poorly
known values of the scattering lengths $a^0_0$, $a^2_0$,
the above relations essentially determine the coefficients $c_1,\ldots,\,c_6$
in terms of these two parameters.

\setcounter{equation}{0}
\section{\hspace{-0.5em}Symmetry breaking in the effective Lagrangian}
\label{sec:eff}

As discussed in section \ref{sec:chiral representation}, 
unitarity fully determines the
scattering amplitude to third order of the chiral expansion, in terms of
the couplings constants occurring in the derivative expansion of the 
effective Lagrangian to $O(p^6)$,
\be {\cal L}_{\eff}={\cal L}_2+{\cal L}_4+{\cal L}_6+\ldots\ee
The leading term ${\cal L}_2$ only contains $F$ and $M^2\equiv 2\,m\,B$.
The vertices relevant for $\pi\pi$ scattering involve the coupling
constants $\ell_1,\ell_2,\ell_3,\ell_4$ from ${\cal L}_4$, and
${\cal L}_6$ generates 6 further couplings: $r_1,\ldots\,,r_6$.
We need to distinguish two different categories of coupling constants:
\begin{description}\item {\it a. Terms that survive in
the chiral limit.} Four of the coupling constants that enter the two loop
representation of the scattering amplitude belong to this category:
$\ell_1,\,\ell_2,\,r_5,\,r_6$.
\item {\it b. Symmetry breaking terms.}
The corresponding vertices are proportional to a power
of the quark mass and involve the coupling constants $\ell_3$, $\ell_4$,
$r_1$, $r_2$, $r_3$, $r_4$.\end{description}
The constants of the first category show up in the momentum dependence of 
the scattering amplitude, so that these couplings may be determined
phenomenologically. The symmetry breaking terms, on the other hand,
specify the dependence of the amplitude on the quark masses. Since these 
cannot
be varied experimentally, information concerning the second category
of coupling constants can only be obtained from sources other than $\pi\pi$
scattering. In part, we are relying on theoretical estimates here.
Although these are
rather crude, the uncertainties do not significantly affect our results, for
the following reason.

The quark masses $m_u,\,m_d$, which are responsible for the symmetry 
breaking
effects, are very small compared to the intrinsic scale $\Lambda$ of the
theory, which is of order 500 MeV or 1 GeV.
The group SU(2)$_\R\times$SU(2)$_\L$ therefore represents a nearly perfect
symmetry of the QCD Hamiltonian.
In the isospin limit, the symmetry breaking effects are
controlled by the ratio
$m/\Lambda$, with $m=\frac{1}{2}\,(m_u+m_d)$.
In view of $m \simeq 5
\,\mbox{MeV}$, the expansion parameter is of the order of
$10^{-2}$, indicating that the expansion converges very rapidly.

In the framework of the effective theory, it is convenient to replace
powers of $m$ by powers of $M_\pi^2$ and to identify the intrinsic scale
$\Lambda$ with $4\,\pi\, F_\pi$. The expansion parameter $m/\Lambda$ is
then replaced by \be \xi=\left(\frac{M_\pi}{4\,\pi\,
F_\pi}\right)^{\!\!2}\fs\ee The numerical value\footnote{ Throughout this
paper, we identify $M_\pi$ with the mass of the charged pion and use $
F_\pi=92.4\,\mbox{MeV}$ \cite{Holstein}.}  $\xi=1.445\cdot 10^{-2}$
confirms the estimate just given.

We know of only one mechanism that can upset the above crude order of
magnitude estimate for the symmetry breaking effects: The perturbations
generated by the quark mass term in the QCD Hamiltonian, $m_u\,\ubar u +m_d
\,\dbar d$, may be enhanced by small energy denominators. Indeed, small
energy denominators do occur:

(i) In the chiral limit, the pions are massless, so that the
straightforward expansion in powers of the quark masses leads to infrared
singularities. For a finite pion mass, these singularities are cut off at a
scale of the order of $M_\pi$ and the divergences are converted to finite
expressions that involve the logarithm of $M_\pi$.  The most important
contributions of this type are generated by the vertices contained in the
leading order effective Lagrangian, which are fully determined by $F_\pi$
and $M_\pi$. Accordingly, the coefficients of the leading chiral logarithms
do not involve any unknown constants. In those cases where this coefficient
happens to be large, the symmetry breaking effects are indeed enhanced, so
that the above rule of thumb estimate then fails.

(ii) States that remain massive in the chiral limit may give rise to small
energy denominators if their mass happens to be small. In the framework of
chiral perturbation theory, the occurrence of such states manifests itself
only indirectly, through the fact that some of the effective coupling
constants are comparatively large.  The $\rho$--meson represents the most
prominent example and it is well-known that some of the coupling constants
(for instance $\ell_1$ and $\ell_2$) are dominated by the contribution from
this state \cite{GL 1984}.  In fact, for all of those effective couplings
that have been determined experimentally, the observed magnitude is well
accounted for by the hypothesis that they are dominated by the resonances
seen at low energies \cite{EGPdeR}.

\setcounter{equation}{0}
\renewcommand{\theequation}{\thesection.\arabic{equation}}
\section{Low energy theorems}
\label{sec:let}
As the two loop formulae are rather lengthy, we first discuss the
principle used to arrive at the prediction for the $S$--wave scattering
lengths at one loop level, where the algebra is quite simple.
The first order corrections to the two low energy theorems (\ref{eq:cW}) are
readily obtained from the formulae given in appendix \ref{sec:bcegs}. 
Expressed in terms of the scale invariant effective coupling constants
$\lbar_1,\ldots\,, \lbar_4$ introduced in \cite{GL 1983}, the result
reads:
\bea\label{eq:c12oneloop}
c_1\al=\al -\frac{M_\pi^2}{F_\pi^2}\left\{1+\xi\left(-
\frac{4}{3}\,\lbar_1 +
\frac{1}{2}\,\lbar_3+2\,\lbar_4-\frac{197}{210} \right)
+O(\xi^2)\right\}\co\no 
c_2\al=\al \frac{1}{F_\pi^2}\left\{1+\xi\left(-
\frac{4}{3}\,\lbar_1 + 2\,\lbar_4-\frac{67}{140}\right)
+O(\xi^2)\right\}\fs\eea
The corrections involve both types of couplings: $\ell_1$ is of 
type {\it a.} and can thus be determined from the momentum dependence of the
scattering amplitude, while $\ell_3$ and $\ell_4$ are of type {\it b.}. 
Indeed, both $\ell_1$ and $\ell_2$ show up in the terms
proportional to $s^2$ and $(t-u)^2$:
\bea
c_3\al=\al \frac{1}{(4 \pi F_\pi)^2}\left\{\frac{\lbar_1}{3} +
\frac{\lbar_2}{6}-\frac{47}{84} \right\}
+O(\xi)\,,\hspace{0.5em} 
c_4=\frac{1}{(4 \pi F_\pi)^2}\left\{\frac{\lbar_2}{6} -
\frac{127}{840}\right\}+O(\xi)\,.\nonumber\eea
These formulae show that, up to and including terms of order $\xi$,
the quantities
\bea \label{eq:defC12}
\hspace*{-6em}C_1\al\equiv \al F_\pi^2\left\{c_2+4M_\pi^2(c_3-c_4)\right\},
\hspace{1em}C_2\equiv \frac{F_\pi^2}{M_\pi^2}
\left\{-c_1+4M_\pi^4(c_3-c_4)\right\}\eea
exclusively contain the symmetry breaking couplings $\ell_3$ and $\ell_4$:
\bea \hspace*{-2em}
C_1\al=\al 1+\xi \left\{2\,\bar{\ell}_4-
\frac{887}{420}\right\}+O(\xi^2)\,,\hspace{0.5em}
C_2= 1+\xi \left\{\frac{\bar{\ell}_3}{2}+2\,\bar{\ell}_4-
\frac{18}{7}\right\}+O(\xi^2)\,.\nonumber\eea

In the following, we analyze the low energy theorems for the $S$--wave
scattering lengths by means of the quantities
$C_1$ and $C_2$ defined in eq.~(\ref{eq:defC12}). 
The one for $2a_0^0-5a_0^2$, for instance, 
is obtained by inserting the matching relations (\ref{eq:mc})
in the definition of $C_1$ and solving for the scattering
lengths. The result reads
\bea\label{eq:a20C} 2a^0_0-5a_0^2\al =\al \frac{3M_\pi^2}{4\pi F_\pi^2}\,C_1
+
M_\pi^4\,\alpha_1+O(M_\pi^8)\co\eea
where $\alpha_1$ collects the contributions from the phenomenological
moments,
\bea \label{eq:alpha1}
\alpha_1\al =\al 16\,M_\pi^2 (8\,\Ibar^0_1+9\,\Ibar^1_1-11\,\Ibar^2_1
-36 M_\pi^2\,\Ibar^2_2-6\, H)\fs\eea
The analogous low energy theorems for $a_0^0$ and $a_0^2$ 
read
\bea \label{eq:aC}
a^0_0\al =\al \frac{7M_\pi^2}{32\pi F_\pi^2}\,C_0 +M_\pi^4\,\alpha_0
+O(M_\pi^8)\co\\
a^2_0\al =\al- \frac{M_\pi^2}{16\pi F_\pi^2}\,C_2 +M_\pi^4\,\alpha_2
+O(M_\pi^8)\co\nonumber
\eea
where $C_0$ is a combination of $C_1$ and $C_2$,
\be\label{eq:C0} C_0=\frac{1}{7}\,(12\, C_1-5 \,C_2)\co\ee
while $\alpha_0$, $\alpha_2$ again stand for a collection of moments
\bea\label{eq:alpha02}
\alpha_0\al=\al\frac{4}{3}\,(5\,\Ibar^0_0+10\,\Ibar^2_0+
28M_\pi^2\Ibar^0_1
+24M_\pi^2\Ibar^1_1-16 M_\pi^2\Ibar^2_1-96M_\pi^4 \Ibar^2_2+6M_\pi^2 H)\co
\no
\alpha_2\al=\al\frac{8}{3}\,(\Ibar^0_0+2\,\Ibar^2_0-4M_\pi^2\Ibar^0_1
-6M_\pi^2\Ibar^1_1+10 M_\pi^2\Ibar^2_1+24M_\pi^4 \Ibar^2_2+6M_\pi^2 H)\fs
\eea
The relations (\ref{eq:defC12})--(\ref{eq:alpha02}) specify the $S$--wave 
scattering lengths in terms of $C_1$, $C_2$ and the phenomenological moments
$\Ibar^I_n$ and $H$. Note that these contain infrared singularities. 
Their chiral expansion starts with the contributions generated
by the square of the tree level amplitudes:
\bea\label{eq:Ibar chiral} \Ibar^0_1\al=\al\frac{101}{M_\pi^2 K}+ O(1)\,,
\hspace{2em}\Ibar^0_2=\frac{227}{14M_\pi^4 K}+ O(M_\pi^{-2})\,,\no
\Ibar^1_1\al=\al\frac{2}{M_\pi^2 K}+O(1)\,,\hspace{2em}
\Ibar^1_2=\frac{1}{7M_\pi^4 K}+ O(M_\pi^{-2})\,,\\
\Ibar^2_1\al=\al\frac{14}{M_\pi^2 K}+O(1)\,,\hspace{2em}
\Ibar^2_2=\frac{13}{7M_\pi^4 K}+ O(M_\pi^{-2})\,,\no
\rule{0em}{1.2em}
H\al=\al O(1)\,,\hspace{6em} K\equiv 61440\, \pi^3 F_\pi^4\,.
\nonumber\eea

The evaluation of the moments requires phenomenological information. 
Since the behaviour of the
imaginary parts near threshold is sensitive to the scattering lengths we
are looking for, the same applies for these moments. In the narrow range of
interest, the dependence is well described by the quadratic formulae in
appendix \ref{sec:moments}, which yield
\bea\label{eq:alphanum} M_\pi^4\alpha_0\al=\al .0448 
+.30\, \Delta a_0^0
-.37\, \Delta a_0^2
+ .5\, (\Delta a_0^0)^2
-1.2\, \Delta a_0^0 \Delta a_0^2
+1.8\, (\Delta a_0^2)^2\,\no
M_\pi^4\alpha_1\al=\al .0619
+.48\, \Delta a_0^0
-.26\, \Delta a_0^2
+ .86\, (\Delta a_0^0)^2
-1.7\, \Delta a_0^0 \Delta a_0^2 
+.3\, (\Delta a_0^2)^2\,\no
M_\pi^4\alpha_2\al=\al .00553
+.023\, \Delta a_0^0
-.095\, \Delta a_0^2
-.1\, \Delta a_0^0 \Delta a_0^2
+.7\, (\Delta a_0^2)^2\co
\eea
with $\Delta a_0^0=a_0^0-0.225$, $\Delta a_0^2=a_0^2+0.03706$.

\setcounter{equation}{0}
\section{The coupling constants {\boldmath $\ell_3$\unboldmath} and
  {\boldmath$\ell_4$\unboldmath}} \label{sec:l34}

The representation of the $S$--wave scattering lengths derived in the 
preceding
section splits the correction to Weinberg's leading order formulae
into two parts: a correction factor $C_n$, which at first nonleading order
only involves the coupling constants $\ell_3$ and $\ell_4$
and a term $\alpha_n$ that can be determined on phenomenological grounds. 

The significance of the coupling constants  $\ell_3$ and $\ell_4$
is best seen in the expansion of $M_\pi$ and $F_\pi$
in powers of the quark mass. The relation of Gell-Mann, Oakes and Renner 
\cite{GMOR} states that the expansion of $M_\pi^2$ starts with a term 
linear in $m$. The coupling constant $\ell_3$ determines the first order 
correction:
\bea\label{eq:Mpiexp} M_\pi^2\al=\al M^2\,\{1-\mbox{$\frac{1}{2}$}\,
\xi\,\lbar_3
+O(\xi^2)\}\co\hspace{2em}M^2\equiv 2\, B\,m\fs\eea
The constant $B$ stands for the value of $|\lvac \ubar u \rvac|/F_\pi^2$
in the chiral limit.
Note that $\lbar_3$ contains a chiral logarithm, 
$\lbar_3= -\ln M_\pi^2+O(1)$.  
The coupling constant $\lbar_4$, which also contains
a chiral logarithm with unit coefficient, 
$\lbar_4= -\ln M_\pi^2+O(1)$, is the analogous term in the expansion of 
the pion decay constant, 
\bea\label{eq:Fpiexp} F_\pi=F\,\{1+\xi\,\lbar_4+O(\xi^2) \}\co\eea
where $F$ is the value of $F_\pi$ in the chiral limit. 

The same two coupling constants also show up in the
scalar form factor
\bea \langle\pi(p')\,|\, m_u \,\ubar u + m_d \,\dbar d
\,|\,\pi(p) \rangle = \sigma_\pi
\,\{1+\mbox{$\frac{1}{6}$}\,\rs\,t +O(t^2) \} \fs\eea
The value of the matrix element at $t=0$ is the pion $\sigma$--term. 
According to the Feynman-Hellman theorem, it is given by 
$\sigma_\pi=m\,\partial M_\pi^2/\partial m$. The relation
(\ref{eq:Mpiexp}) thus shows that $\ell_3$ also determines the
$\sigma$--term to first nonleading order:
\bea \sigma_\pi= M_\pi^2\,\{1-\mbox{$\frac{1}{2}$}\,\xi\,(\lbar_3-1)+
\xi^2\,\Delta_\sigma+O(\xi^3) \}\fs\eea
Moreover, chiral symmetry implies that the same coupling constant that 
determines the difference between $F_\pi$ and $F$ also fixes the 
scalar radius at leading order of the chiral expansion \cite{GL 1984}:
\bea\label{eq:rs} \rs = \frac{3}{8\pi^2 F_\pi^2}
\left\{\lbar_4-\frac{13}{12}+\xi\,\Delta_r+O(\xi^2)\right\}\fs
\eea
We may therefore eliminate $\ell_4$ in favour of the scalar radius
and rewrite the correction factors in the form
\bea\label{eq:letcr} 
C_0 \al=\al 1+
\frac{M_\pi^2}{3}\,\rs-\frac{5\,\xi}{14}\left\{\lbar_3 -\frac{563}{525}
\right\}+\xi^2 \Delta_0+O(\xi^3)\co\no
C_1 \al=\al 1+
\frac{M_\pi^2}{3}\,\rs+\frac{23\,\xi}{420}+\xi^2 \Delta_1+O(\xi^3)\co\\
C_2\al=\al
1+\frac{M_\pi^2}{3}\,\rs+ \frac{\xi}{2}\left\{\lbar_3-
\frac{17}{21}\right\} +\xi^2 \Delta_2+O(\xi^3)\nonumber \co\eea
with $\Delta_0\equiv (12\,\Delta_1-5\,\Delta_2)/7$.
The first order corrections are then determined by $\rs$ and $\ell_3$, 
while
$\Delta_0$, $\Delta_1$ and $\Delta_2$ represent the two loop contributions.
The scalar form factor is also known to two loops 
\cite{Bijnens Colangelo Ecker}. The explicit expressions for 
the second order corrections are given in appendix \ref{sec:NNL}.

For the numerical value of the scalar radius, we rely
on the dispersive evaluation of the scalar form factor described in 
ref.~\cite{DGL}.
We have repeated that calculation with the information about the
phase shift $\delta_0^0(s)$ obtained in ref.~\cite{ACGL}.
In view of the strong final state interaction in the 
$S$--wave, the scalar radius is significantly larger than the electromagnetic
one, $\langle r^2\rangle_{\ind  e.m.}= 0.439\pm 0.008 \,\mbox{fm}^2$ 
\cite{Amendolia}. The result reads
\be\label{eq:rsnum} \rs=0.61\pm 0.04\,\mbox{fm}^2\co\ee 
where the error is our estimate of
the uncertainties to be attached to the dispersive calculation.
The number confirms the value given in ref.~\cite{DGL} 
and is consistent with earlier 
estimates of the low energy constant $\ell_4$, based on the
symmetry breaking seen in $F_K/F_\pi$ or on the decay $K\rightarrow
\pi\ell\nu$ \cite{GL 1985}, but is more accurate. It corresponds
to $\frac{1}{3}M_\pi^2\,\rs= 0.102 \pm 0.007$, so that
the contribution from the scalar radius represents a correction of  
order 10\%, in $C_0$, $C_1$, as well as in $C_2$. 

The crucial parameter that distinguishes the standard framework from the 
one 
proposed in ref.\ \cite{Knecht Moussallam Stern Fuchs} is $\ell_3$. 
The value of this coupling constant is not known accurately. 
Numerically, however, a significant change in the prediction for the 
scattering lengths can only arise if the crude estimate
\be\label{eq:l3barnum} \lbar_3=2.9\pm 2.4\ee 
given in ref.~\cite{GL 1984} should turn out to be entirely wrong: 
With this estimate, the contribution from $\ell_3$ to $a^0_0$ and $a_0^2$
is of order $0.002$ and $0.001$, respectively.
We do not make an attempt at reducing the uncertainty
in $\ell_3$ within the standard framework, because it barely affects our 
final result. Instead, we will explicitly 
display the sensitivity of the outcome to this coupling 
constant.

\setcounter{equation}{0}
\section{Results for {\boldmath$a_0^0$\unboldmath} and {\boldmath$a_0^2$
\unboldmath} at one loop level}
\label{sec:one loop}

We first drop the two loop corrections $\Delta_n$. Inserting the values
$\rs=0.61 \,\mbox{fm}^2$ and $\lbar_3=2.9$, the low energy theorems
(\ref{eq:letcr}) yield
\be\label{C one loop} 
C_0=1.092\co\hspace{2em}C_1=1.103\co\hspace{2em}C_2=1.117\fs\ee
The correction factor $C_1$ is fully determined by the contribution from
the scalar radius. The numerical values of $C_0$ and $C_2$
differ little from $C_1$: The estimate (\ref{eq:l3barnum}) implies that the 
contributions from the coupling constant $\ell_3$ are very small, so that
these terms are also dominated by the scalar radius.
Inserting the values (\ref{eq:alphanum}), (\ref{C one loop}) 
in the relations (\ref{eq:aC}) and solving for
$a_0^0,a_0^2$, we then get
\bea\label{eq:one loop} a_0^0    =  0.2195\,,\hspace{0.5em}a^2_0 = -0.0446\,,
\hspace{0.5em}
2a_0^2-5a^2_0=0.662\,.\eea

These numbers are somewhat different from those obtained in \cite{GL 1983},
which are also based on the one loop representation of the
scattering amplitude. In fact,
even if the two loop corrections $\Delta_n$  
are dropped, the formulae (\ref{eq:aC}) for 
the $S$--wave scattering lengths differ from those given in 
ref.~\cite{GL 1983}.
In the case of $a_0^0$, for example, the formula given there reads
\bea a^0_0=\frac{7M_\pi^2}{32\pi F_\pi^2}\left\{1+\frac{M_\pi^2}{3}\,\rs
-\frac{5\, \xi}{14}\left(\lbar_3-\frac{353}{15}\right)\right\} +
\frac{25}{4} M_\pi^4(a^0_2+2a^2_2)+O(M_\pi^6),\nonumber
\eea 
where $a^0_2$ and $a^2_2$ are the $D$--wave scattering lengths.
As far as the contributions proportional to $\rs$ and $\ell_3$ are 
concerned,
the expression is the same, but 
instead of the phenomenological moments contained in $\alpha_0$, the
above formula contains the term
\bea \alpha_0\;\longleftrightarrow\;\frac{25}{4}\,(a^0_2+2a^2_2)+
\frac{737}{6720\,\pi^3F_\pi^4}\fs\eea
Indeed, the $D$--wave scattering lengths may be expressed in terms of
moments, up to and including contributions of first nonleading order.
Projecting the phenomenological representation (\ref{eq:phenrep}) onto the
$D$--waves, we find that the functions $\Wbar^I(s)$
do not contribute to the scattering lengths, while the contribution from
the background polynomial reads
\bea\label{eq:aDI}
\hspace*{-2em} a^0_2\al=\al \frac{16}{45}\,\left\{\Ibar^0_0+3\Ibar^1_0
+5\Ibar^2_0-4M_\pi^2(\Ibar^0_1-3\Ibar^1_1+5\Ibar^2_1)+30 M_\pi^2 H\right\}
+O(M_\pi^4)\,,\\
\hspace*{-2em} a^2_2\al=\al \frac{8}{45}\,\left\{2\Ibar^0_0-3\Ibar^1_0
+\Ibar^2_0-4M_\pi^2(2\Ibar^0_1+3\Ibar^1_1+\Ibar^2_1)+24 M_\pi^2 H\right\}
+O(M_\pi^4)\,.\nonumber\eea
The comparison with the exact representation for the $D$--wave scattering
lengths given in \cite{ACGL} shows that the contributions from
the imaginary parts of the $S$-- and $P$--waves 
can be represented in terms of the moments and the coefficients
agree with those above. The formula (\ref{eq:aDI}) includes the contributions
from the higher partial waves, up to and including corrections of first
nonleading order.
In the difference,
\bea \Delta a_0^0=  M_\pi^4\left\{\alpha_0-\frac{25}{4}\,(a^0_2+2a^2_2)-
\frac{737}{6720\,\pi^3F_\pi^4}\right\}\co\nonumber\eea
the leading moments cancel, but the terms with $\Ibar^I_1$, $\Ibar^I_2$ and
$H$ remain:
\bea \Delta a_0^0 = -
\frac{737\,M_\pi^4}{6720\,\pi^3F_\pi^4}
+8\,M_\pi^6\,(8\,\Ibar^0_1+4\,\Ibar^1_1+4\,\Ibar^2_1-H)
-128\,M_\pi^8\Ibar^2_2\fs\eea
The low energy expansion of the moments of eq.~(\ref{eq:Ibar chiral})
shows that the contributions of $O(M_\pi^4)$
in $\Delta a_0^0$ indeed cancel out, demonstrating that
the formula given in ref.~\cite{GL 1983} agrees with our representation,
up to terms that are beyond the algebraic accuracy of that formula.

Numerically, however, the leading order terms represent a rather poor
approximation for the moments, so that there is a numerical difference:
The numerical values of the moments are given in appendix \ref{sec:moments}.
Inserting these in 
(\ref{eq:aDI}), we obtain
$a^0_2=1.76\cdot 10^{-3}\,M_\pi^{-4}$, $a^2_2=0.171 \cdot
10^{-3}\,M_\pi^{-4}$, so that the one loop formula of ref.~\cite{GL 1983}
yields $a_0^0=0.205$, instead of the value $a_0^0= 0.2195$ 
given above. 
The difference arises because we are matching the chiral and 
phenomenological
representations differently: We represent the amplitude in terms of 
three functions of a single variable $s$ and match the coefficients of 
the Taylor expansion at $s=0$. In ref.~\cite{GL 1983}, the one loop formulae 
for the various scattering lengths were obtained by directly evaluating the 
chiral representation at threshold -- in other words, the matching was 
performed at $s=4M_\pi^2$ rather than at $s=0$. 

We emphasize that the above discussion in the framework of the one loop
approximation only serves to explicitly demonstrate that the choice of the
matching conditions is not irrelevant. Admittedly, in our final analysis,
where we will be working at two loop accuracy, the noise 
due to that choice is significantly smaller. 
 
\section{Infrared singularities}\label{sec:infrared singularities}
{}From a purely algebraic point of view, the manner in which the matching is 
done is irrelevant, as long as it is performed in the common region 
of validity of the chiral and phenomenological representations. We could
also match the two loop representation to the phenomenological one
at threshold and would then obtain a formula analogous to
the one given in \cite{GL 1983}, but now valid to next--to--next--to--leading 
order. Alternatively, we could match the two representations of the 
scattering amplitude 
at the center of the Mandelstam triangle -- the result would only differ by
contributions that are beyond the accuracy of the chiral representation. 

There is a good reason for preferring the procedure specified above to a
matching at threshold:
The branch cut required by unitarity starts there. The
modifications of the tree level result generated by the higher order
effects are quite large at threshold, because they are enhanced by a small
energy denominator. Indeed, $a_0^0$
contains a chiral logarithm with an unusually large coefficient:
\bdm a_0^0=\frac{7\,M_\pi^2}{32\,\pi\,F_\pi^2}\left\{1+
\frac{9}{2}\,\ell_\chi
+\ldots\right\}\co\hspace{2em}\ell_\chi\equiv
\left(\frac{M_\pi}{4\hspace{0.05em}\pi\hspace{0.05em}F_\pi}\right)^{\!\!2}
\ln\! \left(\!\frac{\mu^2}{M_\pi^2}\!\right)\fs\edm
The phenomenon gives rise to an exceptionally large correction that violates
the rule of thumb of section \ref{sec:eff} by an order of
magnitude: The one-loop correction increases the tree level prediction
by about 25\% !

At the center of the Mandelstam triangle, the amplitude also contains a
chiral
logarithm ($s_0=\frac{4}{3}\,M_\pi^2)$:
\bdm A(s_0,s_0,s_0)=\frac{M_\pi^2}{3\,F_\pi^2}\left\{1
 +\frac{11}{6}\,\ell_\chi+\ldots\right\}\fs\edm
The coefficient is less than half as big as the one in $a_0^0$, but it
still represents a sizeable correction.

In our matching procedure, we replace $a_0^0$ and $a_0^2$ by 
$C_0$ and $C_2$ and at the same time also eliminate $\ell_4$ in favour of
the scalar radius. What matters for the
convergence properties of the quantities appearing in our matching conditions
are the infrared singularities contained in
\bea C_0-\frac{M_\pi^2}{3}\,\rs=1-\frac{5}{14}\,
\ell_\chi+\ldots\co\hspace{1em}
C_2-\frac{M_\pi^2}{3}\,\rs=
1+\frac{1}{2}\,\ell_\chi+\ldots\fs\nonumber\eea
The coefficients occurring here are remarkably small.
The term $C_1-\frac{1}{3}M_\pi^2 \rs$ does not contain a chiral
logarithm at all. We can therefore expect that,  
for the quantities that are relevant for the determination of the 
$S$--wave scattering lengths, the perturbation series
converges very rapidly, much more so than for a matching at threshold or at
the center of the Mandelstam triangle. 
As we will see, this is indeed born out by the
numerical analysis.

\setcounter{equation}{0}
\section{Estimates for symmetry breaking at \boldmath{$O(p^6)$}}
\label{sec:symmetry breaking}

We now extend the
analysis to next--to--next--to--leading order. For that purpose, we need an
estimate for the symmetry breaking couplings $r_1\,\ldots\,,r_4$ and 
$\rS$ of 
${\cal L}_6$, which enter 
the low energy theorems for $C_0,C_1,C_2$ at order $M_\pi^4$, as well as the 
relation between the scalar radius and the coupling constant
$\ell_4$. The corresponding 
correction terms $\Delta_0,\Delta_1,\Delta_2$ are listed in
(\ref{eq:Delta12}). In the normalization used there,
the resonance estimates of refs.~\cite{BCEGS,Bijnens Colangelo Talavera,
Hannah:1997ux} amount to 
\bea\label{eq:rnum} 
\rtilde_1\simeq -1.5\co\hspace{1em}  \rtilde_2\simeq 3.2\co\hspace{1em}
\rtilde_3\simeq-4.2\co\hspace{1em}  \rtilde_4\simeq -2.5\co\hspace{1em}
\rStilde\simeq -0.7\fs\eea
Inserting these numbers in (\ref{eq:Delta12}), we obtain a shift in 
$C_0,C_1,C_2$ by $-0.3$, $-0.5$ and $-0.8$ permille,
respectively. This
confirms the expectation that the effects due to the symmetry breaking
coupling constants $r_n$ are tiny. Since the scale is set 
by the scalar or pseudoscalar 
non--Goldstone states contributing to the relevant sum rules, $M_s\simeq
1\,\mbox{GeV}$, the corresponding corrections are of order 
$M_\pi^4/M_s^4\simeq 4 \cdot 10^{-4}$. 
In the SU(2) framework we are using here, the  
$K\bar{K}$ continuum also contributes to the effective coupling constants,
but
in view of $4M_K^2\simeq M_s^2$, the corresponding scale is even somewhat 
larger. In the following, we assume that the estimates in equation
(\ref{eq:rnum}) are valid to within a factor of two. 

In the case of $r_1,\ldots\,,r_4$, the main uncertainty stems  
from the $\pi\pi$ continuum underneath the resonances, that is from the 
chiral logarithms. 
Since the formulae (\ref{eq:Delta12}) are quadratic in 
these, the scale dependence of those coupling constants is rather
pronounced. This can be seen by varying the scale $\mu$, at which the running
coupling constants
are assumed to be saturated by the resonance contributions.
For $0.5\,\mbox{GeV}<\mu< 1\,\mbox{GeV}$, the corrections vary in the range
\bdm 
0.002\,\lsim\,\xi^2\Delta_0\,\lsim\, 0.005\co\hspace{0.3em}-0.001\, \lsim\,
\xi^2\Delta_1\,\lsim\, 0.003\co
\hspace{0.5em}-0.005 \,\lsim\, \xi^2\Delta_2\,\lsim\, 0.001\fs\edm 

In the representation (\ref{eq:rs}) for the scalar radius, the two loop
correction $\Delta_r$ represents an effect of first order. Estimating the
magnitude in the same manner as for $\Delta_0,\Delta_1,\Delta_2$, 
the result varies in the range $ 0.18\,\lsim\,\xi\Delta_r\,\lsim\, 0.28$.
The correction thus shifts the scalar radius by $0.04\pm0.01\,\mbox{fm}^2$. 

In the following, the central values are calculated
by using the resonance estimates (\ref{eq:rnum}) at the scale $\mu=M_\rho$.
For some of the quantities analyzed in the present paper, the result is
insensitive to the uncertainties inherent in these estimates, but in some 
cases, they even dominate our error bars -- we will discuss the sensitivity
of the various results in detail.

\setcounter{equation}{0}
\section{Final results for {\boldmath$a_0^0$\unboldmath} and
{\boldmath$a_0^2$
\unboldmath}}
\label{sec:final results}

We are now in a position to describe the determination of
$a_0^0$ and $a_0^2$ at two loop accuracy. Our matching conditions identify
two
different representations for the coefficients $c_1,\,\ldots\,c_6\,$:
the chiral representation specified in equation
(\ref{eq:ci}) and the phenomenological one in (\ref{eq:mc}).
For the evaluation of the $S$--wave scattering lengths, only the first four
coefficients are relevant. For these, the chiral representation involves the 
effective coupling constants $\ell_1$, $\ell_2$, $\ell_3$, $\ell_4$,
$r_1$, $r_2$, $r_3$, $r_4$, while the
phenomenological representation contains only the two parameters $a_0^0$ and 
$a_0^2$, which enter explicitly as well as implicitly, through the 
moments $\pbar_1,\,\ldots\,,\,\pbar_4$. In principle, we solve the four
conditions 
for the four variables $a_0^0$, $a_0^2$, $\ell_1$, $\ell_2$,
treating the symmetry breaking coupling constants $\ell_3$, $\ell_4$, 
$r_1,\ldots \,,r_4$ as known. 

The constant $\ell_3$ is varied in the
range specified in (\ref{eq:l3barnum}). Concerning $\ell_4$, we rely on the
result for the scalar radius given in (\ref{eq:rsnum}), thus in effect
replacing the input variable $\ell_4$ by $\langle r^2\rangle_s$. 
The analysis then involves a fifth condition: the relation (\ref{eq:rs}), 
which expresses
the scalar radius in terms of effective coupling constants.

If all of the input variables are taken at their central values, the
representation for 
the moments given in appendix \ref{sec:moments} can be used. The solution of
the resulting system of numerical equations occurs at
the values quoted in table \ref{tab:fix point}, first row.
\begin{table}
\begin{center}
\begin{tabular}{|c|r|r|r|r|r|r|r|}
\hline
\rule[-0.5em]{0em}{1.55em}&$a_0^0\rule{0.8em}{0em}$&$a_0^2\rule{1.2em}{0em}$&
$\lbar_1\rule{0.7em}{0em}$&$\lbar_2\rule{0.4em}{0em}$&
$\lbar_4\rule{0.4em}{0em}$&
$\rtilde_5\rule{0.2em}{0em}$&$\rtilde_6\rule{0.2em}{0em}$\\
\hline 
 &$0.220$&$-0.0444$&$-0.36$&$4.31$&$4.39$&3.8&1.0\\
\hline\hline
\rule{0em}{1em}$\rs$&0.002&0.0003&0.04&0.02&0.19&0.05&0.03
\\
\hline
\rule{0em}{1em}$\ell_3$& 0.004&0.0009&
0.01& 0.00&0.02&0.01&0.00  \\
\hline
\rule{0em}{1em}$r_n$&0.001&0.0002&0.51&0.10 &0.10 &1.04&0.10 \\
\hline
\rule{0em}{1em}exp &0.001&0.0002&0.29&0.04&0.03&0.12&0.02\\
\hline
\hline
\rule{0em}{1em}tot&0.005&0.0010&0.59&0.11 &0.22&1.05&0.11\\
\hline
\end{tabular}
\end{center}
\caption{\label{tab:fix point}Solution of the matching conditions. 
The first row contains the central values. The next four rows indicate the
uncertainties in this result, arising from the one in $\langle r^2\rangle_s$,
$\ell_3$, $r_n$ and in the experimental input used in the Roy equations.
The last row is obtained by adding these up in quadrature.}
\end{table}
The next four rows indicate the sensitivity to the input used for 
$\langle r^2\rangle_s$, $\ell_3$,
to the uncertainties in the symmetry breaking coupling constants $r_n$ of
$O(p^6)$, and to those in the 
experimental information used when solving the Roy equations. 
The details of the error analysis that underlies these numbers are described 
in appendix \ref{sec:error analysis}.

The table shows that the uncertainties in the prediction for $a_0^0$ and
$a_0^2$ are dominated by those from $\ell_3$. In particular,
the result for the $S$--wave scattering lengths is not sensitive to the
contributions from the coupling constants occurring at $O(p^6)$. 
Adding up the uncertainties due to these and to the experimental input in the
Roy equations, we arrive at
\bea\label{l3rdependence} a_0^0\al=\al \hspace{0.8em}0.220\pm 0.001
 + 0.027\, \Delta_{r^2}-0.0017\,\Delta \ell_3\,,\\
a^2_0\al=\al -0.0444 \pm 0.0003 
-0.004\, \Delta_{r^2}-0.0004\,\Delta \ell_3\,,\nonumber\eea
where $\Delta_{r^2}$ and $\Delta\ell_3$ are defined by
\bdm \rs=0.61\,\mbox{fm}^2(1+\Delta_{r^2})\co
\hspace{2em}\lbar_3=2.9+\Delta \ell_3 \fs\edm 
Our final result for the $S$--wave scattering
lengths follows from this representation with the estimates for
$\Delta_{r^2}$, $\Delta\ell_3$  given in (\ref{eq:l3barnum}),
(\ref{eq:rsnum}), and reads
\bea\label{eq:final result} 
\al\al a_0^0= 0.220\pm 0.005\,,\hspace{5.2em} a^2_0=-0.0444\pm
0.0010\,,\\\al\al 
2a_0^0-5a_0^2= 0.663\pm 0.006\,,\hspace{2em} a^0_0-a^2_0= 0.265\pm0.004\,.
\nonumber\eea
Expressed in terms of the coefficients $C_0,C_1,C_2$, this 
result corresponds to 
\bea\label{eq:Cnum} C_0=1.096\pm0.021\co\hspace{1em}
C_1=1.104\pm0.009\co\hspace{1em}C_2=1.115\pm0.022\fs\eea

\setcounter{equation}{0}
\section{Discussion} 
\label{sec:discussion}

The terms omitted in the chiral perturbation series represent an inherent
limitation of our calculation. The matching must be done in such a manner
that
these are small. In contrast to a matching at threshold --
that is, to the straightforward
expansion of the scattering lengths --
our method fulfills this criterion remarkably well: We are using the
expansion
in powers of the quark masses only for the coefficients $C_0$, 
$C_1$ and $C_2$, while
the curvature generated by the unitarity cut is evaluated phenomenologically.
As discussed in section \ref{sec:infrared singularities}, 
the infrared singularities occurring in the expansion of these quantities 
have remarkably small residues. Indeed, truncating 
the expansion of $C_n$ at order $1$, $m$ and $m^2$,
respectively and solving equation (\ref{eq:aC}) in the corresponding
approximation, we obtain  
\bea\label{eq:sequence 1} 
a_0^0\al=\al\hspace{0.8em} 0.197\hspace{0.5em}\rightarrow
\hspace{0.8em}0.2195\hspace{0em}\rightarrow\hspace{0.8em} 
0.220\co\no
a_0^2\al=\al-0.0402\rightarrow -0.0446\rightarrow-0.0444\co\\
2\,a_0^0-5\,a_0^2\al=\al\hspace{0.8em} 0.594\hspace{0.5em}\rightarrow
\hspace{0.8em}0.662\hspace{0.5em}
\rightarrow\hspace{0.8em}0.663\nonumber\co\eea
indicating that the series converges very rapidly.
For this reason, we expect the contributions from yet
higher orders to be entirely negligible.

The rapid convergence of the series is a virtue of the specific method used
to match the chiral and phenomenological representations. To 
demonstrate  
this, we briefly discuss the alternative approach used in refs.~\cite{GL
1983,BCEGS}, where the results for the various scattering lengths and
effective ranges are obtained by directly evaluating the chiral
representation of the scattering amplitude at threshold.  
Keeping the values of the effective coupling constants fixed at
the central values and truncating the series at order $m$, $m^2$ and $m^3$,
we
obtain the sequence
\bea\label{eq:threshold matching} 
a_0^0\al=\al\hspace{0.8em} 0.159\hspace{0.5em}\rightarrow
\hspace{0.8em}0.200\hspace{0.5em}\rightarrow\hspace{0.8em} 
0.216\co\no
a_0^2\al=\al-0.0454\rightarrow -0.0445\rightarrow-0.0445\co\\
2\,a_0^0-5\,a_0^2\al=\al\hspace{0.8em} 0.545\hspace{0.5em}\rightarrow
\hspace{0.8em}0.624\hspace{0.5em}
\rightarrow\hspace{0.8em}0.654\nonumber\fs\eea
The first terms on the right correspond to Weinberg's formulae.
The second and third terms are in agreement with the old one loop results of
ref.~\cite{GL 1983} and the two loop results of
ref.~\cite{BCEGS,ABT_Ke4,Nieves:1999zb}, respectively. 
As indicated by the difference between the second and third terms, 
\begin{figure}[thb]
\psfrag{l3bar}{\raisebox{-1em}{\Large $\lbar_3$}}
\psfrag{a0}{\Large $a_0^0$}
\psfrag{a2}{\Large $a_0^2$}
\psfrag{-0.025}{\hspace{0.9em}$-0.025$}
\psfrag{-0.03}{\hspace{0.3em}$-0.030$}
\psfrag{-0.035}{\hspace{0.9em}$-0.035$}
\psfrag{-0.04}{\hspace{0.3em}$-0.040$}
\psfrag{-0.045}{\hspace{0.9em}$-0.045$}
\psfrag{-0.05}{\hspace{0.3em}$\!-0.050$}
\psfrag{-0.055}{\hspace{0.9em}$-0.055$}
\psfrag{0.26}{\hspace{0.2em}$0.26$}
\psfrag{0.24}{\hspace{0.2em}$0.24$}
\psfrag{0.22}{\hspace{0.2em}$0.22$}
\psfrag{0.18}{\hspace{0.2em}$0.18$}
\psfrag{0.16}{\hspace{0.2em}$0.16$}
\leavevmode
\vspace{-1em}
\begin{center}
\includegraphics[width=14cm]{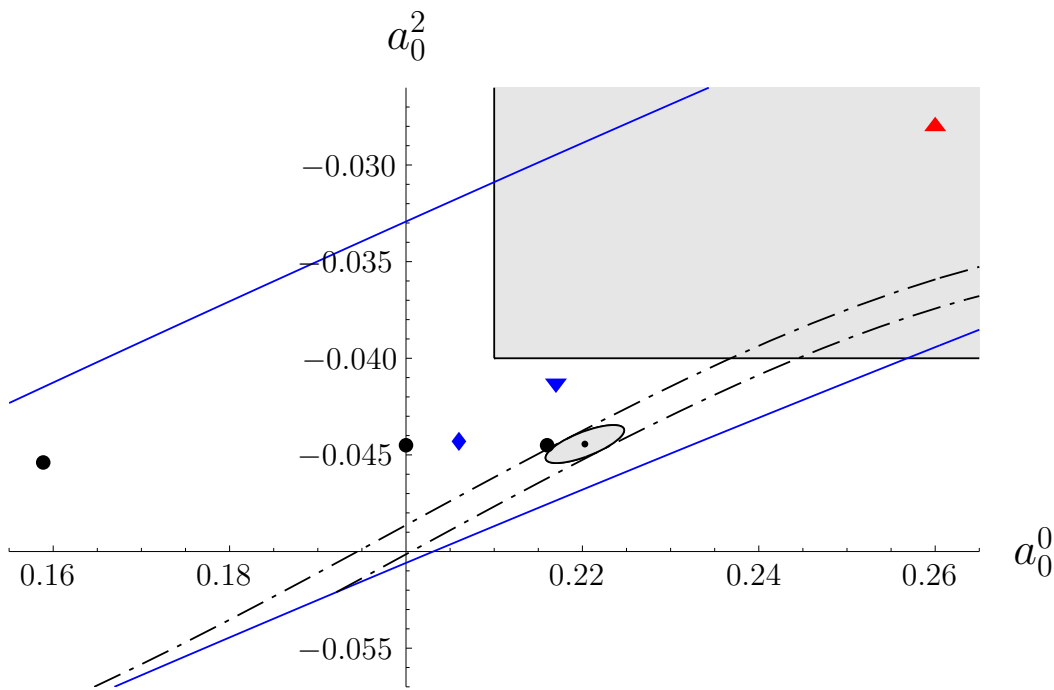}
\end{center}
\caption{\label{fig:aellipse} Constraints imposed on the $S$--wave scattering
lengths by chiral symmetry. The three full circles 
illustrate the convergence of the
chiral perturbation series at threshold, according to eq.~(12.2). 
The one at the left
corresponds to Weinberg's leading order formulae. 
The error ellipse represents our final result. The other elements of the
figure are specified in the text.}
\end{figure}
the corrections of $O(p^6)$ are by no means negligible for a matching at 
threshold. This is illustrated in fig.~\ref{fig:aellipse}, where the
three full circles correspond to the sequence (\ref{eq:threshold matching}). 
The triangle at the right and the shaded rectangle
indicate the central values and the uncertainties quoted in the 1979 
compilation of 
ref.~\cite{Nagels}. The triangle and the diamond near the 
center of the figure correspond to set I and set II of ref.~\cite{BCEGS}, 
respectively. The ellipse represents the 68\% confidence contour of our 
final result in eq.~(\ref{eq:final result}). The details of the
error analysis that underlies this result are described in appendix
\ref{sec:error analysis}. 

The reason why the straightforward expansion of the scattering lengths in
powers of the quark masses converges rather slowly is that these represent
the values of the amplitude at threshold, that is at the place where the 
branch cut required by unitarity starts. 
The truncated chiral representation does not describe that 
singularity well enough, particularly at one loop, where the relevant
imaginary parts stem from the tree level approximation. 

If the effective coupling constants are the same, the only 
difference between our method and a matching at threshold is the one between 
the functions $\Wbar^I(s)$ and $U^I(s)$. In particular, the results 
for $a_0^0$, $a_0^2$ only differ  
because the numerical values of 
$\Wbar^I(s)$ and $U^I(s)$ at $s=4M_\pi^2$ are not the same.
As mentioned above, the difference between the two sets of functions affects 
the scattering amplitude only at $O(p^8)$ and beyond. Numerically,
however, it is not irrelevant which one of the two is used to describe the
effects generated by the unitarity cuts: While the functions $\Wbar^I(s)$
account for the imaginary parts of the $S$- and $P$-waves to the accuracy to
which these are known, the quantities $U^I(s)$ represent a comparatively
crude approximation, obtained by evaluating the imaginary parts
with the one-loop representation. 

\setcounter{equation}{0}
\section{Correlation between \boldmath{$a_0^0$ and $a_0^2$}}
\label{sec:GCHPT}
As mentioned earlier, the main difference between Generalized Chiral
Perturbation Theory and the standard one used in the present paper resides in
the coupling constant $\ell_3$. Apart from that, the formulae are identical
--
only the bookkeeping for the chiral power of the quark mass matrix
is different.\footnote{If $\ell_3$ is 
large, the symmetry breaking effects generated by the quark masses are larger
than in the standard framework, so that a reordering of the series that gives
these more weight is called for.} In particular, the relation between the 
scalar radius and the
coupling constant $\ell_4$ also holds in that framework, but there is no
prediction for 
the $S$--wave scattering lengths $a_0^0$ and $a_0^2$, because these 
involve the coupling constant $\ell_3$. 
The fact that $\ell_4$
is strongly constrained by the value of the scalar radius implies, however,
that there is a strong correlation between $a_0^0$ and $a_0^2$, independently
of whether the quark condensate is the leading order parameter: Apart from
higher order corrections, both of these are controlled by the same
parameter $\ell_3$. The dependence is approximately described by the
parabolae
\bea\label{eq:corral3} a_0^0\al=\al 
0.225-1.6\cdot 10^{-3}\,\lbar_3-1.3 \cdot 10^{-5}\,(\lbar_3)^{2}\co\\
a_0^2\al=\al -0.0433-3.6\cdot 10^{-4}\,\lbar_3-4.3\cdot 10^{-6}\,
(\lbar_3)^{2}\fs\nonumber\eea
\begin{figure}[thb]
\psfrag{l3bar}{\raisebox{-1em}{\large $\lbar_3$}}
\psfrag{a00}{}
\psfrag{a20}{}
\psfrag{-0.035}{\hspace{0.2em}$-0.035$}
\psfrag{-0.04}{\hspace{0.2em}$\!-0.040$}
\psfrag{-0.045}{\hspace{0.2em}$-0.045$}
\psfrag{-0.05}{\hspace{0.2em}$\!-0.050$}
\psfrag{-0.055}{\hspace{0.2em}$-0.055$}
\psfrag{0.26}{$0.26$\hspace{2.3em}
\raisebox{1.5em}{\large $a_0^0$}
\hspace{10em}\raisebox{1.5em}{\large$a_0^2$}}
\psfrag{0.22}{$0.22$}
\psfrag{0.18}{$0.18$}
\psfrag{40}{$40$}
\psfrag{20}{$20$}
\psfrag{0}{$0$}
\leavevmode
\vspace{-1em}
\begin{center}
\includegraphics[width=10cm]{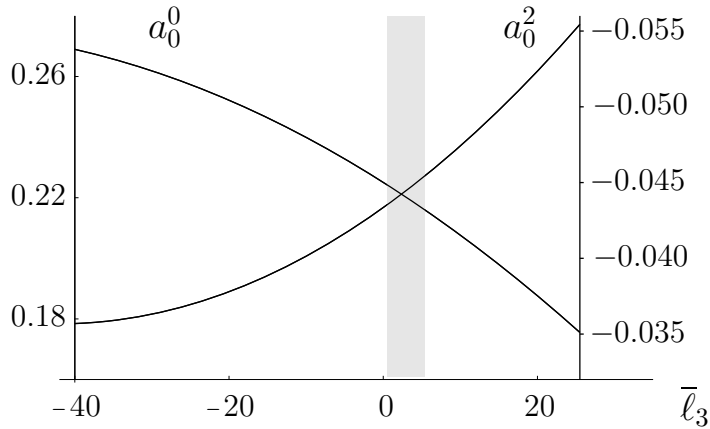}
\end{center}
\caption{\label{fig:al3} $S$--wave scattering lengths as functions of 
$\ell_3$.}
\end{figure}
which are displayed in fig.~\ref{fig:al3}.
Note that the interval shown far exceeds the range relevant for
the standard picture, which is indicated by the vertical bar.

Eliminating the parameter $\ell_3$, we obtain a correlation between $a_0^0$ 
and $a_0^2$:
\bea\label{eq:corra0a2} 
a_0^2\al=\al-.0444\pm .0008 +.236\, (a_0^0-.22)\\\al\al-.61\,
(a_0^0-.22)^2 -9.9\, (a_0^0-.22)^3\fs\nonumber\eea  
The error given accounts for the various sources of uncertainty in our input
-- evaluating these as described in appendix
\ref{sec:error analysis}, 
we find that they are nearly independent
of $a_0^0$. The correlation is indicated
in fig.~\ref{fig:aellipse}: The values of $a_0^0$ and $a_0^2$ are
constrained to the region between the two dash--dotted lines
that touch the error ellipse associated with the standard picture.
As discussed in ref.~\cite{ACGL}, a qualitatively similar correlation also 
results from the Olsson sum rule \cite{olsson sum rule}  -- the two conditions
are perfectly compatible, but the one above is considerably more
stringent. Fig.~\ref{fig:aellipse} also shows that
for $a_0^0< 0.18$, or $ \lbar_3> 25$, the 
center of the region allowed by the correlation falls outside the universal 
band, which is indicated by the tilted lines. The same happens on the opposite
side, for $a_0^0> 0.28$, $\lbar_3< -54$. 
Since the Roy equations only admit solutions if the two
subtraction constants $a_0^0$ and $a_0^2$ are in the universal band,
exceedingly large values of $\lbar_3$ are thus excluded. 
Note also that the correlation implies an upper bound on the
$I=2$ scattering length: $a^2_0< -0.035$.

The correlation between $a_0^2$ and $a_0^0$ can be used, for instance, to
analyze the information about  
the phase difference $\delta_0^0-\delta_1^1$ obtained from  the decay
$K\rightarrow \pi\pi e \,\nu$. At the low energies occurring there, 
this difference is dominated by the contribution
$\propto a_0^0$ from the $I=0$  $S$--wave scattering length. 
The relation (\ref{eq:corra0a2}) allows us to correct for the higher order
terms of the threshold expansion: The phase 
difference can be expressed in terms of the energy and the value of $a_0^0$,
up to very small uncertainties. This is illustrated
in fig.~\ref{fig:deltaKl4}: 
\begin{figure}[thb]
\psfrag{0.18}{\hspace{0.5em}\raisebox{-0.2em}{$0.18$}}
\psfrag{0.22}{\hspace{0.5em}\raisebox{-0.2em}{$0.22$}}
\psfrag{0.26}{\hspace{0.5em}\raisebox{-0.2em}{$0.26$}}
\leavevmode

\vspace{2em}\centering
\includegraphics[width=10cm]{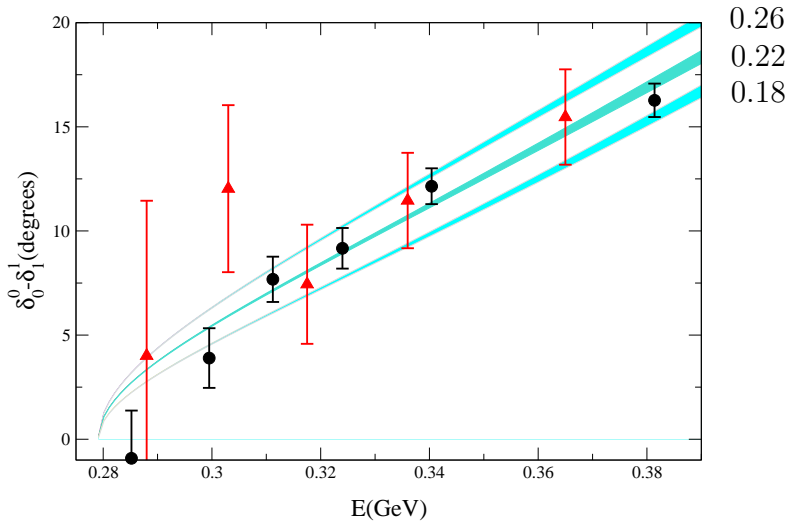}
\caption{\label{fig:deltaKl4} Phase relevant for the decay $K\rightarrow
  \pi\pi e\nu$. The three bands correspond to the 
three indicated values of the $S$--wave scattering length $a_0^0$. The
uncertainties are dominated by those from the experimental input used
in the Roy equations. The triangles are the data points of Rosselet et 
al.~\cite{rosselet}, while the full circles represent the preliminary
E865 results \cite{Truol}.}
\end{figure}
The center of the three narrow bands shown is obtained by fixing the value of
$a_0^2$ with the correlation (\ref{eq:corra0a2}) and inserting the result in
the numerical parametrization
of the phase shifts in appendix D of ref.~\cite{ACGL}. At a given value of 
$a_0^0$, the uncertainties in the result for the phase difference
$\delta_0^0(s)-\delta_1^1(s)$ are dominated by the one in
the experimental input used for the $I=0$ $S$--wave. Near threshold,
the uncertainties are proportional to $(s-4M_\pi^2)^{3/2}$ -- in the range 
shown, they amount to less than a third of a degree. 
While the data of Rosselet et al.~\cite{rosselet} are consistent with all 
three of the indicated values of $a_0^0$, the preliminary results of the E865
experiment at Brookhaven \cite{e865,Truol} are not. Instead they beautifully
confirm the 
prediction (\ref{eq:final result}): The best fit to these data is obtained
for $a_0^0=0.218$, with $\chi^2= 5.7$ for 5 degrees of freedom. As pointed out
in ref.~\cite{CGL letter2}, the correlation (\ref{eq:corra0a2}) can be 
used to convert data on the phase difference into data on the scattering
lengths. For a detailed discussion of the consequences for the value
of $a_0^0$, we refer to \cite{CGL letter2,e865final}.

\setcounter{equation}{0}
\section{\boldmath{Results for $\ell_1$ and $\ell_2$}} 
\label{sec:l12}

\begin{figure}[thb]
\psfrag{l1bar}{\large $\lbar_1$}
\psfrag{l2bar}{\large $\lbar_2$}
\psfrag{OneLoop}{\hspace{-0.1em}\raisebox{0.2em}{\footnotesize GL}}
\psfrag{ABT}{\hspace{-0.3em}\raisebox{-0.1em}{\footnotesize ABT}}
\psfrag{Girlanda}{\hspace{6.1em}\raisebox{2em}{\footnotesize GKMS}}
\psfrag{x}{$\times$}
\leavevmode
\begin{center}
\includegraphics[width=10cm]{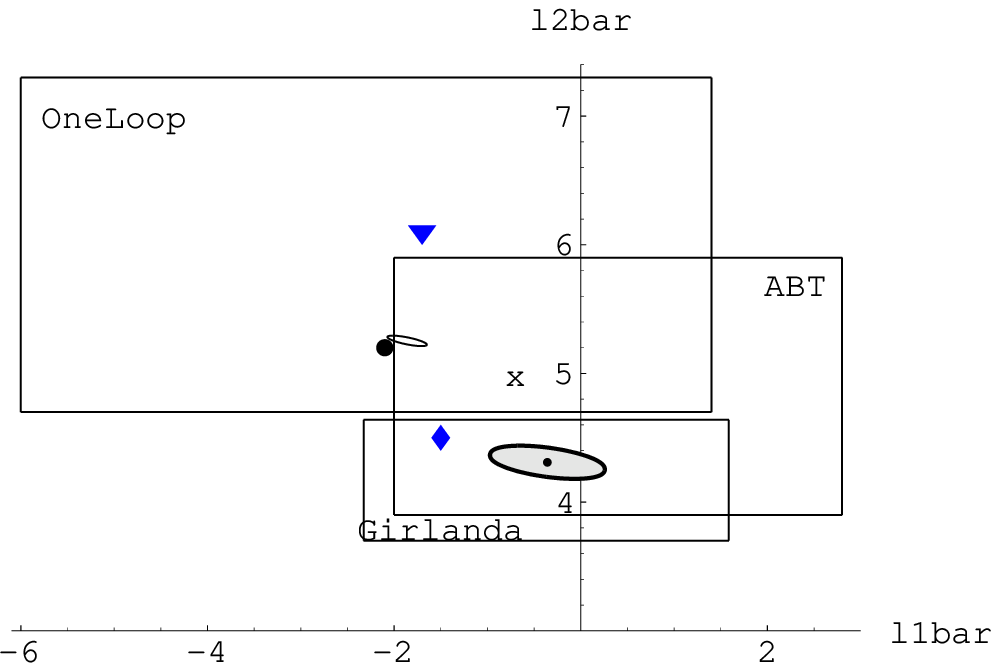}
\end{center}
\caption{\label{fig:l1l2}Values of the coupling constants $\ell_1$ and 
$\ell_2$. The shaded ellipse shows} the result of our calculation. 
The rectangles indicate the ranges quoted in
refs.~\cite{GL 1984}, \cite{ABT_Ke4} and \cite{Girlanda:1997ed}. The triangle
and the diamond correspond to set I and set II of \cite{BCEGS}, respectively.
The cross represents the resonance saturation estimate of 
ref.~\cite{Ecker CD97}. The full circle is the result obtained by matching
at one loop and the thin ellipse close to it represents the uncertainties in
the effective one loop couplings $\ell^\eff_1$, $\ell^\eff_2$.  

\end{figure}

The effective coupling constants of ${\cal L}_4$ enter the chiral 
perturbation theory
representation of the scattering amplitude and of the scalar form factor
only as corrections, so that our results for these are
subject to significantly larger uncertainties than those for $a_0^0$,
$a_0^2$. According to table \ref{tab:fix point}, we obtain
\bea\label{eq:l12num} 
\lbar_1=-0.4\pm 0.6\co\hspace{2em}\lbar_2=4.3\pm 0.1\fs\eea
The noise in the symmetry breaking
couplings $r_n$ of ${\cal L}_6$ and the one in the Roy equation input
yield comparable contributions, while those from the other entries are 
negligibly small. The corresponding error ellipse is shown in 
fig.~\ref{fig:l1l2}.

In order to investigate the uncertainties due to the neglected higher order
terms, we again compare this with what is found if the phenomenological
representation is matched to the one loop approximation of the chiral
perturbation series.  For the central values of the input parameters, the
solution of the matching conditions then occurs at $\lbar_1=-1.8$,
$\lbar_2=5.4$ : The two loop effects shift the one loop result by about
$+1.4$ and $-1.1$ units, respectively.  The shift arises from the fact that
the expansion of the coefficients $c_3$ and $c_4$ contains very strong
infrared singularities at first nonleading order. Analogous contributions
also occur in $c_1$ and $c_2$, at next--to--next--to--leading order, but in the
combinations $C_0$, $C_1$, $C_2$ that matter for the determination of the
scattering lengths, these singularities only generate very small effects:
In these quantities, the contributions of order $p^4$ amount to less than
1\%.  We conclude that, unlike the result for $a_0^0$, $a_0^2$, where the
uncertainties from the neglected higher order terms are tiny, the one for
$\ell_1$ and $\ell_2$ is sensitive to these.  Although we expect the
corresponding contributions to be small compared to the first order shift
given above, they might be of the same order as those from the
uncertainties in our input -- we do not offer a quantitative guess.

The couplings  $\ell_1$ and $\ell_2$ are quark mass independent,
whereas the physical quantities used to estimate their values
incorporate quark mass effects. As a result of this, it is
problematic to rely on phenomenological 
determinations based on the one loop approximation when analyzing quantities
at two loop order. 
The large infrared singularities that accompany the
contributions from $\ell_1$ and $\ell_2$ are automatically accounted for
in the two loop representation, but are missing in the framework of a one 
loop calculation -- in the phenomenological analysis,
their contributions are lumped into those from the coupling
constants. As an illustration, we mention the set I of couplings introduced
in
\cite{BCEGS}, that uses the one-loop values for $\ell_1$ and
$\ell_2$, but leads to $D$-wave scattering lengths that do not agree
well with the values extracted from experiment, as was first pointed
out in ref.~\cite{Girlanda:1997ed}.
For a detailed discussion of this issue, we refer to \cite{Ecker CD97}.

We now show that, once the shift in the values of $\ell_1$, $\ell_2$ is 
accounted for, the one and two loop representations for the coefficients
$c_1,\ldots\,,c_4$ become nearly the same, so that the results obtained by
matching the phenomenological representation with the chiral one at two loop
level nearly coincide with those found in the one loop approximation.
The infrared singularities responsible for that shift are those 
contained in the coefficients $b_3$, $b_4$. If we solve the
expressions for these coefficients in one loop approximation, we obtain
\bea \ell^\eff_1\equiv 3\,(\bbar_3-\bbar_4)+\frac{4}{3}\co
\hspace{2em} \ell^\eff_2\equiv 6\,\bbar_4+\frac{5}{6}\fs\eea 
The expansion of these quantities in powers of the quark masses starts with
$\ell_n^\eff=\lbar_n+O(\xi)$. The infrared singularities
generated by the two loop graphs show up in the terms of order
$\xi$, in particular through contributions proportional to
$\Ltilde^2=\ln^2(\mu^2/M_\pi^2)$, which are very important 
numerically. Accounting for the uncertainties in our input, we obtain
\bea \label{eq:leff12num}\ell^\eff_1=-1.9\pm 0.2\co\hspace{2em}
\ell^\eff_2=5.25\pm 0.04\fs\eea 
The comparison with the values  
$\lbar_1=-1.8$, $\lbar_2=5.4$, found when matching at one loop, shows that
the couplings relevant in the context of the one loop approximation may
indeed
be characterized in this manner (compare
fig.~\ref{fig:l1l2}, where the values for $\lbar_1$, $\lbar_2$ obtained
at one loop are indicated by the full circle, while
the thin ellipse corresponds to the above numerical result
for $\ell^\eff_1$, $\ell^\eff_2$).  

Now comes the point we wish to make: We may also evaluate the one loop
formulae (\ref{eq:c12oneloop}) for $c_1,\,c_2$, replacing 
$\lbar_1$, $\lbar_2$ by the above effective values. 
The outcome differs from what is obtained with the two loop formulae only
by a fraction of a percent -- the difference is in the noise of the two loop
result. In this sense, the main effect of the infrared singularities
in the two loop graphs amounts to a shift in the values of the coupling
constants $\lbar_1$, $\lbar_2$. This explains why the matching conditions 
used in the present paper yield very accurate results for the $S$--wave
scattering lengths already at one loop, while the corresponding results for 
these two couplings are off.

The literature contains
quite a few determinations of the coupling constants $\ell_1$ and $\ell_2$
that are based on the one
loop approximation of chiral perturbation theory 
\cite{GL 1984} -- \cite{AB},
starting with the estimates $\lbar_1=-2.3\pm 3.7$,
$\lbar_2=6.0\pm 1.3$ given in ref.~\cite{GL 1984}, which are perfectly
consistent with our result for 
$\ell^\eff_1$, $\ell^\eff_2$. Note that, in the case of $\lbar_2$,  
the shift generated by the two loop graphs takes the result 
outside the quoted range (as stated in ref.~\cite{GL 1984}, that range 
only measures the accuracy to which the first order corrections can be
calculated and does not include an estimate of contributions due to higher
order terms).

The results for the effective coupling constants obtained by 
Girlanda et al.~\cite{Girlanda:1997ed} read 
$\lbar_1=-0.37 \pm 0.95\pm 1.71$, $\lbar_2=4.17 \pm 0.19\pm 0.43$.
The first error comes from the evaluation of the integrals over the
imaginary parts, while the second 
reflects the uncertainties in the contributions from the couplings of
${\cal L}_6$. Our results in eq.~(\ref{eq:l12num}) confirm these numbers,
with
substantially smaller errors -- we repeat, however, that these only account
for the noise seen in our calculation. 

Amoros, Bijnens and Talavera \cite{ABT_Ke4} have extracted values for the
coupling constants of ${\cal L}_4$ from their two loop analysis of the 
$K_{e_4}$ form factors -- 
which is based on SU(3)$_{\ind R}\times$SU(3)$_{\ind L}$ chiral
perturbation theory -- and obtain $\lbar_1=0.4\pm 2.4$,
$\lbar_2=4.9\pm 1.0$. Fig.~\ref{fig:l1l2} shows that these are
perfectly consistent with ours. As these authors are relying on the one loop
relations between the coupling constants $L_n$ of that framework and the 
couplings $\ell_n$ relevant for SU(2)$_{\ind R}\times$SU(2)$_{\ind L}$,
the results are accompanied by comparatively
large errors. 

\setcounter{equation}{0}
\section{\boldmath{Values of $\ell_4$, $r_5$ and $r_6$ }}
\label{sec:l4}
For the central values of the input, the matching conditions lead to
$\lbar_4=4.39$ (first row in table \ref{tab:fix point}). The uncertainties
in this number due to the various sources of error are dominated by the
one in the scalar radius and the noise in the symmetry breaking
coupling constants $r_1$, $r_2$,  $r_3$, $r_4$,
$\rS$ of ${\cal L}_6$. In order to estimate the uncertainties
due to the higher order effects that our calculation neglects, we compare
the above two loop result with the value $\lbar_4=4.60$, obtained 
by truncating the chiral representation 
for the scalar radius at leading order. The comparison shows that the shift
generated by the two loop contributions is of the same size as the one due to
the uncertainty in the scalar radius. 
Those from yet higher orders are expected to be
significantly smaller, so that the uncertainty in the final result is
dominated by the sources of error listed in the table. The net result reads 
\bea\label{eq:l4num} \lbar_4=4.4 \pm 0.2\fs\eea
The number is consistent with the one loop
estimate $\lbar_4=4.3\pm 0.9$, given in ref.~\cite{GL 1984}. The infrared
singularities that accompany the coupling constant $\lbar_4$
are much weaker than those occurring together with $\lbar_1$, $\lbar_2$.
The same is true also for $\lbar_3$, where the uncertainties
are much too large for such effects to matter at all.
 
The above result confirms the 
value $\lbar_4=4.4\pm 0.3$, obtained by Bijnens, Colangelo and Talavera
\cite{Bijnens Colangelo Talavera},
from a comparison of the two loop representation with
the dispersive result of the scalar radius, but this was to be expected,
because the input used in the two evaluations is nearly the same. 

In the framework of the calculation mentioned in section \ref{sec:l12}, 
Amoros, Bijnens and
Talavera \cite{ABT_Ke4} obtain $\lbar_4=4.2\pm 0.18$, also consistent with 
our result (as emphasized by these authors, the error bar does not
account for the uncertainties due to higher order effects, which in their
approach are quite substantial).

The coupling constants $\rtilde_n\equiv (4\pi)^4\, r_n^r(\mu)$ are scale
dependent. We could introduce corresponding scale independent quantities,
analogous to the terms $\lbar_n$ used for the coupling constants of 
${\cal L}_4$. The scale dependence is rather complicated, however,
because it is quadratic in $\ln \mu$. We instead quote the
values obtained for $\mu=M_\rho=0.77\,\mbox{GeV}$. 
Our analysis does not shed any light 
on the symmetry breaking coupling constants $r_1,\ldots\,,r_4$,
which belong to the input of our calculation, but we can determine 
$r_5$ and $r_6$, from the matching conditions for $c_5$ 
and $c_6$ -- we did not yet make use of these. Numerically, we find:
\bea\label{eq:r56num} 
\rtilde_5 =3.8\pm 1.0\co\hspace{2em} \rtilde_6=1.0\pm 0.1\fs\eea
Table \ref{tab:fix point} shows that the noise seen in our calculation
is dominated by the one
in the estimates for the 
symmetry breaking coupling constants $r_1,\ldots\,,r_4$. Note that
the error bars do not account for the uncertainties due to higher order 
contributions -- our evaluation
does not give us any handle on these. 

The resonance estimates of refs.~\cite{BCEGS,Bijnens
  Colangelo Talavera,Hannah:1997ux} offer a test: They lead to
\be\rtilde_5\simeq 2.7\co\hspace{2em}\rtilde_6\simeq 0.75\co\ee
and thus corroborate 
the outcome of our analysis, both in sign and in magnitude. In fact,
as pointed out by Ecker \cite{Ecker CD97}, 
the estimates 
\be \lbar_1\simeq -0.7\co\hspace{2em}
\lbar_2\simeq 5.0\co\hspace{2em}\lbar_3\simeq 1.9\co\hspace{2em}
\lbar_4\simeq 3.7\co\ee
obtained from resonance saturation of sum rules \cite{EGPdeR},
are perfectly consistent with the numbers found at two loop accuracy.
We conclude that there is good evidence for the 
picture drawn in ref.~\cite{GL 1984} to be valid: The values of all of
the effective coupling constants
encountered in the two loop representation
of the scattering amplitude are consistent with the assumption that these
are dominated by the contributions from the singularities due to the
exchange of the lightest non--Goldstone states. Admittedly, this assumption
does not lead to very sharp values, because the separation of the resonance
contributions from the continuum underneath is not unique. The problem
manifests itself in the scale dependence of the coupling constants --
resonance
saturation can literally hold only at one particular scale. Also, 
it is not a straightforward matter to formulate the resonance saturation
hypothesis for singularities due to the
exchange of particles of spin two or higher 
\cite{Toublan,Ananthanarayan:1998hj}. 
Even so, we consider it important that the values found for the coupling
constants are within the noise inherent in the assumption that, once the
poles
and cuts due to the Goldstone bosons are removed, the low energy behaviour
of the scattering amplitude is dominated by the 
singularities due to the remaining states. Since these remain massive
in the chiral limit, their contributions to the chiral expansion are
suppressed by powers of momenta or quark masses, but they do show up at
nonleading orders.  

\setcounter{equation}{0}
\section{\boldmath{The coefficients $b_1,\,\ldots,\,b_6$}}
\label{sec:bbar}
The matching conditions (\ref{eq:mc}) express the coefficients $c_n$ of the
chiral representation in terms of the $S$--wave scattering lengths and
moments of the imaginary parts. Inserting the numerical representation 
for the dependence of the moments on the scattering lengths
and comparing the result with eq.~(\ref{eq:ci}), we obtain the following
representation 
for the coefficients introduced in ref.~\cite{BCEGS}:
\bea\label{eq:bbaralg}
\bbar_1\al=\al\!\hspace{0.3em}  -.1 \pm .1 -21\,\dao+1670\,\dat
+9\,(\dao)^2+96\,\dao \dat -972\,(\dat)^2 ,\no 
\bbar_2\al=\al\!\hspace{0.5em} 8.2\pm .4
+179\,\dao-602\,\dat-135\,(\dao)^2 +315\,\dao \dat 
-65\,(\dat)^2 ,\no
\bbar_3\al=\al\!-.41\pm .06 
+3.5\,\dao-12.9\, \dat+7\, (\dao)^2-30\,\dao \dat
+40\,(\dat)^2 ,\no
\bbar_4\al=\al\!\hspace{0.7em} .71 \pm .01
 +1.3\,\dao -4.1\,\dat-\,(\dao)^2-4\,\dao
\dat+25\,(\dat)^2 , \hspace{-1em}\\
\bbar_5\al=\al\!\hspace{0.2em}  2.99 \pm .35 
+32.6 \hspace{0.07em}\dao -97.0\hspace{0.07em} \dat+104\hspace{0.07em}
(\dao)^2-451\hspace{0.07em} 
 \dao \dat
+602\hspace{0.07em} (\dat)^2 ,\no
\bbar_6\al=\al\! \hspace{0.3em}   2.18\pm .01
+7.2\,\dao-28.4\,\dat-3\,(\dao)^2+9\,\dao \dat-62\,(\dat)^2 ,
\nonumber\eea
with $\Delta a_0^0\equiv a_0^0-0.225$, $\Delta a_0^2\equiv a_0^2+0.03706$.
\begin{figure}[thb]
\psfrag{-0.6}{\hspace{-0.4em}\raisebox{-0.2em}{$-0.6$}}
\psfrag{-0.5}{\hspace{-0.4em}\raisebox{-0.2em}{$-0.5$}}
\psfrag{-0.4}{\hspace{-0.4em}\raisebox{-0.2em}{$-0.4$}}
\psfrag{-0.3}{\hspace{-0.4em}\raisebox{-0.2em}{$-0.3$}}
\psfrag{-0.2}{\hspace{-0.4em}\raisebox{-0.2em}{$-0.2$}}
\psfrag{-0.1}{\hspace{-0.4em}\raisebox{-0.2em}{$-0.1$}}
\psfrag{0.8}{\hspace{-0.5em}$0.80$}
\psfrag{0.78}{$0.78$}
\psfrag{0.76}{$0.76$}
\psfrag{0.74}{$0.74$}
\psfrag{0.72}{$0.72$}
\psfrag{0.7}{\hspace{-0.5em}$0.70$}
\psfrag{b3bar}{\hspace{0.5em}\large$\bbar_3$}
\psfrag{b4bar}{\raisebox{0.5em}{\large$\bbar_4$}}
\psfrag{SetII}{\footnotesize BCEGS II}
\psfrag{W}{\footnotesize W}
\psfrag{GKMS}{\footnotesize GKMS}
\psfrag{0.28}{\hspace{0.5em}280}
\psfrag{0.3}{\hspace{0.2em}300}
\psfrag{0.32}{\hspace{0.5em}320}
\psfrag{0.34}{\hspace{0.5em}340}
\psfrag{0.36}{\hspace{0.5em}360}
\psfrag{0.38}{\hspace{0.5em}380}
\leavevmode

\vspace{1em}\hspace{6em}
\includegraphics[width=10cm]{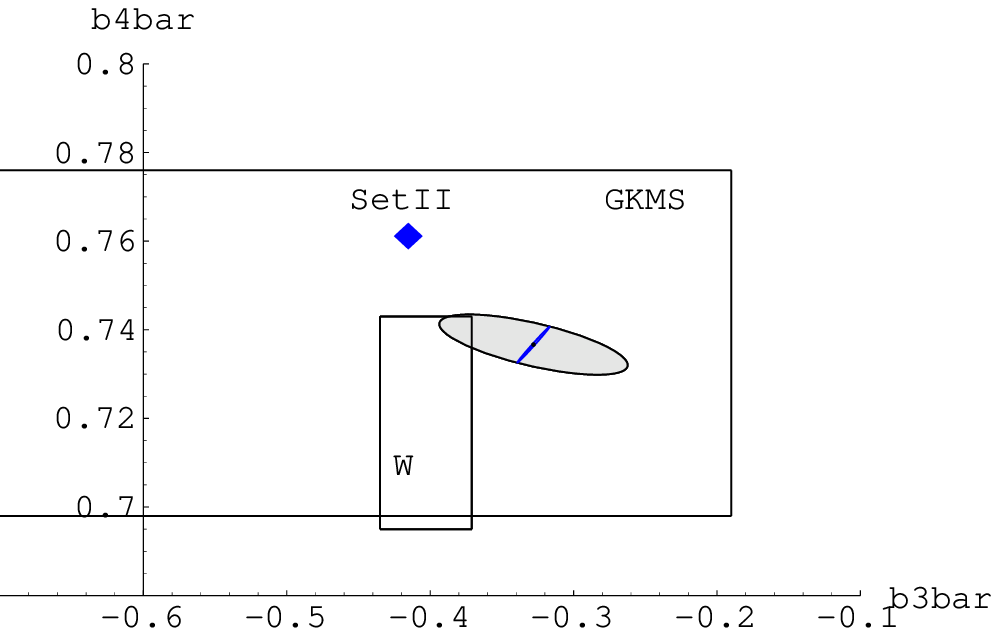}
\caption{\label{fig:b34} Result for $b_3$ and $b_4$. The errors in
our result are dominated by those} in the experimental input used when
solving
the Roy equations: The nearly degenerate ellipse indicates the result
obtained
if these could be ignored. The rectangles correspond to the values quoted in
refs.~\cite{Wanders 1997b} and \cite{Girlanda:1997ed}, while the diamond 
marks the one obtained in ref.~\cite{BCEGS}, set II. 
\end{figure}
The error bars indicate the uncertainties in the outcome due to those in the
experimental input used when solving the Roy equations. 
The representation holds for arbitrary values of the scattering lengths in
the
vicinity of the point of reference. Inserting our results from (\ref{eq:final
  result}) and adding errors quadratically, we finally obtain
\bea\label{eq:bbarnum}
\bbar_1\al=\al  -12.4 \pm 1.6\,,\hspace{1.7em} 
\bbar_2=    11.8 \pm 0.6\,,\hspace{1.5em}
\bbar_3=-0.33 \pm 0.07\,,\no
\bbar_4\al=\al  0.74 \pm 0.01\,,\hspace{2em}
\bbar_5=    3.58 \pm 0.37\,,\hspace{1em}
\bbar_6=    2.35 \pm 0.02\,.\eea
We emphasize that the error bars only indicate the noise seen in our
evaluation. In $\bbar_1,\ldots\,,\bbar_4$, the two loop representation
does account for the contributions of next--to--leading order, but in the 
case of $\bbar_5,\bbar_6$, it only yields the  
leading terms -- these quantities are particularly sensitive to the
neglected higher orders.  
\begin{figure}[thb]
\psfrag{2.375}{}
\psfrag{2.35}{$2.35$}
\psfrag{2.325}{}
\psfrag{2.3}{\hspace{-0.5em}$2.30$}
\psfrag{2.275}{}
\psfrag{2.25}{$2.25$}
\psfrag{2.225}{\hspace{0.5em}\raisebox{-1.5em}{$2.20$}}
\psfrag{2.35}{$2.35$}
\psfrag{3.2}{\raisebox{-0.2em}{$3.2$}}
\psfrag{3.4}{\raisebox{-0.2em}{$3.4$}}
\psfrag{3.6}{\raisebox{-0.2em}{$3.6$}}
\psfrag{3.8}{\raisebox{-0.2em}{$3.8$}}
\psfrag{4}{\hspace{-0.5em}\raisebox{-0.2em}{$4.0$}}
\psfrag{4.2}{\raisebox{-0.2em}{$4.2$}}
\psfrag{b5bar}{\hspace{0.5em}\large$\bbar_5$}
\psfrag{b6bar}{\raisebox{0.5em}{\large$\bbar_6$}}
\psfrag{SetII}{\footnotesize BCEGS II}
\psfrag{W}{\footnotesize W}
\psfrag{W1}{\hspace{0.3em}\footnotesize W}
\psfrag{GKMS}{\footnotesize GKMS}
\psfrag{0.28}{\hspace{0.5em}280}
\psfrag{0.3}{\hspace{0.2em}300}
\psfrag{0.32}{\hspace{0.5em}320}
\psfrag{0.34}{\hspace{0.5em}340}
\psfrag{0.36}{\hspace{0.5em}360}
\psfrag{0.38}{\hspace{0.5em}380}
\leavevmode

\hspace{4em}
\includegraphics[width=10cm]{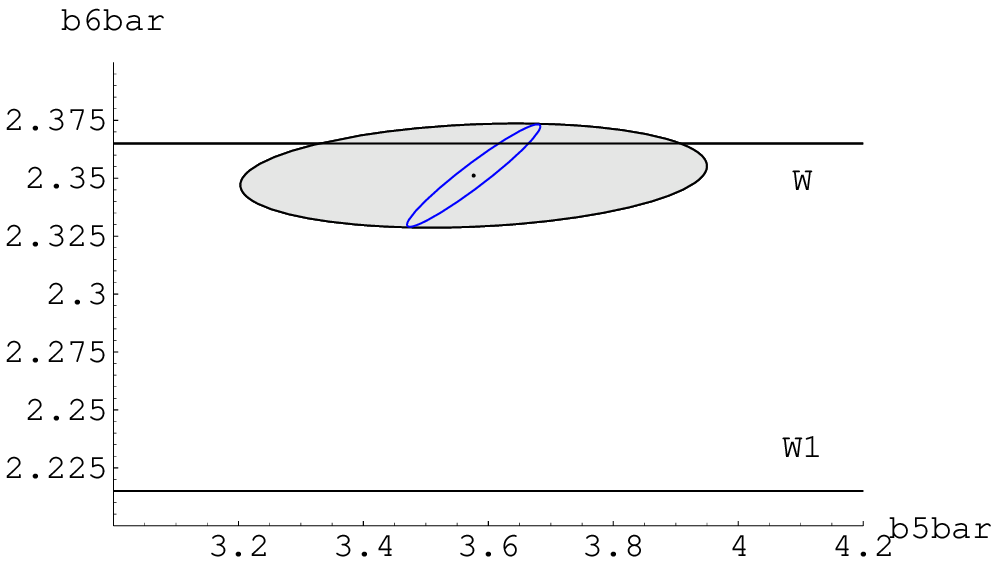}
\caption{\label{fig:b56} Result for $b_5$ and $b_6$. The strip between the 
two horizontal lines corre-}
sponds to the value for $b_6$ of Wanders \cite{Wanders 1997b}. 
\end{figure}

The above results may be compared with the values found in the literature. 
Girlan\-da, Knecht, Moussallam and Stern \cite{Girlanda:1997ed} work within 
generalized chiral perturbation theory and do
not have a prediction for the magnitude of the
coefficients $b_1$ and $b_2$, because the corresponding expressions contain
the two free parameters $\alpha$ and $\beta$. In their framework, the analogs
of the constants 
$b_3,\ldots\,,b_6$ are denoted by $\lambda_1,\ldots\,,\lambda_4$. The
explicit
relation between the two sets of quantities is given in
eq.~(\ref{eq:lambda}).
In our notation, the numerical values of ref.~\cite{Girlanda:1997ed} 
correspond to $\bbar_3=-0.56\pm 0.37$, $\bbar_4=0.737\pm 0.039$,
$\bbar_5=3.25\pm 1.50$, $\bbar_6=2.42\pm 0.22$ and are perfectly
consistent with our results, where the errors are smaller. 
The result for $\bbar_1$ and $\bbar_2$, obtained above within the
standard framework, amounts to a
prediction for the magnitude of $\alpha$ and $\beta$.  
Numerically, we obtain
\bea\label{eq:alphabeta} 
\alpha=1.08 \pm 0.07\co\hspace{2em}\beta=1.12\pm 0.01\fs\eea

Wanders \cite{Wanders 1997b} has obtained values for the coefficients 
$b_3$, $b_4$ and $b_6$ from manifestly crossing symmetric dispersion 
relations. Matching the chiral and dispersive representations at the
center
of the Mandelstam triangle, he obtains the values $\bbar_3=-0.403\pm
0.032$, 
$\bbar_4=0.719\pm0.024$, $\bbar_6=2.29\pm0.075$, which are also
consistent
with our numbers.  Note that the quoted errors only account for the
uncertainties arising from the procedure used in ref.~\cite{Wanders
1997b} and do not cover
those in the input. Fig.~\ref{fig:b34} shows that, in the case of
$\bbar_3$, the experimental input in the Roy equations represents the
dominating source of error.
 
Amoros, Bijnens and Talavera \cite{ABT_Ke4} have determined the
coefficients $b_n$ on the basis of their analysis of the $K_{e_4}$ form 
factors, referred to earlier. The results for the coefficients
$b_3,\ldots\,,b_6$ are accompanied by rather
large errors and we do not list these here, but merely note that 
the central values in eq.~(\ref{eq:bbarnum})
are within the quoted range, in all cases. 
For the first two terms, however, Amoros et al.~arrive at comparatively 
accurate values,
$\bbar_1= -10.8\pm 3.3$, $\bbar_2=10.8\pm3.2$, which are also perfectly
consistent with those in eq.~(\ref{eq:bbarnum}). The fact that, in their
analysis, the remaining coefficients are subject to large uncertainties,
also manifests itself in column C of table \ref{tab:threshold}: The 
error bars in the first five rows of the table, $a_0^0\, \cdots\, a_1^1$,
are much smaller than those in the remainder. 
 
\setcounter{equation}{0}
\section{{\boldmath $S$-- and $P$--wave phase shifts\unboldmath}}
\label{sec:num rep}

\begin{figure}[thb]
\vspace{0.8cm}
\leavevmode
\begin{center}
\includegraphics[width=14cm]{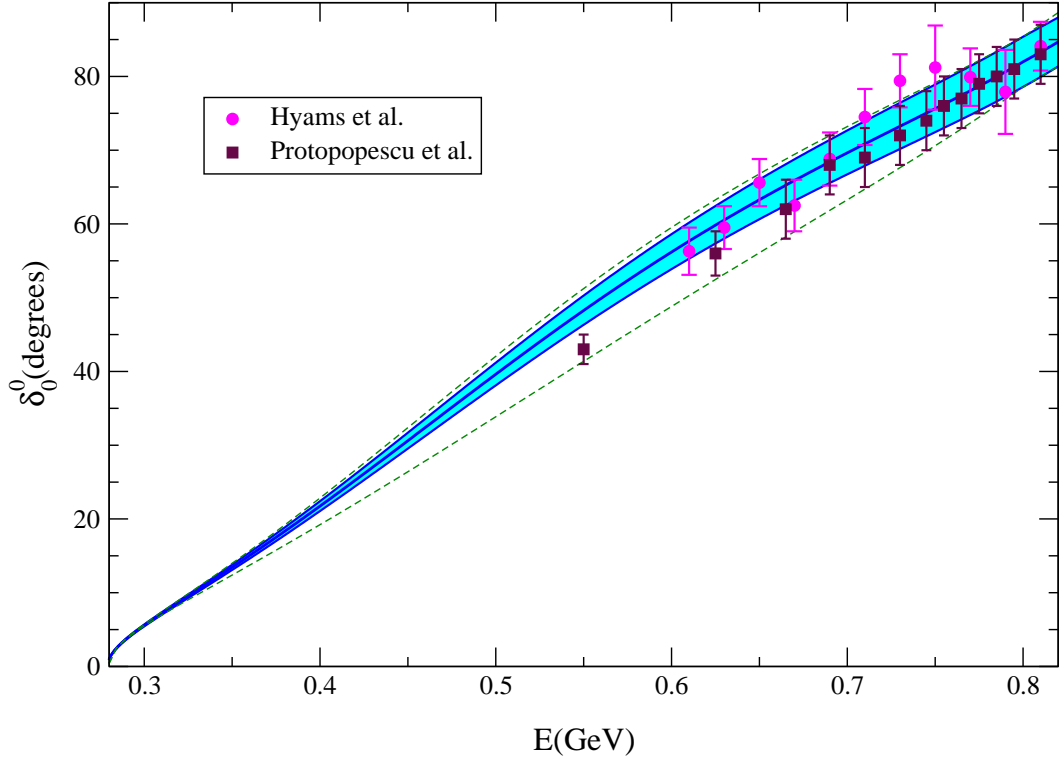}
\end{center}
\caption{\label{fig:Swave}
$I=0$ $S$--wave phase shift. The full line
results with the central values of the scattering
lengths and of the
experimental input used in the Roy equations. The
shaded region corresponds to the uncertainties of the result. The dotted
lines
indicate the boundaries of the region allowed if the constraints imposed by
chiral symmetry are ignored \cite{ACGL}. The data points are from
refs.~\cite{hyams} and \cite{Protopopescu}.}
\end{figure}

\begin{figure}[thb]
\leavevmode
\begin{center}
\includegraphics[width=14cm]{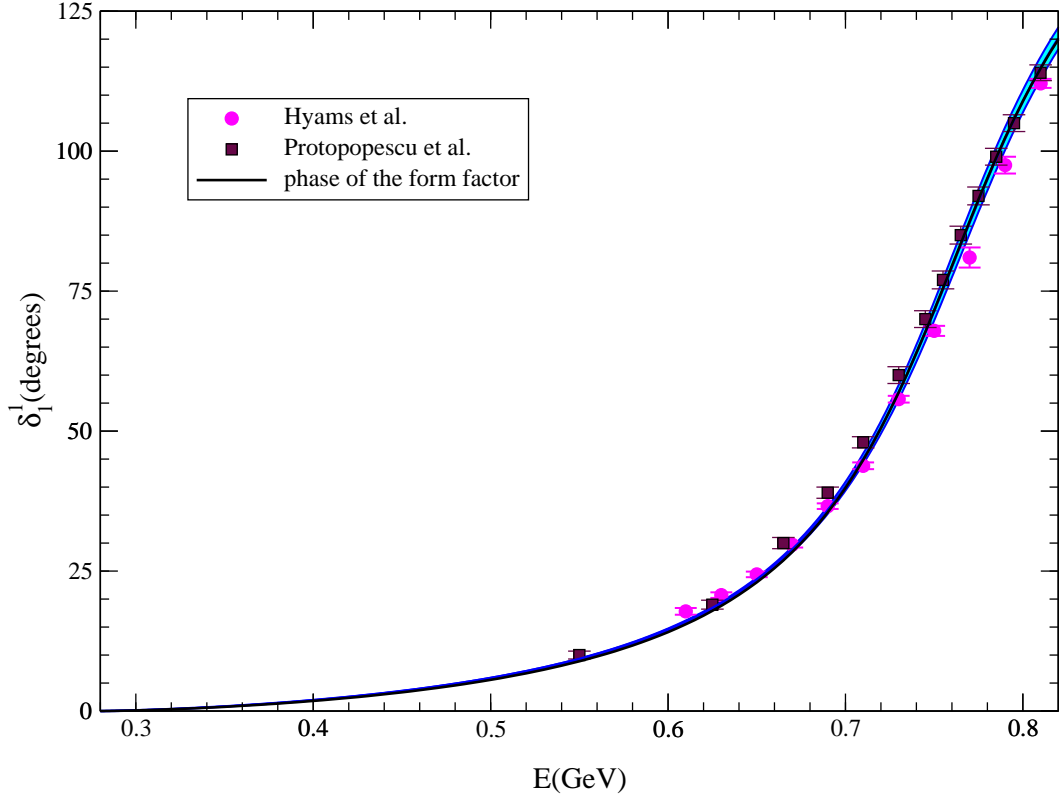}
\end{center}
\caption{\label{fig:Pwave}
$P$--wave phase shift. The phase of the pion form factor is also shown, but
it
can barely be distinguished from the central result of our analysis. 
The data points are
from refs.~\cite{hyams} and \cite{Protopopescu}.} 

\end{figure}

\begin{figure}[thb]
\vspace{0.8cm}
\leavevmode
\begin{center}
\includegraphics[width=14cm]{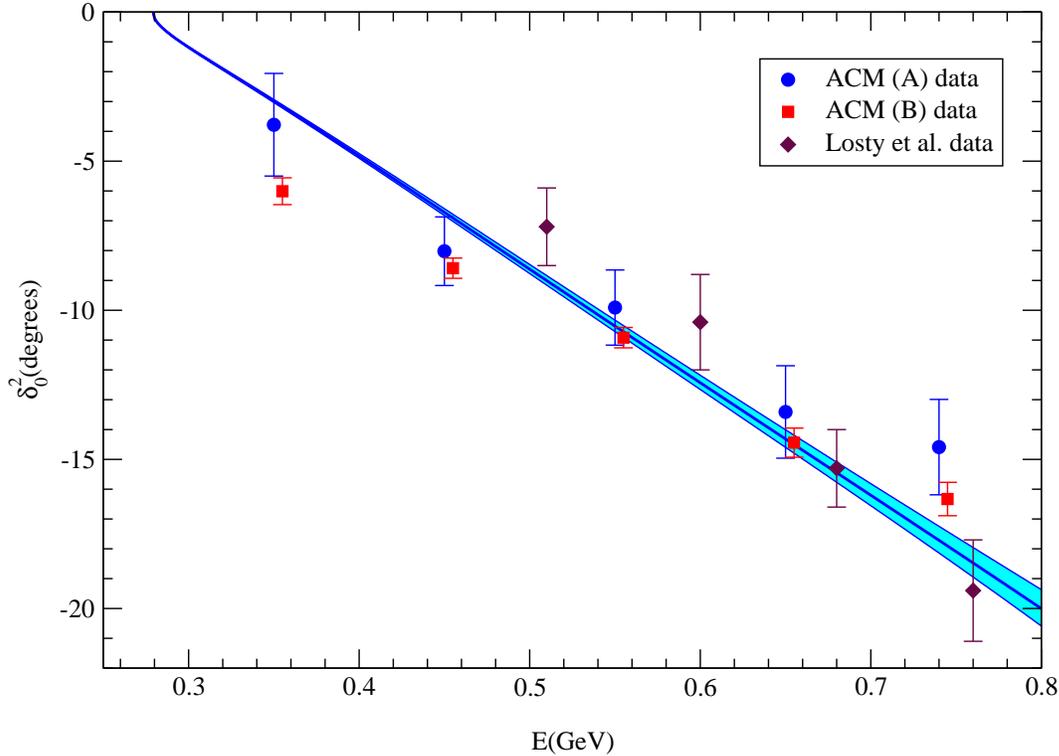}
\end{center}
\caption{\label{fig:S2wave}
$I=2$ $S$--wave phase shift. The full line
results with the central values of the scattering
lengths and of the
experimental input used in the Roy equations. The
shaded region corresponds to the uncertainties of the result. The data points
represent the two phase shift representations of the
Aachen-Cern-Munich collaboration \cite{hoogland} and the one of
Losty et al.~\cite{losty}}
\end{figure}

For the reasons discussed in detail in ref.~\cite{ACGL}, the two $S$--wave
scattering lengths are the essential parameters in the low energy domain.
The result in eq.~(\ref{eq:final result}) specifies these
to within very small uncertainties. In particular, we can now work out 
the phase shifts of the $S$-- and $P$--waves on this basis, using the Roy
equation analysis of \cite{ACGL}. The available experimental information for 
the imaginary parts above $\sqrt{s_0}=0.8 \,\mbox{GeV}$, 
as well as the scattering
lengths $a_0^0$, $a_0^2$ are used as an input, while the
output of the calculation 
consists of the phases $\delta_0^0(s)$, $\delta_1^1(s)$
and $\delta_0^2(s)$, in the region below $s_0$. 
In view of the two
subtractions occurring in the Roy equations, the behaviour of the
imaginary parts above 1 GeV has very little influence on the behaviour of the
solutions below $0.8 \,\mbox{GeV}$. Also, there is a consistency
check: In the region above $s_0$, the output must agree
with the input. For the values of the scattering lengths required by chiral
symmetry, this condition is indeed met. In fact, the solution of the Roy
equations closely follow the input, within the rather broad range of 
variations allowed for the imaginary parts in ref.~\cite{ACGL}.
This also means that the Roy equations do not strongly constrain the
behaviour
of the phases above 0.8 GeV.
 
The result is shown in figs.~\ref{fig:Swave}, \ref{fig:Pwave}
and \ref{fig:S2wave}. For comparison, these figures also show the
data points of the phase shift analyses given by Hyams et al.~\cite{hyams},
Protopopescu et al.~\cite{Protopopescu}, the solutions A and B of Hoogland et
al.~(ACM) \cite{hoogland} and the one of  
Losty et al.~\cite{losty}, 
as well as the $P$--wave phase extracted from the data
on the reactions $e^+e^-\rightarrow\pi^+\pi^-$ and $\tau\rightarrow\nu\pi\pi$.
For further information on the $S$--wave phase shifts, we refer the reader to
\cite{Takamatsu:2001ew,GAMS}. 

The three central curves are described by the parametrization \cite{Schenk}
\begin{equation}
\tan \delta_\ell^I = \sqrt{1-{4 M_\pi^2 \over s}}\; q^{2\, \ell} 
\left\{A^I_\ell
+ B^I_\ell q^2 + C^I_\ell q^4 + D^I_\ell q^6 \right\} \left({4
  M_\pi^2 - s^I_\ell \over s-s^I_\ell} \right) \; ,
\end{equation} 
\newpage\noindent
with $s=4\,(M_\pi^2+q^2)$. The numerical values of the coefficients are:
\bea \label{eq:Schenk 1}
A_0^0\al=\al .220\co\hspace{4.5em}
A^1_1=.379\cdot 10^{-1}\co\hspace{2.2em}
A^2_0= -.444\cdot 10^{-1}\co\no
 B_0^0\al=\al .268\co\hspace{4.5em}
B^1_1=  .140\cdot 10^{-4}\co\hspace{2em}
B^2_0= -.857\cdot 10^{-1}\co\\
C_0^0\al=\al -.139\cdot 10^{-1}\co\hspace{1em}
C^1_1= -.673\cdot 10^{-4}\co\hspace{1.3em}
C^2_0= -.221\cdot 10^{-2}\co\no
D_0^0\al=\al -.139\cdot 10^{-2}\co\hspace{1em}
D^1_1=  .163\cdot 10^{-7}\co\hspace{2em}
D^2_0= -.129\cdot 10^{-3}\co\nonumber\eea
in units of $M_\pi$.
In particular, the constants $A_\ell^I$ represent the scattering lengths
of the three partial waves under consideration, while the $B_\ell^I$ are
related to the effective ranges.

The parameters $s_\ell^I$ specify the value of $s$ 
where $\delta_\ell^I(s)$ passes through $90^\circ$:
\bea\label{eq:Schenk 2} s_0^0=  36.77\,M_\pi^2\co\hspace{1em}
s^1_1=  30.72\,M_\pi^2\co\hspace{1em}
s^2_0= -21.62\,M_\pi^2\fs\eea
In the channels with $I=0,1$, the corresponding energies are 
$846\, \mbox{MeV}$ and 
$774\, \mbox{MeV}$, respectively 
(the negative sign of $s_0^2$ indicates that in the $I=2$ channel, which is
exotic, the phase remains below $90^\circ$). 

The value of
 the phase difference $\delta^0_0-\delta^2_0$ at $s=M_K^2$
is of special interest, in connection with the decays $K\rightarrow\pi\pi$.
In particular, the phase of $\epsilon'/\epsilon$ is determined by
that phase difference. Our representation of the scattering amplitude allows
us to pin this quantity down at the 3\% level of acuracy:
\bea \delta_0^0(M_{K^0}^2)-\delta^2_0(M_{K^0}^2)=
47.7^\circ \pm 1.5^\circ\fs\eea

We add two remarks concerning the comparison with the $P$--wave phase
  shift extracted 
from the $e^+e^-$ and $\tau$ data. First, we note that the agreement 
at 0.8 GeV is enforced by our approach: In the Roy equation
analysis, the value of the phase shift at that energy
represents an input parameter and we have made use
of those data to pin it down. Once that is done, however, 
the behaviour of the phase shift at lower energies is  
unambiguously fixed: Chiral symmetry determines the two
subtraction constants, so that the solution of the Roy equations becomes
unique.  In other words -- disregarding the small effects due to the
uncertainties  in the input of our analysis, which are shown in
fig.~\ref{fig:Pwave} -- there is only one 
interpolation between threshold and 0.8 GeV that is consistent with 
the constraints imposed by analyticity, unitarity and
chiral symmetry. Figure \ref{fig:Pwave} 
shows that the predicted curve indeed very closely
follows the phase extracted from the $e^+e^-$ and $\tau$ data. This confirms
the conclusions reached in ref.~\cite{Pich Portoles}.  

Actually, the figure conceals a discrepancy in the threshold region, where
the phase is too small for the effect to be seen by eye: Evaluating the
$P$--wave scattering length with the Gounaris--Sakurai parametrization of
the form factor given in ref.~\cite{CLEO} (the curve shown in the figure),
we obtain a result that is smaller than the value for $a_1^1$ in table
\ref{tab:threshold}, by about 10\%, that is by many standard deviations of
our prediction.  The discrepancy is in the noise of the data on the form
factor: There is little experimental information in the threshold region,
so that the behaviour of the form factor is not strongly constrained
there. Indeed, there are alternative parametrizations that also fit the
data, but have a distinctly different behaviour near and below
threshold. Even parametrizations with unphysical singularities at $s=0$,
such as the one proposed in \cite{Kuehn Santamaria}, provide decent fits in
the experimentally accessible region.  In this respect, the present work
does add significant information about the $P$--wave phase shift, as it
predicts the behaviour near threshold, within very narrow limits.

\setcounter{equation}{0}
\section{Poles on the second sheet}
The partial wave amplitude $t^1_1(s)$ contains a pole
on the second sheet. Denoting the pole position by 
$s=(M_\rho-\frac{i}{2}\,\Gamma_\rho)^2$, we obtain
\bea\label{eq:Mrho} M_\rho=762.4\pm 1.8\,\mbox{MeV}\co\hspace{1em}
\Gamma_\rho=145.2\pm 2.8\,\mbox{MeV}\fs
\eea
Note that the
values quoted for the ``mass'' often represent the energy where the real part
of the amplitude vanishes -- in contrast to the position of the 
pole, that value is not independent of the process considered. As the
scattering is approximately elastic there, the corresponding mass is the
  energy where the phase shift goes through $90^\circ$. For the
$P$--wave, this happens at
\bea m_\rho=773.5\pm 2.5\,\mbox{MeV} \fs\nonumber\eea
The real part of the pole position is smaller 
than the energy where the phase
shift passes through $90^\circ$, by about 10 MeV. The uncertainty in
$\Gamma_\rho$ is significantly smaller than the error bar quoted in
\cite{ACGL}: The constraints imposed on the scattering amplitude by the low
energy theorems of chiral symmetry also allow a better determination of the
width. 

The  $I=0$ $S$--wave also contains a pole on the second sheet. 
The uncertainties in the pole position are considerably larger than in the
case of the $\rho$, because the singularity is far from 
the real axis. Also, the uncertainties in the phase shift are somewhat larger
here. Varying the input parameters as well as the analytic form of the 
representation used for $t_0^0$, we find that
the pole occurs in the region 
$\sqrt{s}=(470\pm 30)-\,i\, (295\pm 20)\,\mbox{MeV}$,
while the phase passes through $90^\circ$ at $\sqrt{s}=844 \pm 13
\,\mbox{MeV}$. 

There is no harm in
calling this an unusually broad resonance, but that sheds  
little light on the low energy structure of the scattering amplitude. 
In particular, it should not come as a surprise if
the values for the mass and width of the resonance, obtained on the basis 
of the assumption that the pole represents the most important feature in this
channel, are very different from the real and imaginary parts of the
energy at which the amplitude actually has a pole  -- there is more
to the physics of the $S$--wave than the occurrence of a pole far from the
real axis. A collection of numbers concerning the pole 
position is given in \cite{PDG} and for a recent review of the abundant 
literature on the subject, we refer to \cite{sigma}. 
A recent discussion in the framework of the $NN$
interaction is given in \cite{Oset}.

We add a remark concerning the physics behind the pole in $t^0_0$ --
admittedly, the reasoning is of qualitative nature. 
In the chiral limit, current algebra predicts
$t^0_0=s/16 \pi F_\pi^2$: The amplitude vanishes at threshold, but the real
part grows quadratically with the energy, so that the imaginary part rises
with the fourth power. The rapid growth signals the occurrence of a strong
final state interaction.  In order to
estimate the strength of the corresponding branch cut, we invoke the inverse 
amplitude method, replacing the above formula by 
$t_0^0=s/(16\pi F_\pi^2-i\, s)$. The virtue of this operation is that, while
it retains the algebraic accuracy of the current algebra approximation, 
it yields an expression that does obey elastic unitarity. The formula shows
that, in this approximation, the amplitude contains a pole at
$\sqrt{s}=\sqrt{-i\,16\,\pi}\,F_\pi=463 -i\, 463 \,\mbox{MeV}$, indeed 
not far away from the place where the full amplitude has a pole. 

The physics of the $P$--wave is very different, because the unitarity
cut generated by 
low energy $\pi\pi$ intermediate states 
is very weak. Repeating the above exercise for $t^1_1$, one again finds a
pole
with equal real and imaginary parts, 
but it is entirely fictitious, as it occurs at $1.1-i\,1.1$ GeV, far 
beyond the region where current
algebra provides a meaningful approximation. The occurrence of a pole 
near the real axis cannot be understood on the basis of chiral symmetry
and unitarity alone. 

In the framework of the effective theory, the difference
manifests itself as follows. While the unitarity corrections account
perfectly well for the low energy behaviour of the imaginary parts, the
presence of the $\rho$ only shows up in the values of the effective
coupling constants $\ell_1,\ell_2$ and $\ell_6$. There is no such pole
in $t^1_1$, for instance, if the underlying theory is identified with the 
linear $\sigma$--model, and the values of those coupling constants are then
very different \cite{GL 1984}. In this sense, the pole in $t^1_1$ reflects a
special property of QCD, while the one in $t^0_0$ 
can be understood on the basis of the fact that chiral symmetry
predicts a strong unitarity cut: The pole position is related to the
magnitude
of $F_\pi$. 

\setcounter{equation}{0}
\section{Threshold parameters}
\label{sec:other thp}
\begin{table}[thb]
\hspace*{-0.5em}\begin{tabular}{|r||r||r|r||r|r||l|}
\hline
&A\rule{2.5em}{0em}&B\rule{.95em}{0em}&C\rule{2.5em}{0em}&D\rule{1.8em}{0em}&
E\rule{1.7em}{0em}&\rule{1em}{0em}units
\\\hline\hspace{-0.3em}\rule{0em}{1em}
$a_0^0$&$ .220\pm .005$&$.216$&$.220\pm .005$&$.24\pm.06$&$.26\pm .05$&
\\\rule[-0.3em]{0em}{0em}
$b^0_0$&$.276\pm .006$
&$.268$&$.280\pm .011$&$.26\pm.02$&$.25\pm.03$&\rule{2em}{0em}$M_\pi^{-2}$
\\\hline\hspace{-0.3em}\rule{0em}{1em}
$a_0^2$&$-.444\pm .010$&$-.445$&$-.423\pm .010$&$-.36\pm.13$&$-.28\pm .12$&
$10^{-1}$
\\\hspace{-0.3em}\rule[-0.5em]{0em}{0em}
$b_0^2$&$-.803\pm .012$
&$-.808$&$-.762\pm .021$&$-.79\pm .05$&$-.82\pm .08$&$10^{-1}M_\pi^{-2}$
\\\hline\hspace{-0.3em}\rule{0em}{1em}
$a_1^1$&$.379 \pm .005$&$.380$&$.380\pm .021$
&$.37\pm .02$&$.38\pm .02$&$10^{-1}M_\pi^{-2}$
\\\hspace{-0.3em}\rule[-0.5em]{0em}{0em}
$b_1^1$&$.567\pm .013$&$.537$&
$.58\pm.12$\g&$.54\pm.04$&&$10^{-2}
M_\pi^{-4}$
\\\hline\hspace{-0.3em}\rule{0em}{1em}
$a^0_2$&$.175\pm .003$
&$.176$&$.22\pm .04$\g&$.17\pm.01$&
$.17\pm .03$&
$10^{-2}M_\pi^{-4}$
\\\hspace{-0.3em}\rule[-0.5em]{0em}{0em}
$b_2^0$&$-.355\pm .014$&$-.343$&
$-.32\pm .10$\g&$-.35\pm.06 $&&$10^{-3}M_\pi^{-6}$
\\\hline\hspace{-0.3em}\rule{0em}{1em}
$a^2_2$&$.170\pm .013$
&$.172$&$.29\pm.10$\g&$.18\pm.08$&$.13\pm .3$\g&$10^{-3}M_\pi^{-4}$
\\\hspace{-0.3em}\rule[-0.5em]{0em}{0em}
$b_2^2$&$-.326\pm .012$&$-.339$&
$-.36\pm .9$\g\g&$-.34\pm.07$&
&$10^{-3}M_\pi^{-6}$
\\\hline\hspace{-0.3em}\rule{0em}{1em}
$a_3^1$&$.560\pm .019$&$.545$&$.61\pm.11$\g
&$.58\pm.12 $
&$.6\pm .2$\g&$10^{-4}M_\pi^{-6}$
\\\hspace{-0.3em}\rule[-0.5em]{0em}{0em}
$b_3^1$&$-.402\pm .018$&$-.312$&$-.36\pm .02$\g&$-.44\pm.14$
&&$10^{-4}M_\pi^{-8}$
\\\hline
\end{tabular}
\caption{\label{tab:threshold}
Threshold parameters.  
Our results are listed in column A. The numbers in the next two columns 
are obtained by evaluating the chiral representation at threshold:
The entries under B follow from our values
of the effective coupling constants, while those under C are taken from 
ref.~\cite{ABT_Ke4}. Column D gives the outcome of a Roy equation analysis
that does not invoke chiral symmetry \cite{ACGL}, while E contains the old
``experimental'' values \cite{Nagels}.}
\end{table}

The scattering lengths of the partial waves with $\ell\geq1$,
as well as the effective ranges (also of those of the $S$--waves)  
can be expressed in terms of sum rules over the imaginary
parts \cite{Wanders sum rules}. The corresponding numerical values are listed
in the table \ref{tab:threshold}, together with 
the $S$--wave scattering lengths.
Column A indicates our final results, obtained by matching the
phenomenological and chiral representations in the subthreshold
region and using the Roy equations to evaluate the amplitude and its
derivatives at threshold. In column B,
we quote the numbers obtained from a direct evaluation of the two loop
representation at threshold, using our central
values for the effective coupling constants -- 
this amounts to truncating the expansion of the threshold parameters
in powers of $m_u$ and $m_d$. Column C lists the
results of ref.~\cite{ABT_Ke4}, where the amplitude is also expanded at
threshold, but the coupling constants are 
determined on the basis of a two loop analysis of the $K_{e4}$
form factors (see the next section for a comment concerning these entries). 
The comparison of the columns A, B and C 
clearly shows that two loop chiral
perturbation theory works very well in describing both $\pi \pi$
scattering and $K_{e4}$ decays. For the reasons given in section
\ref{sec:infrared singularities}, the method described in the present paper
yields the smallest error bars. 
In fact, it is quite remarkable that the results for
the effective range $b_3^1$ in columns B and C represent a decent 
estimate:
In this case, only the infrared singularities occurring in the
expansion in powers of the quark masses contribute. For comparison,  
column D lists the values of ref.~\cite{ACGL}, which are obtained 
by analyzing 
the available data with the Roy equations and do not invoke chiral
perturbation theory. Finally, column E contains the values of the
compilation in refs.~\cite{Nagels}.

\setcounter{equation}{0}
\section{\boldmath{Quark mass dependence of $M_\pi^2$ and $F_\pi$}}
\label{sec:dep qm}

The dependence of physical quantities 
on the quark masses is of interest, in particular, for the following
reason: By now, dynamical quarks with a mass of the order of 
the physical value of $m_s$ are within reach on the lattice, 
but it is notoriously 
difficult to equip the two lightest quarks with their proper masses. 
Invariably, the numerical results obtained for the physical values of $m_u$
and $m_d$ rely on an extrapolation of numerical data. For a recent,
comprehensive review of lattice work on the light quark masses, we
refer to \cite{Lubicz:2001ch}.  

In this connection,
chiral perturbation theory 
may turn out to be useful, because it predicts the mass dependence in terms
of
a few constants: the coupling constants of the effective Lagrangian. Above,
we
have determined some of these and we now discuss the consequences for the
dependence of $M_\pi^2, F_\pi, a_0^0, a_0^2$ on the masses of the two
lightest quarks: We keep $m_s$ fixed at the physical value and 
set $m_u=m_d=m$, 
but vary the value of $m$ in the range $0< m<\frac{1}{2}\,m_s$ (at
the upper end of that range, the pion mass is about 500 MeV).

In the preceding sections, we have expressed all of the quantities in terms
of the physical pion mass and the physical decay constant, using the ratio
$\xi$ as an expansion parameter. Also, the logarithmic infrared
singularities were normalized at the scale $M_\pi$. In particular, the
coupling constants $\lbar_n$ contain a chiral logarithm with unit
coefficient, so that they may be represented as 
\bea
\lbar_n\equiv\ln\frac{\Lambda_n^{\,2}}{M_\pi^2}\co
\eea 
where $\Lambda_n$ is
the intrinsic scale of $\ell_n$ and is independent both of $m$ and of the
running scale $\mu$.  In order to explicitly exhibit the quark mass
dependence, we replace $M_\pi^2$ by the variable $M^2\equiv 2\,m\,B$ and
also normalize the chiral logarithms at the scale $M$, trading the
quantities $\xi$ and $\lbar_n$ for 
\bea 
x\equiv\frac{M^2}{16\,\pi^2
F^2}\co\hspace{2em} \lhat_n\equiv\ln\frac{\Lambda_n^{\;2}}{M^2}\co
\eea
respectively. According to (\ref{eq:Mpiexp}), (\ref{eq:Fpiexp}) the two
sets of variables are related by \bea
x\al=\al\xi\{1+\mbox{$\frac{1}{2}$}\,\xi\,(\lbar_3+4\lbar_4)+
O(\xi^2)\}\co\hspace{2em}\lhat_n=\lbar_n-
\mbox{$\frac{1}{2}$}\,\xi\,\lbar_3+O(\xi^2)\\ \xi\al=\al
x\{1-\mbox{$\frac{1}{2}$}\,x\,(\lhat_3+4\lhat_4)+O(x^2)\}
\co\hspace{1.8em}\lbar_n=\lhat_n+ \mbox{$\frac{1}{2}$}\,x\,\lhat_3+O(x^2)
\fs\nonumber\eea

The expansions of $M_\pi^2$ and $F_\pi$ in powers of $m$
are known to next--to--next--to--leading order 
\cite{Buergi,Colangelo double logs,BCEGS}. 
In the above notation, the explicit expressions may be written in the 
form \cite{Leutwyler Bangalore}:
\bea\label{eq:MF} M_\pi^2\al=\al M^2\,\{ 
1-\mbox{$\frac{1}{2}$}\,x\,\lhat_3+
\mbox{$\frac{17}{8}$}\,x^2 \lhat_{\!\ind M}^{\;2}+
x^2 k_{\ind M}+O(x^3)\co\no
F_\pi\al=\al F\,\{1+x\,\lhat_4-\mbox{$\frac{5}{4}$}\,x^2
\lhat_{\!\ind F}^{\;2}
+x^2 k_{\ind F}+O(x^3)\co\no
\lhat_{\!\ind M}\al=\al \frac{1}{51}\,(28\,\lhat_1+32\,\lhat_2-
9\,\lhat_3+49)\co\\
\lhat_{\!\ind F}\al=\al \frac{1}{30}\,(14\,\lhat_1+16\,\lhat_2+
6\,\lhat_3-6\,\lhat_4+23)\co
\nonumber\eea 
In this representation, the infrared singularities are hidden in 
the scale invariant quantities $\lhat_1,\ldots\,,\lhat_4$. 
Those generated by the two loop graphs have been completed to 
a square. The normalization of the auxiliary quantities
$\lhat_{\!\ind M}$, $\lhat_{\!\ind F}$ is chosen such that their mass
dependence is also of the form
\bea \lhat_{\!\ind M}=\ln\frac{\Lambda_{\!\ind M}^{\,2}}{M^2}\co\hspace{2em}
\lhat_{\!\ind F}=\ln\frac{\Lambda_{\!\ind F}^{\,2}}{M^2}\fs\eea
The constants $k_{\!\ind M}$, $k_{\!\ind F}$ collect the analytic
contributions at order $x^2$ and are independent of 
$m$ and $\mu$. By completing the logarithms at 
oder $x^2$ to a square, 
we have in effect chosen a particular running scale: the one where the
coefficient of the term linear in $\ln M^2$ vanishes. This simplifies the
representation, but is without physical significance -- the decomposition
into an infrared singular part arising from the Goldstone bosons and  
a regular remainder is not unique. In ref.~\cite{Bijnens Colangelo Talavera},
a somewhat different representation for the analytic terms of order $x^2$ is
used, which involves the two parameters $r_{\ind M}$, $r_{\ind F}$ instead
of $k_{\!\ind M}$, $k_{\!\ind F}$. 

\setcounter{equation}{0}
\section{\hspace{-0.5em}Numerical results for quark mass dependence}
\label{sec:quark mass dependence}
\begin{figure}[thb] 
\vspace{2em}
\centering
\hspace*{0em}\mbox{\epsfysize=6.5cm \epsfbox{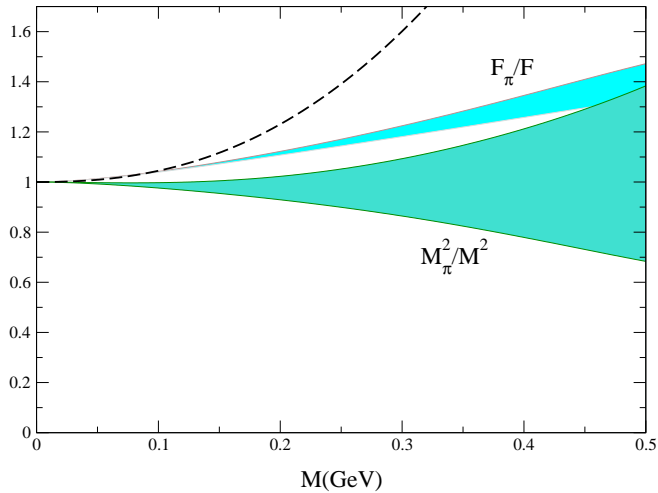} }

\caption{\label{fig:MF}Dependence of the ratios $F_\pi/F$ and $M_\pi^2/M^2$
on 
the mass of the two lightest quarks. The variable $M$
is defined by $M^2=(m_u+m_d)\,B$ and $F$ is the value of the pion decay
constant for $m_u=m_d=0$. The strange quark mass is 
held fixed at the
physical value.} 
\end{figure}

In addition to the coupling constants $\ell_1,\ldots,\ell_4$ that also govern
the low energy properties of the $\pi\pi$ scattering amplitude, the ratios
$M_\pi^2/M^2$ and $F_\pi/F$ contain 
the two fourth order constants $k_{\!\ind M}$ and $k_{\!\ind F}$. We
expect that the contributions from these terms are of order 
$M_\pi^4/M_{\ind S}^4$ and can just as well be dropped --
unless  $m$ is taken much larger than in nature. We did not make an attempt
at
quantifying the uncertainties associated with these terms, because
they are small compared to those from the coupling constants
$\ell_1,\ldots,\ell_4$. The bands shown in fig.~\ref{fig:MF} 
correspond to $r_{\ind M}=r_{\ind F}=0$. If we were to instead
set the constants $k_{\!\ind M}$, $k_{\!\ind F}$ equal to zero, the
boundaries 
would be slightly
shifted, but the shifts are small compared to the width of the bands.

For small values of $m$, the contributions of $O(m^2)$ dominate. These
are determined by the two scales $\Lambda_3$ and $\Lambda_4$.
As discussed in section \ref{sec:l34}, the information about 
the first one is meagre -- the crude estimate (\ref{eq:l3barnum}) amounts
to $0.2\,\mbox{GeV}<\Lambda_3<2\,\mbox{GeV}$. For the second one, however,
the value for $\lbar_4$ obtained in section \ref{sec:final results} yields
a rather decent determination: 
\bea \Lambda_4=1.26\pm0.14\,\mbox{GeV}\fs\eea
The two parameters $\Lambda_3,\Lambda_4$ play the same role as the
coefficients
$c_{\ind M}$ and $c_{\ind F}$ in the polynomial approximations 
$M_\pi^2=2\,m B\,(1+c_{\ind  M} \,m)$, $F_\pi=F\,(1+c_{\ind F} \,m)$, that
are
sometimes used to perform the 
extrapolation of lattice data. In contrast to these
approximations, the formulae (\ref{eq:MF}) do account for the infrared
singularities of the functions
$M_\pi(m)$ and $F_\pi(m)$ -- to the order of the expansion in powers of $m$
considered, that representation is exact. 

Consider first the ratio $F_\pi/F$, for which the poorly known scale 
$\Lambda_3$ only enters at next--to--next--to--leading order. 
The upper one of the two shaded regions in fig.~\ref{fig:MF} 
shows the behaviour of this
ratio as a function of $M$, according to formula (\ref{eq:MF}).
The change in $F_\pi$ occurring 
if $M$ is increased from the physical value to $M_K$ is of 
the expected size, comparable to the difference between $F_K$ and $F_\pi$.
The curvature makes it evident
that a linear extrapolation in $m$ is meaningless. 
The essential para\-meter here is the
scale $\Lambda_4$ that determines the magnitude of the term of order $M^2$.
The corrections of order $M^4$ are small -- the scale relevant for these
is $\Lambda_F\simeq 0.5\,\mbox{GeV}$. 

In the case of the ratio $M_\pi^2/M^2$, on the other hand, the
dominating contribution is determined by the scale $\Lambda_3$ --
the corrections of $O(M^4)$ are small also in this case (the relevant scale
is 
$\Lambda_M\simeq 0.6\,\mbox{GeV}$). The fact that the information about the
value of $\ell_3$ is very meagre shows up through very large uncertainties.
In
particular, 
with $\Lambda_3\simeq 0.5 \,\mbox{GeV}$, the ratio $M_\pi^2/M^2$ would remain
very close to 1, on the entire interval shown. Note that outside the range
of values for $\ell_3$ considered in the present paper,
the dependence of $M_\pi^2$ on the 
quark masses would necessarily exhibit strong curvature. This is illustrated 
with the dashed line that indicates the behaviour of the ratio 
$M_\pi^2/M^2$ for $\lbar_3=-10$. According to fig.~\ref{fig:al3} this value
corresponds to $a_0^0\simeq 0.24$.

The above discussion shows
that brute force is not the only way to reach the very small values of
$m_u$ and $m_d$ observed in nature on the lattice. 
It suffices to equip the strange quark with the physical value of $m_s$ 
and to measure the dependence
of the pion mass on $m_u,m_d$ in the region where $M_\pi$ is comparable to
$M_K$. Since the dependence on the quark masses is known rather
accurately in terms of the two constants $B$ and $\Lambda_3$, a fit to the
data based on eq.~(\ref{eq:MF}) should provide an extra\-po\-lation to the 
physical quark masses that is under good control. Moreover, the resulting 
value for $\Lambda_3$ would be of considerable interest, because that scale
also shows up in other contexts, in the $\pi\pi$ scattering lengths, 
for example. For recent lattice work in this direction, we refer
to \cite{Heitger:2000ay}. 
A measurement of the mass dependence 
of $F_\pi$ in the same region would be useful too, because
it would provide a check on the dispersive analysis of the scalar radius
that underlies our determination of $\Lambda_4$ -- in view of the strong
unitarity cut in the scalar form
factor, a direct evaluation of the scalar radius on the lattice 
is likely more difficult. Chiral logarithms also occur in the quenched
approximation \cite{qCHPT,CP}, but since the coefficients differ from those
in the full theory, a naive comparison of the above formulae with quenched
lattice data is not meaningful.
\begin{figure}[thb] 
\vspace{2em}
\centering
\hspace*{1.4em}\mbox{\epsfysize=7.5cm \epsfbox{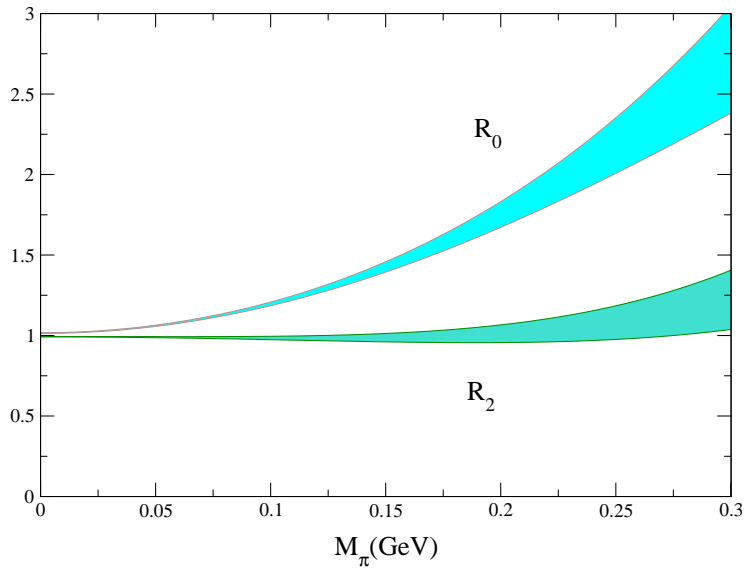} }
  
\caption{\label{fig:a0a2}
Size of the corrections to Weinberg's leading order predictions 
for the $\pi\pi$ $S$--wave scattering lengths, as a function of the pion
mass.}
\end{figure}

In section \ref{sec:infrared singularities}, we noted that the expansion of
the scattering length $a_0^0$ in powers of the quark mass
contains an unusually large infrared singularity
at one loop level. We can now complete that discussion with an evaluation of
the contributions arising at next--to--next--to--leading order, repeating the
analysis of ref.~\cite{Colangelo 1997} with the information about the
coupling
constants available now. The result is
shown in figure \ref{fig:a0a2}, where we indicate the behaviour of 
the correction factors $R_0,R_2$, defined by
\bea a_0^0=\frac{7\,M_\pi^2}{32\, \pi F_\pi^2}\,R_0\co\hspace{2em}
a_0^2=-\frac{M_\pi^2}{16\, \pi F_\pi^2}\,R_2\co\nonumber\eea
as a function of $M_\pi$. The reason for choosing the variable $M_\pi$
rather than $M$ is that the uncertainties in
$\Lambda_3$ then affect the result less strongly. 
The comparison with figure \ref{fig:MF}, where a larger mass range is shown, 
demonstrates that $R_0$ grows much
more rapidly with the quark masses than $F_\pi$.   
The effect arises from the
chiral logarithms associated with the unitarity cut -- the coefficient
of the leading infrared singularity in $a_0^0$ exceeds the one in $F_\pi$ 
by a factor $\frac{9}{2}$. Note that the 
chiral perturbation theory formulae underlying the figure are meaningful only
in the range where the corrections are small.
The shaded regions exclusively account for the 
uncertainties in the values of the coupling constants. In the case of $R_0$,
those due to the terms of order $M_\pi^6$ are by no
means negligible for $M_\pi>0.2\,\mbox{GeV}$, so that it matters what exactly
is plotted. The curves shown in the figure 
are obtained by expressing $R_0$, $R_2$ in terms
of the coupling constants $F,\,\ell_1,\ldots$ and of $M_\pi$, 
expanding the result in powers of $M_\pi$ and truncating the series at order
$M_\pi^4$. On the left half of the figure, the behaviour of $R_0$ obtained, 
for instance, by truncating the expansion in powers of $M$ instead of the one
in $M_\pi$ is practically the same, but on
the right half, there is a substantial difference, indicating that the
chiral perturbation series is out of control there.

We conclude that in the case of the $I=0$ scattering length, a meaningful
extrapolation of lattice data to the physical values of $m_u$ and $m_d$
requires significantly smaller quark
masses than in the case of $M_\pi$ or $F_\pi$. In the $I=2$ channel,
the effects are much smaller, because this channel is
exotic: The final state interaction is weak and repulsive. 
The lattice result, 
$a_0^2=-0.0374\pm 0.0049$ \cite{lattice I2}, corresponds to
$R_2=0.82\pm0.11$.
It is on the low side, but not
inconsistent with our prediction: $a_0^2=-0.0444\pm 0.0010$,
$R_2=0.98 \pm0.02$. As in the case of $F_\pi$ and $M_\pi$, the comparison
between the lattice result and our prediction is not really meaningful,
because that result relies on the quenched approximation. 
The evaluation of the scattering lengths to one loop in the
quenched approximation \cite{CP,BG} has shown that the infrared
singularities are different from those in full QCD. Moreover, as pointed out
by Bernard and Golterman \cite{BG}, the very method used to extract the
infinite volume scattering lengths 
from finite volume observables \cite{luscher} is affected: In addition to
the purely statistical error, the numbers in \cite{lattice I2} have
a sizeable systematic error.

\setcounter{equation}{0}
\section{Summary and conclusion}
\label{sec:conclusion}

\rule{1em}{0em} 1. The Roy equations determine the $\pi\pi$ scattering
amplitude in terms of the imaginary parts at intermediate energies, except
for two subtraction constants: the $S$--wave scattering lengths $a_0^0$,
$a_0^2$. At low energies, the contributions from the imaginary parts are
small, so that the current experimental information about these suffices,
but the one about the scattering lengths is subject to comparatively large
uncertainties.

2. The low energy theorems of chiral symmetry provide the missing element:
They predict the values of the two subtraction constants. In the chiral
limit, where the pions are massless, the $S$--wave scattering lengths
vanish. The breaking of the symmetry generated by the quark masses $m_u$
and $m_d$ leads to nonzero values, for $M_\pi$ as well as for $a_0^0$ and
$a_0^2$.  The Gell-Mann-Oakes-Renner relation (\ref{eq:Mpi}) shows that the
leading term in the expansion of $M_\pi^2$ in powers of the quark masses is
determined by the quark condensate and by the pion decay constant.
Weinberg's low energy theorems (\ref{eq:weinberg}) demonstrate that the
same two constants also determine the leading term in the expansion of the
scattering lengths.  Ignoring the higher order contributions, these
relations predict $a_0^0=0.159$, $a_0^2=-.0454$.

3. Chiral perturbation theory allows us to analyze the higher order terms
of the expansion in a systematic manner. In the isospin limit, $m_u=m_d=m$,
the perturbation series of the $\pi\pi$ scattering amplitude has been
worked out to next--to--next--to--leading order (two loops). The result, in
particular, specifies the expansion of $a_0^0$ and $a_0^2$ in powers of $m$
up to and including $O(m^3)$. The isospin breaking effects due to $m_u\neq
m_d$ have also been been studied \cite{isospin breaking}. These effects
only show up at nonleading orders of the expansion and are small -- in
contrast to the kaons or the nucleons, the pions are protected from isospin
breaking. In the present paper, we have ignored these effects altogether.

4. Chiral symmetry does not fully determine the higher order contributions,
because it does not predict the values of the coupling constants occurring
in the effective Lagrangian. There are two categories of coupling
constants: terms that survive in the chiral limit and symmetry breaking
terms proportional to a power of $m$. The former show up in the momentum
dependence of the scattering amplitude, so that these couplings can be
determined phenomenologically.  For the coupling constants of the second
category, which describe the dependence on the quark masses, we need to
rely on sources other than $\pi\pi$ scattering.

5. The higher order terms of the expansion are dominated by those of
next--to--leading order, which involve the coupling constants $\ell_1$,
$\ell_2$ from the first category and $\ell_3$, $\ell_4$ from the second.
We rely on the dispersive analysis of the scalar pion form factor to pin
down the coupling constant $\ell_4$. Crude theoretical estimates indicate
that the contributions from $\ell_3$ are very small, but the uncertainties
of these estimates dominate the error in our final result for the $S$--wave
scattering lengths.  We also rely on theoretical estimates for the symmetry
breaking coupling constants $r_1,\ldots,,r_4$ of next--to--next--to--leading
order. These indicate that the contributions to $a_0^0$ and $a_0^2$ from
those constants are tiny and could just as well be dropped.

6. The expansion in powers of the quark masses contains infrared
singularities -- the chiral logarithms characteristic of chiral
perturbation theory. In the case of $a_0^0$, for instance, these
singularities enhance the magnitude of the corrections quite
substantially. The origin of the phenomenon is well understood: The final
state interaction in the $I=0$ $S$--wave generates a strong branch cut. For
this reason, the straightforward expansion of $a_0^0$ in powers of $m$
converges only rather slowly. We exploit the fact that, in the subthreshold
region, the expansion of the scattering amplitude converges much more
rapidly: In our approach, the chiral and phenomenological representations
of the scattering amplitude are matched there. With this method, even the
one loop approximation of the chiral perturbation series yields values for
the scattering lengths that are within the errors of our final result,
which reads 
\bea 
\al\al a_0^0= 0.220\pm 0.005\,,\hspace{5.2em}
a^2_0=-0.0444\pm 0.0010\,,\no
\al\al 2a_0^0-5a_0^2= 0.663\pm 0.007\,,\hspace{2em} 
a^0_0-a^2_0= 0.265\pm0.004\,.  \nonumber\eea

7. We have worked out the implications for the phase shifts of the $S$--
and $P$--waves. As shown in figs.~\ref{fig:Swave}--\ref{fig:S2wave}, chiral
symmetry and the existing experimental information constrain these
to a rather narrow range. The corresponding predictions for the
scattering lengths of the $D$-- and $F$--waves, as well as for the effective
ranges are listed in table \ref{tab:threshold}.

8. Our representation of the scattering amplitude, in particular, also yields
an accurate prediction for the phase of $\epsilon'/\epsilon$: The result for
the phase difference between the two $S$--waves at $s=M_{K^0}^2$ reads
\bea \delta^0_0(M_{K^0}^2)-\delta^2_0(M_{K^0}^2)=47.7^\circ\pm
1.5^\circ\fs\eea

9. The mass and the width of
the $\rho$--meson can be calculated to within remarkably small uncertainties:
\bea M_\rho\al=\al 762.4\pm 1.8\,\mbox{MeV}\co\hspace{2em}
\Gamma_\rho=145.7\pm 2.6\,\mbox{MeV}\fs\eea
Also, we confirm that the $I=0$ $S$--wave contains a pole 
far away from the real axis, at  
$\sqrt{s}=(470\pm 30) -i\,(295\pm 20)\,\mbox{MeV}$. The phenomenon is 
related to the fact that chiral symmetry requires the scattering 
amplitude to be very
small at threshold and then to grow with the square of the energy.

10. The consequences for the coupling constants $\ell_1$, $\ell_2$,
$\ell_4$, $r_5$ and $r_6$ of the effective Lagrangian were studied in
detail.  Our results are in good agreement with previous work, but are more
accurate.  In particular, we have shown that $\ell_1$ and $\ell_2$ are
accompanied by strong infrared singularities generated by the two loop
graphs, which shift the numerical values obtained in one loop
approximation, quite substantially.  The effective couplings relevant at
one loop level are given in eq.~(\ref{eq:leff12num}). We have shown that,
with these values, a very decent representation of the scattering amplitude
is obtained by matching the Roy equations with the one loop approximation
of chiral perturbation theory: The result can barely be distinguished from
the representation that underlies the present paper.

11. The results for the various quantities of interest are strongly
correlated. We have examined the correlations in detail, not only for the
threshold parameters and coupling constants, but also for the coefficients
$b_1,\ldots\,,b_6$ of the polynomial that occurs in the chiral perturbation
theory representation of the scattering amplitude.

12. The resulting picture for the low energy structure of the scattering
amplitude is consistent with the resonance saturation hypothesis: For all
of the effective coupling constants encountered in the two loop
representation, our results are consistent with the assumption that, once
the poles and cuts due to pion exchange are removed, the low energy
structure of the amplitude is dominated by the singularities due to the
lightest non--Goldstone states \cite{GL 1984}. Since the splitting of a
resonance contribution from the continuum underneath it is not unique, the
saturation hypothesis does not lead to very sharp predictions, but it is by
no means trivial that these are consistent with the values found, both in
sign and magnitude.

13. On the lattice, it is difficult to reach the small values of $m_u$ and
$m_d$ that are realized in nature. We have shown that chiral perturbation
theory can be used to extrapolate the results obtained at comparatively
large values for these masses, in a controlled manner. The method at the
same time also allows a measurement of some of the coupling constants that
occur in the symmetry breaking part of the effective Lagrangian, in
particular, of $\ell_3$. In quite a few cases, the uncertainties in our
results are dominated by those from this term.

14. We emphasize that most of our results rely on the standard picture,
according to which the quark condensate represents the leading order
parameter of the spontaneously broken symmetry, so that the
Gell-Mann-Oakes-Renner relation holds.  The crude theoretical estimates for
the coupling constant $\ell_3$ we are relying on indicate that the higher
order terms in the expansion of $M_\pi^2$ are very small, so that the
square of the pion mass indeed grows linearly with
$m=\frac{1}{2}\,(m_u+m_d)$ -- a curvature only shows up if $m$ is taken
much larger than the physical value.  In ref.~\cite{Knecht Moussallam Stern
Fuchs}, $\ell_3$ is instead treated as a free parameter and is allowed to
take large values, so that the dependence of $M_\pi^2$ on the quark masses
fails to be approximately linear, even in the region below the physical
value of $m$. There is no prediction for the scattering lengths in that
framework.

15. Even if the quark condensate is not assumed to represent the leading
order parameter, a strong correlation between $a_0^0$ and $a_0^2$ emerges,
which originates in the relation between these quantities and the scalar
radius.  The correlation is of interest, in particular, in connection with
the analysis of $K_{e_4}$ data: As shown in fig.~\ref{fig:deltaKl4},
the preliminary results of the E865 experiment at Brookhaven
\cite{e865,Truol}
yield a remarkably good determination of $a_0^0$. The outcome beautifully
confirms the prediction (\ref{eq:final result}): The best fit to these data
is obtained for $a_0^0=0.218$, with $\chi^2= 5.7$ for 5 degrees of freedom.  
For a detailed discussion of the consequences for the value
of $a_0^0$, we refer to \cite{CGL letter2,e865final}.

16. A measurement that aims at determining the lifetime of a $\pi^+\pi^-$
atom to an accuracy of 10\% is currently under way at CERN.  The
interference of the electromagnetic and strong interaction effects in the
bound state and in the decay is now well understood, also on the basis of
chiral perturbation theory \cite{GLR}. The decay rate of the ground state
can be written in the form 
\bea 
\Gamma\al=\al\frac{2}{9}\,\alpha^3
p\,|a_0^0-a_0^2|^2\, (1+\delta) \co\nonumber
\eea 
with
$p=\sqrt{M_{\pi^+}^2-M_{\pi^0}^2-\mbox{$\frac{1}{4}$}\,\alpha^2\,M_{\pi^+}^2}$.
The term $\delta$ accounts for the
corrections of order $\alpha$ as well as those due to $m_u\neq
m_d$. According to ref.~\cite{GLR}, these effects increase the rate by
$\delta=0.058 \pm 0.012$, that is by about 6\%. Inserting our result
(\ref{eq:final result}) for $a_0^0-a_0^2$, we arrive at the following
prediction for the lifetime \cite{GLR1}: 
\bea\label{eq:tau} \tau=(2.90 \pm 0.10)\cdot 10^{-15}\,\mbox{s}\fs
\eea 
Since the decay rate is proportional to $|a_0^0-a_0^2 |^2$, the outcome of
the experiment is expected to lead to a 
determination of $| a_0^0-a_0^2 |$ to an accuracy of 5\%, thereby
subjecting chiral perturbation theory to a very sensitive test.

\subsection*{Acknowledgement} 
We are indebted to S.~Pislak and P.~Tru\"ol for providing
us with preliminary data of the E865 collaboration, and we 
thank H.~Bijnens and G.~Wanders for informative discussions and
comments. 
This work was supported in part by the Swiss National Science
Foundation, and by TMR, BBW-Contract No. 97.0131  and  EC-Contract
No. ERBFMRX-CT980169 (EURODA$\Phi$NE).

\appendix
\setcounter{equation}{0}
\renewcommand{\theequation}{\thesection.\arabic{equation}}
\section{Notation}\label{sec:notation}
We use the following abbreviations:
\bea\label{eq:not1}\al\al 
\xi=\left(\frac{M_\pi}{4\pi F_\pi}\right)^{\!\!2}\co\hspace{1em}
x=\left(\frac{M}{4\pi F}\right)^{\!\!2}\co\hspace{1em}
 \Ltilde=\ln\frac{\mu^2}{M_\pi^2}\co\hspace{1em}
N= 16\pi^2\fs\nonumber\eea
The intrinsic scales of the coupling constants of ${\cal L}_4$ are denoted by
$\Lambda_n$. In terms of these, the standard renormalized
couplings are given by
\bea \ell^r_n(\mu)=\frac{\gamma_n}{32 \pi^2}\,
 \ln\frac{\Lambda_n^{\;2}}{\mu^2}\,,\hspace{2em}
\gamma_1=\mbox{$\frac{1}{3}$}\,,\hspace{1em} 
\gamma_2=\mbox{$\frac{2}{3}$}\,,\hspace{1em} 
\gamma_3=-\mbox{$\frac{1}{2}$}\,,\hspace{1em} 
\gamma_1=2\,,\nonumber\eea 
where $\mu$ is the running scale. The scales relevant in the various
applications are different and the formulae can be simplified considerably 
if the coupling constants are normalized at the appropriate scale.
For this reason, we use three different symbols:
\bea \label{eq:not4} 
\al\al \lbar_n=\ln\frac{\Lambda_n^{\;2}}{M_\pi^2}\co\hspace{2em}
\lhat_n=\ln\frac{\Lambda_n^{\;2}}{M^2}\co\hspace{2em}
\ltilde_n=\ln\frac{\Lambda_n^{\;2}}{\mu^2}\fs\nonumber\eea
The first coincides with the one introduced in \cite{GL 1983}, in 
the framework of a one loop analysis, where there is no need to distinguish 
$\lbar_n$ from $\lhat_n$.
The quantities $\ltilde_n$ differ from the running coupling constants 
$\ell^r_n(\mu)$ only by a numerical factor, which it is convenient to remove
in order to simplify the expressions.
For the same reason, we work with the coefficients 
$\bbar_1,\ldots\,,\bbar_6$, introduced in \cite{BCEGS}, 
\bdm
\bbar_1=N b_1\,,\hspace{1em}
\bbar_2=N b_2\,,\hspace{1em}
\bbar_3=N b_3\,,\hspace{1em} \bbar_4=N b_4\,,\hspace{1em}
\bbar_5=N^2b_5\,,\hspace{1em} \bbar_6=N^2b_6\;.\edm 
and also
stretch the coupling constants of ${\cal L}_6$ by a power of $N=16\, \pi^2$,
\bea\label{eq:not2}
\al\al\rtilde_n=N^2\,r^r_n(\mu)\,,\hspace{3em}(n=1,\ldots\,,8)\fs
\nonumber
\eea  
In Generalized Chiral Perturbation Theory, the coefficients $\bbar_n$ are
replaced by the constants $\alpha$, $\beta$, $\lambda_1,\ldots,\,\lambda_4$:
\bea\label{eq:lambda} \al\al\alpha= 1+\xi\,(3\,\bbar_1+ 4\,
\bbar_2+4\,\bbar_3-4\,\bbar_4)-\mbox{$\frac{11}{36}$}\,\pi^2\,\xi^2-
\mbox{$\frac{152}{9}$}\,\xi^2\co\no
\al\al\beta= 1 +\xi\,(\bbar_2+4\,\bbar_3-4\,\bbar_4)+4\,
\xi^2\, (3\,\bbar_5- \bbar_6)-\mbox{$\frac{13}{72}$}\,\pi^2\,\xi^2+
\mbox{$\frac{152}{9}$}\,\xi^2\co\no
\al\al N \lambda_1= \bbar_3-\bbar_4 +2\,\xi\,(3\,\bbar_5 -\bbar_6)
+\mbox{$\frac{1}{48}$}\,\pi^2\,\xi+\mbox{$\frac{38}{3}$}\,\xi\co\no
\al\al N \lambda_2= 2\,\bbar_4-\mbox{$\frac{1}{3}$}\xi\co\\
\al\al N^2 \lambda_3= \bbar_5-\mbox{$\frac{1}{3}$}\,\bbar_6+
\mbox{$\frac{82}{27}$}\co\no
\al\al N^2 \lambda_4=-\mbox{$\frac{4}{3}$}\,\bbar_6-\mbox{$\frac{14}{27}$}
\fs\nonumber\eea
 Throughout this
paper, we identify $M_\pi$ with the mass of the charged pion and use $
F_\pi=92.4\,\mbox{MeV}$ \cite{Holstein}.

\setcounter{equation}{0}
\section{Polynomial part of the chiral representation}
\label{sec:bcegs}
In this appendix, we convert the representation 
obtained in ref.\ \cite{BCEGS} for
the scattering amplitude to two loops of chiral perturbation theory into
an explicit expression for the coefficients of the polynomial $C(s,t,u)$
defined in eq.\ (\ref{eq:poly}).
As a first step,
we decompose that representation into three functions of a
single variable and a polynomial, according to eq.\
(\ref{eq:chiral decomposition}). The resulting representation for the
functions $U^0(s)$, $U^1(s)$,
$U^2(s)$ contains linear combinations
of the loop integrals $\bar{J}(s),\,\ldots\,,\,K_4(s)$ introduced in
ref.~\cite{BCEGS}. Note that the decomposition is 
not unique: eq.~(\ref{eq:chiral decomposition}) fixes the functions
$U^I(s)$ only up to a polynomial in $s$. 

Next, we expand the loop integrals in powers of $s$, using the
explicit expressions of ref.~\cite{BCEGS}. 
In terms of the dimensionless
variable $\sbar=s/M_\pi^2$, the result reads:
\bea \bar{J}(s)\al=\al
\frac{\sbar}{96\,\pi^2}\left\{1+\frac{\sbar}{10}+\frac{\sbar^2}{70}+
O(\sbar^3)\right\}\co
\no
K_1(s)\al=\al -\frac{\sbar}{256\,\pi^4}\left\{ 1 +
\frac{\sbar}{12}+\frac{\sbar^2}{90} +O(\sbar^3)\right\}\co
\no
     K_2(s)\al=\al -\frac{\sbar}{384\,\pi^4}\left\{ 1 +
\frac{7\,\sbar}{120}+\frac{\sbar^2}{168} +O(\sbar^3)\right\}\co
\\
     K_3(s)\al=\al \frac{\sbar}{1024\,\pi^4}\left\{\frac{\pi^2-6}{3}
+\frac{(2\,\pi^2-15)\,\sbar}{30}+\frac{(6\pi^2-49)\,\sbar^2}{420}+
O(\sbar^3)\right\}\co
\no
     K_4(s)\al=\al -\frac{\sbar}{3072\,\pi^4}\left\{
 \frac{\pi^2-9}{3}+\frac{(2\,\pi^2-19)\,\sbar}{20}
+\frac{(3\pi^2-29)\,\sbar^2}{105}
+O(\sbar^3)\right\}\fs
\nonumber\eea
These representations suffice to determine the Taylor series expansions of
the
functions $U^I(s)$ around $s=0$, up to and including $O(s^3)$ in the case
of $U^0(s)$, $U^2(s)$ and to $O(s^2)$ for $U^1(s)$. 

The ambiguity mentioned above is fixed with the dispersive 
representation (\ref{eq:disp U}),
which requires the first few derivatives of the functions $U^I(s)$ 
to vanish at 
$s=0$. Starting with an arbitrary decomposition, for which this requirement
need not be obeyed, we truncate the Taylor series
for $U^0(s)$, $U^2(s)$ at order $s^3$ and the
one for $U^1(s)$ at order $s^2$. It is straightforward to check 
that the functions obtained by subtracting these terms indeed 
obey the dispersion relations (\ref{eq:disp U}). Absorbing the subtractions
in $C(s,t,u)$, we may then read off the coefficients
of this polynomial:
\bea \label{eq:ci}
c_1\al=\al-\frac{M_\pi^2}{F_\pi^2}\left\{1
+\xi\left( - \bbar_1-{{68}\over {315}}\right)
+\xi^2\left( -   {{8\,\bbar_1}\over {105}}
- {{32\, \bbar_2}\over {63}}\right.\right. \no
\al\al\left.\left.
-   {{464\, \bbar_3}\over {315}} - {{3824\, \bbar_4}\over {315}}
 + {{601\,\pi^2}\over {945}}   -{{17947}\over {2835}}\right)\right\}\co\no
c_2\al=\al\frac{1}{F_\pi^2}\left\{1+\xi\left( \bbar_2
-{{323}\over {1260}} \right)+\xi^2\left(-  {{11\, \bbar_1}\over {70}}
- {{211\, \bbar_2}\over {315}} \right.\right. \no
\al\al\left.\left.  -   {{628\, \bbar_3}\over {315}}
- {{5164\, \bbar_4}\over {315}}
-{{3977}\over {630}} + {{5237\,\pi^2}\over {7560}}  \right)\right\}\co\no
c_3\al=\al \frac{1}{N F_\pi^4}\left\{ \bbar_3 +
{1\over {42}}  + \xi\left( {{18\, \bbar_1}\over {35}} +
  {{59\, \bbar_2}\over {105}} + {{731\, \bbar_3}\over {315}}
\right.\right.\no\al\al\left.\left.\hspace{9em}
+   {{3601\, \bbar_4}\over {315}}
   - {{5387\,{{\pi }^2}}\over {15120}} -{{19121}\over {7560}}\right)
\right\}\co\\
c_4\al=\al \frac{1}{N F_\pi^4}\left\{ \bbar_4
-{{31}\over {2520}} +\xi\left(- {{43\, \bbar_1}\over {420}} -
  {{8\,\bbar_2}\over {63}} + {{23\, \bbar_3}\over {63}}
\right.\right.\no\al\al\left.\left.  \hspace{9em}
+   {{997\,\bbar_4}\over {315}} + {{467\,{{\pi }^2}}\over {7560}}
-{{63829}\over {45360}} \right)
\right\}\co\no
c_5\al=\al \frac{1}{N^2 F_\pi^6}\left\{
{{137}\over {1680\,\xi}}+ {{ \bbar_1}\over {16}} +
  {{379\, \bbar_2}\over {1680}} - {{25\, \bbar_3}\over {28}} -
  {{731\, \bbar_4}\over {180}} +  \bbar_5 \right.\no
\al\al\left. \hspace{17em}
+   {{269\,{{\pi }^2}}\over {15120}} +
{{61673}\over {18144}}  \right\}\co\no
c_6\al=\al  \frac{1}{N^2 F_\pi^6}\left\{
-{{31}\over {1680\,\xi}} + {{ \bbar_1}\over {112}} -
  {{47\, \bbar_2}\over {1680}} - {{65\, \bbar_3}\over {252}} -
  {{547\, \bbar_4}\over {420}} +  \bbar_6 \right.\no\al\al\left.
\hspace{17em} + {{{{\pi }^2}}\over {15120}}
 + {{44287}\over {90720}} \right\}\fs  \nonumber
\eea
The constants $\bbar_n$ represent dimensionless combinations of 
coupling constants, introduced in ref.~\cite{BCEGS}. In the notation of
appendix \ref{sec:notation},
the expressions read:
\bea\label{eq:bbar}
\bbar_1\al=\al-\frac{7 \Ltilde}{6}+\frac{4\, \ltilde_1}{3}-
\frac{\ltilde_3}{2}
-2\,\ltilde_4+\frac{13}{18} -\frac{49\, \xi \Ltilde^2}{6}\no\al\al
\hspace{-1.5em}+
\xi  \Ltilde\left\{-\frac{4\, \ltilde_1}{9}-
\frac{56\, \ltilde_2}{9}-\ltilde_3-\frac{26\, \ltilde_4}{3} -\frac{47}{108}
\right\}
+ \xi\left\{\rtilde_1+\frac{16\, \ltilde_1 \ltilde_4}{3}  
-\frac{\ltilde_3^{\;2}}{2}\right.\no\al \al\hspace{-1.5em}-\left.
3\, \ltilde_3 \ltilde_4 -5\, \ltilde_4^{\;2}+\frac{28\, \ltilde_1}{27} +
\frac{80\, \ltilde_2}{27}-\frac{15\, \ltilde_3}{4}  +\frac{26\, \ltilde_4}{9}
-\frac{34\, \pi^2}{27}+\frac{3509}{1296}\right\}\,,\no
\bbar_2\al=\al\frac{2\,\Ltilde}{3}-\frac{4 \,\ltilde_1}{3} +2\,
\ltilde_4-\frac{2}{9} +\!
\frac{431\,\xi  \Ltilde^2}{36}+\xi  \Ltilde\left\{6\,\ltilde_1+
\frac{124\,\ltilde_2}{9}-\frac{5\,\ltilde_3}{2}+\frac{20\,\ltilde_4}{3} +\!
\frac{203}{54}
\right\}\no\al\al\hspace{-1.5em}+ \xi\left\{\rtilde_2-\frac{16\, \ltilde_1
    \ltilde_4}{3}  
+ \ltilde_3 \ltilde_4 +5\, \ltilde_4^{\;2}-4\, \ltilde_1\!-
\frac{166\,\ltilde_2}{27}+\frac{9\,\ltilde_3}{2}  -\frac{8\,\ltilde_4}{9}
+\frac{317\,\pi^2}{216}\!-
\frac{1789}{432}\right\}\!,\no
\bbar_3\al=\al\frac{\Ltilde}{2}+\frac{\ltilde_1}{3} +\frac{\ltilde_2}{6}
-\frac{7}{12} -
\frac{40\,\xi  \Ltilde^2}{9}+\xi \Ltilde\left\{-\frac{38\,\ltilde_1}{9}-
\frac{20\,\ltilde_2}{3}+2\,\ltilde_4 +\frac{365}{216}
\right\}\no\al\al\hspace{-1.5em}+ \xi\left\{\rtilde_3+\frac{4\, \ltilde_1
    \ltilde_4}{3}  
+\frac{2\, \ltilde_2 \ltilde_4}{3} +\frac{89\,\ltilde_1}{27} +
\frac{38\,\ltilde_2}{9} -\frac{7\,\ltilde_4}{3}-
\frac{311\,\pi^2}{432}+\frac{7063}{864}\right\}\,,\\
\bbar_4\al=\al\frac{\Ltilde}{6}+\frac{\ltilde_2}{6}
-\frac{5}{36} +
\frac{5\,\xi \Ltilde^2}{6}+\xi \Ltilde\left\{\frac{\ltilde_1}{9}+
\frac{8\,\ltilde_2}{9}+\frac{2\,\ltilde_4}{3} -\frac{47}{216}
\right\}\no\al+\al \xi\left\{\rtilde_4 
+\frac{2\, \ltilde_2 \ltilde_4}{3} +\frac{5\,\ltilde_1}{27} +
\frac{4\,\ltilde_2}{27} -\frac{5\,\ltilde_4}{9}+
\frac{17\,\pi^2}{216}+\frac{1655}{2592}\right\}\,,\no
\bbar_5\al=\al\frac{85\,\Ltilde^2}{72}+
\Ltilde\left\{\frac{7\,\ltilde_1}{8}
+\frac{107\,\ltilde_2}{72}-\frac{625}{288}\right\}+\rtilde_5
-\frac{31\,\ltilde_1}{36}-\frac{145\,\ltilde_2}{108}+\frac{7\,\pi^2}{54}
-\frac{66029}{20736}\,,\no
\bbar_6\al=\al\frac{5\,\Ltilde^2}{24}+\Ltilde
\left\{\frac{5\,\ltilde_1}{72}
+\frac{25\,\ltilde_2}{72}-\frac{257}{864}\right\}+\rtilde_6
-\frac{7\,\ltilde_1}{108}-\frac{35\,\ltilde_2}{108}+\frac{\pi^2}{27}-
\frac{11375}{20736}\,.
\nonumber\eea
Note that the quark masses exclusively enter through
$\xi$ and $\Ltilde$ -- the remaining quantities are independent thereof.
The expressions
involve the logarithm of $M_\pi^2$, as well as the square thereof -- in the
chiral limit, the coefficients $\bbar_n$ diverge logarithmically.  
The coefficients of the leading infrared singularities are pure numbers, 
which are determined
by the structure of the symmetry group and the transformation
properties of the symmetry breaking part of the Hamiltonian, that is of
the quark mass term. The scales of the logarithmic divergences, on the other
hand, are not determined by the symmetry, but are fixed by   
the intrinsic scales $\Lambda_1,\ldots\,,\Lambda_4$ of the 
effective coupling constants of ${\cal L}_4$. 

\setcounter{equation}{0}
\section{The corrections {\boldmath $\Delta_0,
\Delta_1,\Delta_2,\Delta_r$\unboldmath}}
\label{sec:NNL} 
The leading terms in the expansion of the scalar radius in powers of the
quark
masses are determined by $\lbar_4$. The next--to--leading order correction
$\Delta_r$, which is defined in (\ref{eq:rs}), 
was calculated in \cite{Bijnens Colangelo Talavera}, on the basis of an 
evaluation of the scalar form factor to two loops. In the
notation introduced above, their result reads:
\bea\label{eq:Deltar}
 \Delta_r\al=\al-\frac{29}{18}\,\Ltilde^2+\Ltilde
\left\{
-\frac{31}{9}\,\ltilde_1-\frac{34}{9}\,\ltilde_2+4\,\ltilde_4-
\frac{145}{216}\right\}+\rStilde
+\ltilde_3 \ltilde_4+ 2\,\ltilde_4^{\;2}
\no\al+\al
\frac{22}{9}\,\ltilde_1+2\,\ltilde_2-\frac{5}{24}\,\ltilde_3-
\frac{13}{6}\,\ltilde_4
-\frac{23\,\pi^2}{72}+\frac{869}{648}
\eea

As discussed in section \ref{sec:let}, the  
quantities $C_0,C_1$ and $C_2$ tend to unity in the chiral limit. According
to
equation (\ref{eq:letcr}), the first order
corrections can be expressed in terms 
of the scalar radius and the coupling constant $\lbar_3$.
To work out the corrections of second order, it suffices to insert
the relations (\ref{eq:ci}), (\ref{eq:bbar}) in the definition
(\ref{eq:defC12}) of $C_1$ and
$C_2$ and read off the coefficients of the terms of order $\xi^2$.
The result reads
\bea \label{eq:Delta12}
\Delta_1\al=\al- \frac{71\,\Ltilde^2}{12}+\Ltilde
\left\{-\frac{40\,\ltilde_1}{9}-\frac{80\,\ltilde_2}{9}-\frac{5\,\ltilde_3}{2}
+4\,\ltilde_4+\frac{5393}{315}\right\}
\no\al+\al\rule{0em}{1.3em} \rtilde_2+4\,\rtilde_3-4\,\rtilde_4-2\,\rStilde-
\ltilde_3 \ltilde_4 +
\ltilde_4^{\;2}\no\al
+\al \rule{0em}{2em}\frac{1826\,\ltilde_1}{315}+
\frac{3118\,\ltilde_2}{315}+\frac{79\,\ltilde_3}{21}-\frac{144\,\ltilde_4}{35}-
\frac{521}{252}\pi^2+\frac{24221}{3024}\co\\
\rule{0em}{1.7em}
\Delta_2\al=\al\rule{0em}{2.2em}- \frac{175\,\Ltilde^2}{18}+\Ltilde
\left\{-10\,\ltilde_1-\frac{148\,\ltilde_2}{9}+\ltilde_3+6\,\ltilde_4
+\frac{9311}{630}\right\}
\no\al-\al\rule{0em}{1.2em} \rtilde_1+4\,\rtilde_3-4\,\rtilde_4-2\,
\rStilde+
\frac{\ltilde_3^{\;2}}{2}+\ltilde_3 \ltilde_4 +
\ltilde_4^{\;2}\no\al
+\al \rule{0em}{1.7em}\frac{556\,\ltilde_1}{63}+
\frac{4372\,\ltilde_2}{315}+\frac{104\,\ltilde_3}{35}-
\frac{125\,\ltilde_4}{21}-
\frac{2939\,\pi^2}{1260}+\frac{43109}{5040}\fs\nonumber\eea
According to (\ref{eq:C0}), the analogous correction in the low energy 
theorem for $C_0$ can be expressed in terms of these:
\bea \Delta_0=\frac{1}{7}\,(12\,\Delta_1-5\,\Delta_2)\fs\eea
 
\setcounter{equation}{0}
\section{Phenomenological representation}
\label{sec:phen}

In the present appendix, we convert the low energy representation of the
scattering amplitude constructed in ref.~\cite{ACGL}
into the form given in section \ref{sec:phenomenological representation}.
That representation consists of a sum of two terms:
\bea\label{eq:ASPd} A(s,t,u)=A(s,t,u)\SP+A(s,t,u)_d\co\eea
The first describes the contributions generated by the imaginary parts of the
$S$-- and $P$--waves below $\sqrt{s_2}=2\,\mbox{GeV}$,
while the background amplitude $A(s,t,u)_d$ collects those from
the higher partial waves and higher energies. 

The explicit expression for the first term involves three functions of a 
single variable:
\bea\label{eq:ASP} A(s,t,u)\SP 
\al=\al32\pi\left\{\mbox{$\frac{1}{3}$}W^0(s)+
\mbox{$\frac{3}{2}$}(s-u)W^1(t)
+\mbox{$\frac{3}{2}$}(s-t)W^1(u)\right.\no
\al\al \left.+\mbox{$\frac{1}{2}$}W^2(t)+\mbox{$\frac{1}{2}$}W^2(u)
-\mbox{$\frac{1}{3}$} W^2(s) \right\}\fs\eea
The functions $W^0(s)$, $W^1(s)$, $W^2(s)$ are determined by the 
imaginary parts of the $S$-- and $P$--waves
and by the two subtraction constants $a_0^0,a_0^2$:
\bea\label{eq:W} W^0(s)\al=\al\frac{a_0^0\, s}{4M_\pi^2} +
\frac{s(s-4M_\pi^2)}{\pi}\int_{4M_\pi^2}^{s_2}
ds'\;\frac{\mbox{Im}\, t_0^0(s')}{s'(s'-4M_\pi^2)(s'-s)}\co \no
      W^1(s) \al=\al \frac{s}{\pi}\int_{4M_\pi^2}^{s_2}
ds'\;\frac{\mbox{Im}\, t_1^1(s')}{s'(s'-4M_\pi^2)(s'-s)}\co\\
       W^2(s)\al=\al\frac{a_0^2\, s}{4M_\pi^2}  +
\frac{s(s-4M_\pi^2)}{\pi}\int_{4M_\pi^2}^{s_2}
ds'\;\frac{\mbox{Im}\, t_0^2(s')}{s'(s'-4M_\pi^2)(s'-s)}\fs \nonumber\eea
These functions are closely related to the quantities $\Wbar^I(s)$ introduced
in section \ref{sec:phenomenological representation}, but there are two
differences: The subtractions are not the same and the range of integration
differs.  

To compare $W^0(s)$ with $\Wbar^0(s)$, we consider the function 
\bdm  w^0(s)=
\frac{s^4}{\pi}\int_{4M_\pi^2}^{s_2}
ds'\;\frac{\mbox{Im}\, t_0^0(s')}{s^{\prime\,4}(s'-s)}\co\edm
which is intermediate between the two: It differs from $\Wbar^0(s)$ only in 
the range of integration and from $W^0(s)$ only by a subtraction polynomial.
The latter may be expressed in terms
of the following moments of the imaginary part:
\bea J^0_n=\frac{1}{\pi}\int_{4M_\pi^2}^{s_2}
\frac{ds\,\mbox{Im}\,t^0_0(s)}{s^{ n+2}(s-4M_\pi^2)}\co \hspace{2em}n=0,1,2
\fs
\eea
The explicit relation between $W^0(s)$ and $w^0(s)$ reads
\bea\label{eq:Ww}\hspace{-1.5em} W^0(s)\al=\al 
w^0(s)+ \frac{a_0^0 s}{4M_\pi^2}+
s(s-4M_\pi^2)\,J_0^0+s^2(s-4M_\pi^2)\,J_1^0+4M_\pi^2s^3
J_2^0\,.\eea 

The difference $\Wbar^0(s)-w^0(s)$, on the other hand, is given by an 
integral over the region
$s>s_2$, which merely generates contributions
of $O(p^8)$, so that 
\bdm w^0(s)=\Wbar^0(s)+O(p^8)\fs\edm 
To the accuracy to which the two loop representation holds, we may
thus replace the term $w^0(s)$ on the r.h.s~of eq.~(\ref{eq:Ww}) 
by $\Wbar^0(s)$.

This shows that, up to terms of $O(p^8)$, the functions 
$\Wbar^0(s)$ and $W^0(s)$ only differ by a polynomial whose coefficients
are determined by the moments $J^0_n$.
The same reasoning also applies to the components with $I=1,2$. The
relevant moments are given by
\bea J^1_n=\frac{1}{\pi}\int_{4M_\pi^2}^{s_2}
\frac{ds\,\mbox{Im}\,t^1_1(s)}{s^{ n+2}(s-4M_\pi^2)}\co \hspace{2em}
J^2_n=\frac{1}{\pi}\int_{4M_\pi^2}^{s_2}
\frac{ds\,\mbox{Im}\,t^2_0(s)}{s^{ n+2}(s-4M_\pi^2)}
\fs\eea
The net result amounts to a representation for $A(s,t,u)\SP$
in terms of the functions $\Wbar^I(s)$ and a set of polnomials
involving the above moments:
\bea W^0(s)\al=\al \Wbar^0(s)\!+ \frac{a_0^0 s}{4M_\pi^2}+
s(s-4M_\pi^2)\,J_0^0+s^2(s-4M_\pi^2)\,J_1^0+4M_\pi^2s^3 J_2^0+O(p^8)\co\no
 W^1(s)\al=\al \Wbar^1(s)-s\,J_0^1-s^2J_1^1+O(p^6)\co\\
W^2(s)\al=\al \Wbar^2(s)\!+ \frac{a_0^2 s}{4M_\pi^2}+
s(s-4M_\pi^2)\,J_0^2+s^2(s-4M_\pi^2)\,J_1^2+4M_\pi^2s^3 J_2^2
+O(p^8)\fs
\nonumber\eea

\setcounter{equation}{0}
\section{Moments of the background amplitude}
\label{sec:moments}
We now turn to the second part in the decomposition (\ref{eq:ASPd}).
The chiral representation shows that the infrared singularities contained
in $A(s,t,u)_d$ start manifesting themselves only at higher orders. Up to and
including $O(p^6)$, the background amplitude is a crossing symmetric 
polynomial of the momenta: 
\bea A(s,t,u)_d\al=\al P(s,t,u)+O(p^8)\\
P(s,t,u)\al=\al p_1+p_2\,s+p_3\,s^2+p_4\,(t-u)^2+
p_5\,s^3+p_6\,s(t-u)^2\,.\nonumber\eea
The coefficients $p_1,\ldots,\,p_6$ may be calculated by expanding the 
dispersion integrals in powers of the momenta. For a detailed discussion, we 
refer to appendix B of ref.~\cite{ACGL}.
By construction, $A(s,t,u)_d$ does not contribute to the
$S$--wave scattering lengths. This condition fixes
$p_1$ and $p_2$ in terms of the
remaining coefficients:
\be p_1= -16M_\pi^4\,p_4\,,\;\;\;p_2 =
4M_\pi^2\,(-p_3 + p_4 - 4M_\pi^2\,p_5)\,. \ee
The explicit expressions for these read 
\bea
\label{eq:pn} p_3\al=\al\frac{8\,\pi}{3}\,(4I^0_0-9I^1_0-I^2_0)
      +\frac{16\,\pi}{3}\,M_\pi^2\,(-8\,I^0_1 - 21\, I^1_1 + 11\,I^2_1
        +12 \,H)\no
p_4\al=\al 8\,\pi \,(I^1_0+I^2_0)+
   16\,\pi\,M_\pi^2\,(I^1_1 +I^2_1)\\
p_5\al=\al \frac{4\,\pi}{3}\,(8\,I^0_1+9\,I^1_1-11\,I^2_1-6\,H)\no
p_6\al=\al 4\,\pi (I^1_1-3\,I^2_1+2\,H)
     \fs\nonumber\eea
The moments $I^I_n$ represent integrals over the 
imaginary parts at $t=0$ (total cross sections), except that the 
contributions from the $S$-- and 
$P$--waves below $\Etwo=2 \,\mbox{GeV}$ are removed. In terms
of the imaginary parts of the partial waves, the explicit expression reads
        \cite{ACGL}:   
\bea\label{eq:In} I_n^I\al=\al
\sum_{\ell=2}^\infty \,\frac{(2l+1)}{\pi}\int_{4M_\pi^2}^{s_2}\!
ds\;\frac{\mbox{Im}\,t^I_\ell(s)}{s^{n+2}(s-4M_\pi^2)}\\
\al+\al  
\sum_{\ell=0}^\infty \,\frac{(2l+1)}{\pi}\int_{s_2}^\infty\!
ds\;\frac{\mbox{Im}\,t^I_\ell(s)}{s^{n+2}(s-4M_\pi^2)}\fs\nonumber\eea
The term $H$ represents an analogous integral over the derivatives
of the imaginary parts with respect to $t$ at $t=0$. 
Since the $S$--wave contributions are independent of $t$,
they drop out. Moreover, on account of crossing
symmetry, the contributions with $I=1$ may be expressed in terms of those
with
$I=0,2$:
\bea H\al=\al
\sum_{\ell=2}^\infty\,
(2l+1)\,\ell(\ell+1)\,\frac{1}{\pi}\int_{4M_\pi^2}^\infty\!
\!ds\;\frac{2\,\mbox{Im}\,t^0_\ell(s)+4\,\mbox{Im}\,t^2_\ell(s)}{9\,s^{3}
(s-4M_\pi^2)}\fs\eea

There is an important difference between the moments relevant for the
background amplitude and those associated with the $S$-- and $P$--waves:
While $I^I_n$ and $H$ remain finite
when the quark masses are sent to zero, 
\bea I^I_0=O(1)\co\hspace{2em}I^I_1=O(1)\co\hspace{2em}
I^I_2=O(1)\co\hspace{2em}H=O(1)\co\nonumber\eea 
the $S$-- and $P$--wave moments with $n\geq 1$ explode in that limit: 
\bea J^I_0=O(1)\co\hspace{2em}J^I_1=O(M_\pi^{-2})\co\hspace{2em}
J^I_2=O(M_\pi^{-4})\fs\nonumber\eea
The phenomenon arises from the manner in which we have chosen to decompose
the amplitude into a contribution from the $S$-- and $P$--waves,
described by the functions $\Wbar^I(s)$, and a polynomial. These functions
develop an infrared singularity in the chiral limit, which cancels the
one occurring in the polynomial -- the full scattering
amplitude approaches a finite limit when the quark masses are turned off.
In fact, the problem does not
occur in the original form of the decomposition, based on the
functions $W^I(s)$: these do have a decent chiral limit. 

The same singularities also show up in the 
chiral representation of the amplitude: As we have normalized 
the unitarity correction by subtracting the dispersion integrals
at $s=0$, they contain a quadratic infrared divergence in the chiral limit. 
Indeed, the relations (\ref{eq:ci}) show that the coefficients $c_5$ and
$c_6$ contain contributions that are inversely proportional to $M_\pi^2$
and precisely cancel this divergence.
The above choice of the decomposition has the advantage that it is scale 
independent and leads to a
simple form of the matching conditions.
The backside
of the coin is that the two pieces do not have a smooth chiral limit. 
We could modify the normalization of
the unitarity correction in such a
way that it remains finite in the chiral limit, at the price of introducing
an
arbitrary scale to normalize the logarithmic infrared singularities.
There is no gain in doing that, however: Anyway, only the sum
of the polynomial and the unitarity correction is relevant, so that different
decompositions
lead to identical results. We stick to the one above.

Finally, we add up the two parts of the amplitude. The result takes the
form of eq.~(\ref{eq:phenrep}), where $\Pbar(s,t,u)$ represents
the sum of the two polynomials associated with the two parts. 
The coefficients $\pbar_1,\ldots,\,\pbar_6$ involve a linear combination
of the various moments introduced above. In fact, up to terms
of $O(p^8)$, the result may be expressed in terms of the 
combinations
\bea \Ibar^0_n=I^0_n+J^0_n\co\hspace{1em}\Ibar^1_n=I^1_n+3
J^1_n\co\hspace{1em} \Ibar^2_n=I^2_n+J^2_n\fs\eea
The contributions from the $S$-- and $P$--wave moments $J^I_n$ precisely 
represent the
pieces needed to complete the sum over the angular momenta -- the factor of 
3 in front of the $P$--wave moments
accounts for the weight $2\ell+1$ that occurs in the
definition (\ref{eq:In}) of the moments $I^1_n$. The result is
independent of the energy $s_2$ used in the decomposition (\ref{eq:ASPd})
of the amplitude and is given in eqs.~(\ref{eq:Inbar}),
(\ref{eq:pbar}). 

The moments are readily evaluated with the information given in
ref.~\cite{ACGL}. Since the angular momentum barrier suppresses the 
higher partial waves near threshold, the terms $I^0_2$, $I^1_2$ and $I^2_2$
are negligibly small. The contributions from the $S$-- and $P$--waves
depend on the values of the two
$S$--wave scattering lengths $a_0^0$ and $a^2_0$. In the narrow range
of interest here, this dependence is well described by
a quadratic interpolation of the form  
\bea \Ibar=u_0+u_1\, \Delta a_0^0+u_2\, \Delta a_0^2+u_3\, (\Delta a_0^0)^2+
u_4\, \Delta a_0^0\, \Delta a_0^2+u_5\, (\Delta a_0^2)^2\co\nonumber\eea
with $\Delta a^0_0 =a^0_0-0.225$, 
$\Delta a^2_0 =a^2_0+0.03706$. 
The numerical values of the coefficients are listed in table
\ref{tab:moments}. 
\begin{table}[H]
\vspace*{-1em}
\begin{center}
\begin{tabular}{|l|r|r|r||r|r|r||r|r|r|}
\hline
&\multicolumn{3}{c||}{$I=0$}&\multicolumn{3}{|c||}{$I=1$}&
\multicolumn{3}{c|}{$I=2$}\\
\cline{2-10}
\rule{0em}{1.2em}&$\Ibar^0_0$\rule{0.6em}{0em} & $\Ibar^0_1$\rule{0.6em}{0em}
& $\Ibar^0_2$\rule{1em}{0em} &$\Ibar^1_0$\rule{0.6em}{0em} 
 & $\Ibar^1_1$\rule{0.6em}{0em} & $\Ibar^1_2$ \rule{0.6em}{0em}&
$\Ibar^2_0$\rule{0.6em}{0em}& $\Ibar^2_1$\rule{0.6em}{0em} &
$\Ibar^2_2$\rule{0.6em}{0em}\\ 
&\hspace*{-0.2em}$ \mbox{\footnotesize GeV}^{-4}$\hspace*{-0.3em} &
\hspace*{-0.2em}$ \mbox{\footnotesize GeV}^{-6}$\hspace*{-0.3em}&
\hspace*{-0.2em}$ \mbox{\footnotesize
  GeV}^{-6}$\rule{0.3em}{0em}\hspace*{-0.3em}& 
\hspace*{-0.2em}$ \mbox{\footnotesize GeV}^{-4}$\hspace*{-0.3em}&
\hspace*{-0.2em}$ \mbox{\footnotesize GeV}^{-6}$\hspace*{-0.3em}&
\hspace*{-0.2em}$ \mbox{\footnotesize GeV}^{-6}$\hspace*{-0.3em}&
\hspace*{-0.2em}$ \mbox{\footnotesize GeV}^{-4}$\hspace*{-0.3em}&
\hspace*{-0.2em}$ \mbox{\footnotesize GeV}^{-6}$\hspace*{-0.3em}&
\hspace*{-0.2em}$ \mbox{\footnotesize GeV}^{-6}$\hspace*{-0.3em}\\
\hline
$u_0$\rule{0em}{1em} &$\!\! 9.44 $&$\!\! 66.7 $&$\!\! 609 $&
$\!\! 1.90 $&$\!\! 3.92 $&$ \!\! 10.6 $&$\!\!   .469 $&$\!\! 2.61
$&$\!\! 21.4 $\\ 
$u_1$ &$\!\! 58.5  $&$\!\! 507 $&$\!\! 4980 $&$\!\! 1.97 $&$\!\! 6.35 $&$\!\!
25.6 $&$\!\! 1.56 $&$\!\! 5.84 $&$\!\! 33.9 $\\ 
$u_2$ &$\!\!-75.2 $&$\!\! -462 $&$\!\! -3300 $&$\!\! -7.99
   $&$\!\! -25.2 $&$\!\! -99.3 $&$\!\! -15.1 $&$\!\! -97.8 $&$\!\! -884$ \\ 
$u_3$ &$\!\! 82.4 $&$\!\! 902 $&$\!\! 9800 $&$\!\! -1.09 $&$\!\! -2.1 $&$\!\!
-2.84 $&$\!\! -.295  $&$\!\! -2.65 $&$\!\! -23.6$\\
$u_4$ &$\!\!-232 $&$\!\! -1920 $&$\!\! -16200 $&$\!\! -2.66 $&$\!\! -18.3 
$&$\!\! -110 $&$\!\!  -14.5 $&$\!\! -54.3 $&$\!\! -317 $\\ 
$u_5$ &$\!\! 322 $&$\!\! 2280 $&$\!\! 16900 $&$\!\! 17.4 $&$\!\! 
  72.6 $&$\!\! 349 $&$\!\! 125 $&$\!\! 932 $&$\!\! 9190$ \\
\hline
\end{tabular}
\caption{\label{tab:moments}Moments of the background amplitude.}
\vspace*{-1em}
\end{center}
\end{table}

The quantity $H$ exclusively receives contributions
from the partial waves with $\ell\geq 2$. As discussed in detail in
\cite{ACGL}, $H$ is dominated by the contribution from the lowest
spin 2 resonance, which is independent of $a_0^0,a_0^2$. 
The behaviour of the integrand in the threshold region 
does depend on the two $S$--wave scattering lengths, because these determine
the threshold parameters of the higher partial waves, but since
the contributions
from that region are very small, we ignore the dependence on $a_0^0,a_0^2$
and use the value given in \cite{ACGL}:
\bea H=0.32 \,\mbox{GeV}^{-6}\fs\eea  

\setcounter{equation}{0}
\section{Error analysis and correlations}
\label{sec:error analysis}

The matching conditions and the two loop representation for the scalar radius
determine the values of the constants
\bdm \vec{x} = 
\{a^0_0,a^2_0,\lbar_1,\lbar_2,\lbar_4, \rtilde_5, \rtilde_6\}\edm
as functions of the parameters occurring in our input,
\bdm \vec{y} = \{\langle r^2\rangle_s,\lbar_3, \rtilde_1, \rtilde_2, 
\rtilde_3, \rtilde_4, \rS,\theta_0,\theta_1\}
\fs\edm
The last two of these characterize the experimental input used
when solving the Roy equations. Strictly speaking, that part of our input
involves three functions, the imaginary parts of the $S$-- and $P$--waves
above 0.8 GeV, but in practice, the solution of the matching conditions is 
sensitive only to two parameters:
the values of the phases $\delta_0^0(s)$ and $\delta_1^1(s)$ at 
$\sqrt{s}=0.8\,\mbox{GeV}$, which we
denote by $\theta_0$ and $\theta_1$, respectively. In ref.~\cite{ACGL},
the uncertainties in these parameters are estimated at 
$\theta_0=82.3^\circ\pm3.4^\circ$, $\theta_1=108.9^\circ\pm 2^\circ$.

The uncertainties in the input
give rise both to uncertainties in the
individual components of $\vec{x}$ and to correlations among these. 
We describe the correlations in terms of a Gaussian
distribution: The probability for $\vec{x}$ to be contained in the volume
element $dx$ is represented  as 
\bea \al\al dP= N\,\exp \,(-\mbox{$\frac{1}{2}$}\,Q)\;dx \co\\
\al\al Q=\sum_{ab}C_{ab}\,
\Delta x^a\,\Delta x^b\co\hspace{2em} \Delta x^a=x^a-\langle x^a\rangle
\fs\nonumber\eea
The uncertainties in
our results are characterized by the coefficients $C_{ab}$
of the quadratic form in the exponential. In the Gaussian approximation, 
the matrix $C$ is given by the inverse of the
correlation matrix $K$,
\bea\label{eq:Kab} 
K^{ab}=\langle \Delta x^a \Delta x^b\rangle\co\hspace{2em}C\cdot K={\bf
  1} \co\eea
so that the error analysis boils down to an evaluation of the 
matrix $K$.

In the small region of interest, the response to a change of the input
variables is approximately linear. We denote the central values of the input
parameters by $y_c$ and set $y^i=y_c^i\,(1+\eta^i)$, with 
$\langle\eta^i\rangle=0$. Linearity then implies that the mean value of $x^a$
coincides with the solution of the matching conditions
that corresponds to our central set of input parameters,
and that the correlation matrix $K$
can be expressed in terms of the matrix $
\langle \eta^i\,\eta^k\rangle$. 
We treat the input variables as statistically independent, so that 
this matrix is diagonal, 
\bea \langle \eta^i\,\eta^k\rangle=\delta^{ik} (\sigma_i)^2\fs\eea 

The result of the dispersive evaluation,
$\langle r^2\rangle_s=0.61\pm 0.04\,\mbox{fm}^2$, implies that the mean
square
deviation in the variable $\eta^1$ is given by 
$\sigma_1=0.04/0.61$. We interpret the estimate 
$\lbar_3=2.9\pm 2.4$ in the same manner: $\sigma_2=2.4/2.9$.
Concerning the variables $r_1,\ldots\,,r_4$ and $\rS$, 
we assume that all values in the interval from $0$ to twice the value 
obtained from resonance
saturation are equally likely, so that, for $n=3,\ldots, 7$, 
the mean square deviation
becomes $\sigma_n=1/\sqrt{3}$. Finally, the uncertainties
in the input parameters $\theta_0$ and $\theta_1$ amount to 
$\sigma_8=3.4/82.3$ and
$\sigma_9=2/108.9$, respectively.

We also need an estimate for the sensitivity of our results to the value of
the scale $\mu$ used when applying the resonance estimates. For the mean 
values, we use $\mu=M_\rho$. 
To estimate the uncertainties due to that choice, we evaluate the
shift occurring in the quantity of interest if the coupling constants $r_n$ 
are held fixed, but the scale is replaced by 
$\mu=1\,\mbox{GeV}$, 
repeat the calculation for $\mu=0.5\,\mbox{GeV}$, and
take the mean square of the two shifts.  Likewise, the correponding 
contribution to the correlation matrix is the average of the two matrices 
associated with those two shifts. Alternatively we could assume that all
values of $\mu$ in the interval between $0.5$ and $1$ GeV are equally
likely. The uncertainties then become somewhat smaller, but it suffices to 
slightly stretch the interval for the outcome to be nearly the same.
\begin{table}[H]
\begin{tabular}{|r|r|r|r|r|r|r|r|}
\hline
\rule{0em}{1em}&$\Delta a_0^0\rule{0.5em}{0em}$&$\Delta
a_0^2\rule{0.5em}{0em}$&$\Delta\lbar_1\rule{0.5em}{0em}$&
$\Delta\lbar_2\rule{0.8em}{0em}$&$\Delta\lbar_4\rule{0.8em}{0em}$&
$\Delta\rtilde_5\rule{0.8em}{0em}$&$\Delta\rtilde_6\rule{0.8em}{0em}$\\
\hline
$\Delta a_0^0$\rule{0em}{1em}&
$\!2.0 \!\cdot\! 10^{-5}\!\! $&$\! 3.2 \!\cdot\! 10^{-6}\!\!  
$&$\! 1.9 \!\cdot\! 10^{-4}\!\!$&$ -1.7 \!\cdot\! 10^{-5}\!\!$&
$ 4.2 \!
\cdot\! 10^{-4}\!\!$&$3.2 \!\cdot\! 10^{-4}\!\!$&$3.3 
\!\cdot\! 10^{-5}\!\!$ \\ 
\hline
$\Delta a_0^2\rule{0em}{1em}$&&
$ 9.7\!\cdot\!10^{-7}\!\!$&$\! 1.6\!\cdot\! 10^{-4}\!\!$&
$\!\! -1.2 \!\cdot\! 10^{-5}\!\!$&
$ -4.2 \!\cdot\! 10^{-6}\!\!$&$\!\!-2.2 \!\cdot\!10^{-4}\!\!$&
$\!\!-2.2 \!\cdot\!10^{-5}\!\!$\\
\hline
$\Delta\lbar_1\rule{0em}{1em}$&
&$ $&$\! 3.5\!\cdot\! 10^{-1}\!\!$&$\!\! -3.3 \!\cdot\! 10^{-2}\!\!$&
$ 6.7\!\cdot\! 10^{-2}\!\!$&$\!\! -5.4 \!\cdot\! 10^{-1}\!\!$&
$\!\! -3.7 \!\cdot\! 10^{-2}\!\!$\\ 
\hline
$\Delta\lbar_2\rule{0em}{1em}$&
&$ $&$ $&$ 1.2\!\cdot\! 10^{-2}\!\!$&$\!\! -7.2\!\cdot\! 10^{-3}\!\!$&
$ 1.1 \!\cdot\! 10^{-2}\!\!$&$\!\! -4.6 \!\cdot\! 10^{-3}\!\!$ \\ 
\hline
$\Delta\lbar_4\rule{0em}{1em}$&
&$ $&$ $&$ $&$ 4.8\!\cdot\! 10^{-2}\!\!$&$\!\! -9.1\!\cdot\! 10^{-2}\!\!$&
$\!\! -2.2 \!\cdot\! 10^{-3}\!\!$\\ 
\hline
$\Delta\rtilde_5\rule{0em}{1em}$&
&$ $&$ $&$ $&$ $&$ 1.1\rule{1.2em}{0em}$&$\g 9.2\!\cdot\! 10^{-2}\!\!$\\
\hline
$\Delta\rtilde_6\rule{0em}{1em}$&
&$ $&$ $&$ $&$ $&$ $&$\g 1.1 \!\cdot\! 10^{-2}\!\!$\\
\hline
\end{tabular}
\caption{\label{tab:correlation matrix} Numerical elements
of the correlation matrix (F.2).}
\end{table}

The elements of the resulting correlation matrix are listed in table
\ref{tab:correlation matrix}. The
off--diagonal elements are of interest only if their numerical values are 
comparable to the product of the square roots 
of the corresponding diagonal entries -- the numbers listed are significant 
only insofar as this condition is met (those below the
diagonal are omitted -- the matrix is symmetric). The errors quoted in the
various rows of table \ref{tab:fix point} 
are the square roots of the corresponding contributions to the diagonal 
elements of the correlation matrix.

If all variables except $a_0^0$ and $a_0^2$ are integrated out, the 
distribution reduces to a Gaussian in these two variables:
\bea dp=n \exp (-\mbox{$\frac{1}{2}$}\,q)\;d a_0^0\, d a_0^2\co\hspace{1em}
 q=c_{11}\,(\Delta a_0^0)^2 + 2\,c_{12}\,\Delta a_0^0\Delta a_0^2+
c_{22}\,(\Delta a_0^0)^2\co\nonumber\eea
where the $2\times 2$ matrix $c$ is the inverse of the submatrix of $K$ that
collects the correlations of $a_0^0$ and $a_0^2$.
The result is illustrated in fig.~\ref{fig:aellipse}: The small ellipse
shows the standard 68\% confidence limit, that is the contour where $q=1$.
For a careful analysis of the errors and correlations associated with the
various numerical evaluations found
in the literature, we refer to \cite{Nieves:1999zb}, where the 
corresponding error ellipses are also shown. 
Note that the radiative corrections in the value of
$F_\pi$ are often not accounted for. At the accuracy under discussion, these 
matter, as they increase the results for the $S$--wave
scattering lengths by about two percent.


\begin{thebibliography}{99}


\bibitem{Weinberg Physica}
S.~Weinberg,
Physica A {\bf 96} (1979) 327.


\bibitem{GMOR}
M.~Gell-Mann, R.~J.~Oakes and B.~Renner,
Phys.\ Rev.\ {\bf 175} (1968) 2195.

\bibitem{Weinberg 1966}
S.~Weinberg,
Phys.\ Rev.\ Lett.\ {\bf 17} (1966) 616.

\bibitem{GL 1983}
J.~Gasser and H.~Leutwyler,
Phys.\ Lett.\ B {\bf 125} (1983) 325.

\bibitem{BCEGS}
J.~Bijnens, G.~Colangelo, G.~Ecker, J.~Gasser and M.~E.~Sainio,
Phys.\ Lett.\ B {\bf 374} (1996) 210
[hep-ph/9511397].

\bibitem{ACGL}
B.~Ananthanarayan, G.~Colangelo, J.~Gasser and H.~Leutwyler,
hep-ph/0005297, Phys. Rep., in press.

\bibitem{Roy}
S.~M.~Roy,
Phys.\ Lett.\ B {\bf 36} (1971) 353.

\bibitem{CGL}
G.~Colangelo, J.~Gasser and H.~Leutwyler,
Phys.\ Lett.\  B {\bf 488} (2000) 261
[hep-ph/0007112].

\bibitem{GL 1984}
J.~Gasser and H.~Leutwyler,
Annals Phys.\ {\bf 158} (1984) 142.

\bibitem{Donoghue Ramirez Valencia pipi}
J.~F.~Donoghue, C.~Ramirez and G.~Valencia,
Phys.\ Rev.\ D {\bf 38} (1988) 2195.

\bibitem{Borges} 
J.~Sa Borges,
Phys.\ Lett.\ B {\bf 149} (1984) 21;\\
J.~Sa Borges, J.~Soares Barbosa and V.~Oguri,
Phys.\ Lett.\ B {\bf 393} (1997) 413;\\
I.~P.~Cavalcante and J.\ S\'{a} Borges,
hep-ph/0101037,
hep-ph/0101104.

\bibitem{Bijnens:1990mr}
J.~Bijnens,
Nucl.\ Phys.\ B {\bf 337} (1990) 635.

\bibitem{Riggenbach:1991zp}
C.~Riggenbach, J.~Gasser, J.~F.~Donoghue and B.~R.~Holstein,
Phys.\ Rev.\ D {\bf 43} (1991) 127.

\bibitem{Pennington Portoles}
M.~R.~Pennington and J.~Portol\'es,
Phys.\ Lett.\ B {\bf 344} (1995) 399
[hep-ph/9409426].

\bibitem{ATW}
B.~Ananthanarayan, D.~Toublan and G.~Wanders,
Phys.\ Rev.\ D {\bf 51} (1995) 1093 [hep-ph/9410302];
{\it ibid.} D {\bf 53} (1996) 2362
[hep-ph/9510254].

\bibitem{AB}
B.~Ananthanarayan and P.~B\"uttiker,
Phys.\ Rev.\ D {\bf 54} (1996) 1125 [hep-ph/9601285];
{\it ibid.}  D {\bf 54} (1996) 5501 [hep-ph/9604217];
Phys.\ Lett.\ B {\bf 415} (1997) 402 [hep-ph/9707305].

\bibitem{Wanders 1997a}
G.~Wanders,
proposal,''
Helv.\ Phys.\ Acta {\bf 70} (1997) 287
[hep-ph/9605436].

\bibitem{Wanders 1997b}
G.~Wanders,
Phys.\ Rev.\ D {\bf 56} (1997) 4328
[hep-ph/9705323].

\bibitem{Toublan}
D.~Toublan,
Phys.\ Rev.\ D {\bf 53} (1996) 6602
[hep-ph/9509217].

\bibitem{Ananthanarayan:1998hj}
B.~Ananthanarayan,
Phys.\ Rev.\  D {\bf 58} (1998) 036002 [hep-ph/9802338].

\bibitem{Bijnens Colangelo Talavera}
J.~Bijnens, G.~Colangelo and P.~Talavera,
JHEP {\bf 9805} (1998) 014
[hep-ph/9805389].

\bibitem{ABT_Ke4}
G.~Amoros, J.~Bijnens and P.~Talavera,
Nucl.\ Phys.\ B {\bf 585} (2000) 293
[hep-ph/0003258]; 
erratum: LU TP 00-11, Jan. 2001, to be published in Nucl. Phys. B.


\bibitem{Knecht Moussallam Stern Fuchs}
M.~Knecht, B.~Moussallam, J.~Stern and N.~H.~Fuchs,
Nucl.\ Phys.\ B {\bf 457} (1995) 513
[hep-ph/9507319];
{\it ibid.} B {\bf 471} (1996) 445 [hep-ph/9512404].

\bibitem{Girlanda:1997ed}
L.~Girlanda, M.~Knecht, B.~Moussallam and J.~Stern,
Phys.\ Lett.\  B {\bf 409} (1997) 461
[hep-ph/9703448].


\bibitem{roy-num}
J.~L.~Basdevant, J.~C.~Le Guillou and H.~Navelet,
Nuovo Cim.\  A {\bf 7} (1972) 363;\\
M.R.~Pennington and S.D.~Protopopescu,
Phys.\ Rev.\  D {\bf 7} (1973) 1429; {\it ibid.} D {\bf 7} 2591;\\
J.~L.~Basdevant, C.~D.~Froggatt and J.~L.~Petersen,
Phys.\ Lett.\  B {\bf 41} (1972) 173; {\it ibid.} B {\bf 41} 178;
Nucl.\ Phys.\ B {\bf 72} (1974) 413.

\bibitem{Froggatt:1977hu}
C.~D.~Froggatt and J.~L.~Petersen,
Nucl.\ Phys.\ B {\bf 129} (1977) 89.

\bibitem{Nagels}
M.~M.~Nagels {\it et al.},
Nucl.\ Phys.\ B {\bf 147} (1979) 189.

\bibitem{rosselet}
L.~Rosselet {\it et al.},
Phys.\ Rev.\  D {\bf 15} (1977) 574.

\bibitem{e865}
J.~Lowe, in  ref.~\cite{mainz}, p.~375, 
and hep-ph/9711361; 
 in: Proccedings of the {\it Workshop on Physics and Detectors for}
 {DA$\Phi$NE},  Frascati, Nov.~16-19, 1999, p.439 
[http://wwwsis.lnf.infn.it/talkshow/dafne99.htm];\\
S.~Pislak {\it et al.}, ``A new measurement of $K^+\rightarrow
  \pi^+\pi^-e^+\nu$ ($K_{e4}$)'', talk given by S. Pislak 
at  Laboratori Nazionali di
  Frascati, June 22, 2000.

\bibitem{Truol}P.~Truol {\it et al.}  [E865 Collaboration],
hep-ex/0012012.

\bibitem{mainz}
A.M. Bernstein, D. Drechsel and T. Walcher, editors,
{\it Chiral Dynamics: Theory and Experiment}, Workshop held in Mainz,
  Germany, 1-5 Sept.~1997, Lecture Notes in Physics Vol. 513, Springer, 1997.

\bibitem{BCG}
J.~Bijnens, G.~Colangelo and J.~Gasser,
Nucl.\ Phys.\ B {\bf 427} (1994) 427
[hep-ph/9403390].

\bibitem{Batley}
R.~Batley et al. [NA48 Collaboration], CERN/SPSC 2000-003.

\bibitem{Nemenov}
B.~Adeva {\it et al.}, CERN proposal CERN/SPSLC 95-1 (1995);
 available at  http://dirac.web.cern.ch/DIRAC/.

\bibitem{GLR} 
J.~Gasser, V.~E.~Lyubovitskij and A.~Rusetsky,
Phys.\ Lett.\ B {\bf 471} (1999) 244
[hep-ph/9910438];\\
H.~Sazdjian,
Phys.\ Lett.\ B {\bf 490} (2000) 203
[hep-ph/0004226].\\
The literature on the subject may be traced from these references.

\bibitem{Triumf} 
M.~Kermani {\it et al.}  [CHAOS Collaboration],
Phys.\ Rev.\ C {\bf 58} (1998) 3419;
{\it ibid.} C {\bf 58} (1998) 3431.

\bibitem{Holstein}
B.~R.~Holstein,
Phys.\ Lett.\ B {\bf 244} (1990) 83.


\bibitem{EGPdeR}
G.~Ecker, J.~Gasser, A.~Pich and E.~de Rafael,
Nucl.\ Phys.\ B {\bf 321} (1989) 311.


\bibitem{Bijnens Colangelo Ecker}J.~Bijnens, G.~Colangelo and G.~Ecker, 
JHEP {\bf 9902} (1999) 020
[hep-ph/9902437];
Annals Phys.\ {\bf 280} (2000) 100
[hep-ph/9907333].

\bibitem{DGL}
J.~F.~Donoghue, J.~Gasser and H.~Leutwyler,
Nucl.\ Phys.\ B {\bf 343} (1990) 341.


\bibitem{Amendolia}
S.~R.~Amendolia {\it et al.}  [NA7 Collaboration],
Nucl.\ Phys.\  B {\bf 277} (1986) 168.

\bibitem{GL 1985}
J.~Gasser and H.~Leutwyler,
Nucl.\ Phys.\ B {\bf 250} (1985) 465.


\bibitem{Hannah:1997ux}
T.~Hannah,
Phys.\ Rev.\  D {\bf 55} (1997) 5613
[hep-ph/9701389]. 


\bibitem{Nieves:1999zb}
J.~Nieves and E.~Ruiz Arriola,
Eur.\ Phys.\ J.\ A {\bf 8} (2000) 377
[hep-ph/9906437].

\bibitem{olsson sum rule}
M.~G.~Olsson, 
Phys.\ Rev.\ {\bf 162} (1967) 1338.


\bibitem{CGL letter2}
G.~Colangelo, J.~Gasser and H.~Leutwyler
 [hep-ph/0103063].

\bibitem{e865final}Forthcoming paper by the E865 collaboration.

\bibitem{Ecker CD97}
G.~Ecker,
in  \cite{mainz}, p. 337-351 [hep-ph/9710560].


\bibitem{hyams}       
B.~Hyams {\it et al.},
Nucl.\ Phys.\  B {\bf 64} (1973) 134.

\bibitem{Protopopescu}
S.~D.~Protopopescu {\it et al.},
Phys.\ Rev.\  D {\bf 7} (1973) 1279.


\bibitem{hoogland}       
W.~Hoogland {\it et al.},
Nucl.\ Phys.\  B {\bf 126} (1977) 109.

\bibitem{losty}        
M.~J.~Losty {\it et al.},
Nucl.\ Phys.\  B {\bf 69} (1974) 185.

\bibitem{Takamatsu:2001ew}
K.~Takamatsu  [Sigma Collaboration],
data,''
Prog.\ Theor.\ Phys.\ {\bf 102} (2001) E52
[hep-ph/0012324].


\bibitem{GAMS}
D.~Alde {\it et al.}  [GAMS Collaboration],
Phys.\ Lett.\ B {\bf 397} (1997) 350.\\
R.~Bellazzini {\it et al.}  [GAMS Collaboration],
interactions at 450-GeV/c,''
Phys.\ Lett.\ B {\bf 467} (1999) 296.

\bibitem{Schenk}
A.~Schenk,
Nucl.\ Phys.\ B {\bf 363} (1991) 97.


\bibitem{Pich Portoles}
A.~Pich and J.~Portol\'es,
hep-ph/0101194.


\bibitem{CLEO}
S.~Anderson {\it et al.}  [CLEO Collaboration],
hep-ex/9910046.

\bibitem{Kuehn Santamaria}
J.~H.~K\"uhn and A.~Santamaria,
Z.\ Phys.\ C {\bf 48} (1990) 445.

\bibitem{PDG}
D.~E.~Groom {\it et al.},
Eur.\ Phys.\ J.\  C {\bf 15} (2000) 1.


\bibitem{sigma}
S. Ishida et al., editors,
{\it Possible Existence of the $\sigma$-Meson  and its Implications to
Hadron Physics}, workshop held in Kyoto, Japan, 12-14 June~2000,
Soryushiron Kenkyu Volume 102 No.~5 (2001). Reprint available at
http://amaterasu.kek.jp/YITPws/online/index.html.

\bibitem{Oset}
E.~Oset, H.~Toki, M.~Mizobe and T.~T.~Takahashi,
Prog.\ Theor.\ Phys.\ {\bf 103} (2000) 351
[nucl-th/0011008].

\bibitem{Wanders sum rules}
G.~Wanders, Helv.\ Phys.\ Acta {\bf 39} (1966) 228.


\bibitem{Lubicz:2001ch}
V.~Lubicz,
Nucl.\ Phys.\ Proc.\ Suppl.\ {\bf 94} (2001) 116
[hep-lat/0012003].


\bibitem{Buergi}U.~B\"urgi,
Nucl.\ Phys.\ B {\bf 479} (1996) 392
[hep-ph/9602429].

\bibitem{Colangelo double logs}G.\ Colangelo,
Phys.\ Lett.\ B {\bf 350} (1995) 85 [hep-ph/9502285];
B {\bf 361} (1995) 234 (E).


\bibitem{Leutwyler Bangalore}H.~Leutwyler,
Nucl.\ Phys.\ Proc.\ Suppl.\ {\bf 94} (2001) 108
[hep-ph/0011049].


\bibitem{Heitger:2000ay}
J.~Heitger, R.~Sommer and H.~Wittig  [ALPHA Collaboration],
Nucl.\ Phys.\ B {\bf 588} (2000) 377
[hep-lat/0006026].

\bibitem{qCHPT}
S.~R.~Sharpe,
Phys.\ Rev.\ D {\bf 41} (1990) 3233.
Phys.\ Rev.\ D {\bf 46} (1992) 3146
[hep-lat/9205020].\\
C.~W.~Bernard and M.~F.~Golterman,
Phys.\ Rev.\ D {\bf 46} (1992) 853
[hep-lat/9204007].

\bibitem{CP}
G.~Colangelo and E.~Pallante,
Phys.\ Lett.\ B {\bf 409} (1997) 455
[hep-lat/9702019]; 
Nucl.\ Phys.\ B {\bf 520} (1998) 433
[hep-lat/9708005].

\bibitem{Colangelo 1997}G.~Colangelo, 
Phys.\ Lett.\ B {\bf 395} (1997) 289
[hep-ph/9607205].

\bibitem{lattice I2}
S.~Aoki {\it et al.}  [JLQCD Collaboration],
Nucl.\ Phys.\ Proc.\ Suppl.\ {\bf 83} (2000) 241
[hep-lat/9911025].

\bibitem{BG}
C.~W.~Bernard and M.~F.~Golterman,
Phys.\ Rev.\ D {\bf 53} (1996) 476
[hep-lat/9507004].

\bibitem{luscher}
M.~Luscher,
Commun.\ Math.\ Phys.\ {\bf 105} (1986) 153.

\bibitem{isospin breaking}
K.~Maltman and C.~E.~Wolfe,
Phys.\ Lett.\ B {\bf 393} (1997) 19
[nucl-th/9610051], ibid. B {\bf 424} (1998) 413.\\
U.~Meissner, G.~Muller and S.~Steininger,
Phys.\ Lett.\ B {\bf 406} (1997) 154
[hep-ph/9704377]; {\it ibid.} B {\bf 407} (1997) 454.\\
M.~Knecht and R.~Urech,
Nucl.\ Phys.\ B {\bf 519} (1998) 329
[hep-ph/9709348].

\bibitem{GLR1}
A.~Gall {\it et al.}, in preparation.


\end{thebibliography}
\end{document}